\documentclass[12pt,twoside]{article}
\usepackage{amssymb}
\usepackage{amsmath}
\usepackage{latexsym}
\usepackage{longtable}
\usepackage{epsfig}
\usepackage{graphicx,bbm,psfrag}
\graphicspath{{images/}}

\newtheorem{theorem}{Theorem}[section]

\newtheorem{proposition}[theorem]{Proposition}
\begin{document}
\begin{center}
{\Large {\bf The Phonon Boltzmann
Equation, Properties and\medskip\\Link to Weakly Anharmonic Lattice Dynamics\bigskip\bigskip\\}}
{\large{Herbert Spohn}}\footnote{{\tt spohn@ma.tum.de}}\medskip\\
Zentrum Mathematik and Physik Department, TU M\"{u}nchen,\\
D - 85747 Garching, Boltzmannstr. 3, Germany
\end{center}\bigskip\bigskip\bigskip\bigskip
\setcounter{figure}{0}
%%%%%%%%%%%%%%%%%%%%%%%%%%%%
{\bf Abstract:} For low density gases the validity of the
Boltzmann transport equation is well established. The central
object is the one-particle distribution function, $f$, which in
the Boltzmann-Grad limit satisfies the Boltzmann equation. Grad
and, much refined, Cercignani argue for the existence of this
limit on the basis of the BBGKY hierarchy for hard spheres. At
least for a short kinetic time span, the argument can be made
mathematically precise following the seminal work of Lanford. In
this article a corresponding program is undertaken for weakly
nonlinear, both discrete and continuum, wave equations. Our
working example is the harmonic lattice with a weakly nonquadratic
on-site potential. We argue that the role of the Boltzmann
$f$-function is taken over by the Wigner function, which is a very
convenient device to filter the slow degrees of freedom. The
Wigner function, so to speak, labels locally the covariances of
dynamically almost stationary measures. One route to the phonon
Boltzmann equation is a Gaussian decoupling, which is based on the
fact that the purely harmonic dynamics has very good mixing
properties. As a further approach the expansion in terms of
Feynman diagrams is outlined. Both methods are extended to the
quantized version of the weakly nonlinear wave equation.

The resulting phonon Boltzmann equation has been hardly studied on
a rigorous level. As one novel contribution we establish that the
spatially homogeneous stationary solutions are precisely the
thermal Wigner functions. For three phonon processes such a result
requires extra conditions on the dispersion law. We also outline
the reasoning leading to Fourier's law for heat conduction.
\newpage
\tableofcontents

\section{Goals and Introduction}\label{sec.1}
\setcounter{equation}{0}

Dielectric crystals, as Si and GaAs, have their electronic bands
completely filled and separated by an energy gap from the
conduction band. Therefore electronic energy transport is
suppressed and the dominant contribution to heat transport is due
to the vibrations of the atoms around their mechanical equilibrium
position. Below room temperature these deviations are small,
typically only a few percent of the lattice constant, hence by
necessity weakly anharmonic. As envisioned by R.~Peierls in 1929
\cite{Pe}, the obvious theoretical option is to regard the
anharmonicities as a, in a certain sense, small perturbation to
the perfectly harmonic crystal, which at the very end leads to a
kinetic description of an interacting ``gas of phonons" in terms
of a nonlinear Boltzmann transport equation. The actual
computation of the thermal conductivity of dielectric crystals is
then based on the phonon Boltzmann equation. Through the work of
many, for example see \cite{Le,Gu,Sr,Ca}, it has become apparent
that such a program can be made to work resulting in a reliable
prediction over a considerable temperature range.  Only recently
the kinetic description has been augmented by molecular dynamics,
which numerically solves the classical equations of motion, see
for example \cite{GaKa}. To determine the thermal conductivity one
computes either the Green-Kubo formula in an equilibrium system at
a fixed temperature or the average energy flux in the steady state
with a temperature difference imposed at the boundaries.

In this note I focus on the step from the weakly anharmonic
lattice dynamics to the kinetic equation. As an aside, I discuss a
few basic properties of the phonon Boltzmann equation, mostly to
provide some indication on the physics which persists on the
kinetic level but also to advertise an evolution equation which
apparently has received little attention.

\begin{figure}
  \begin{center}
 % \subfigure[]{
  \includegraphics[width=6.5cm,height=7cm]{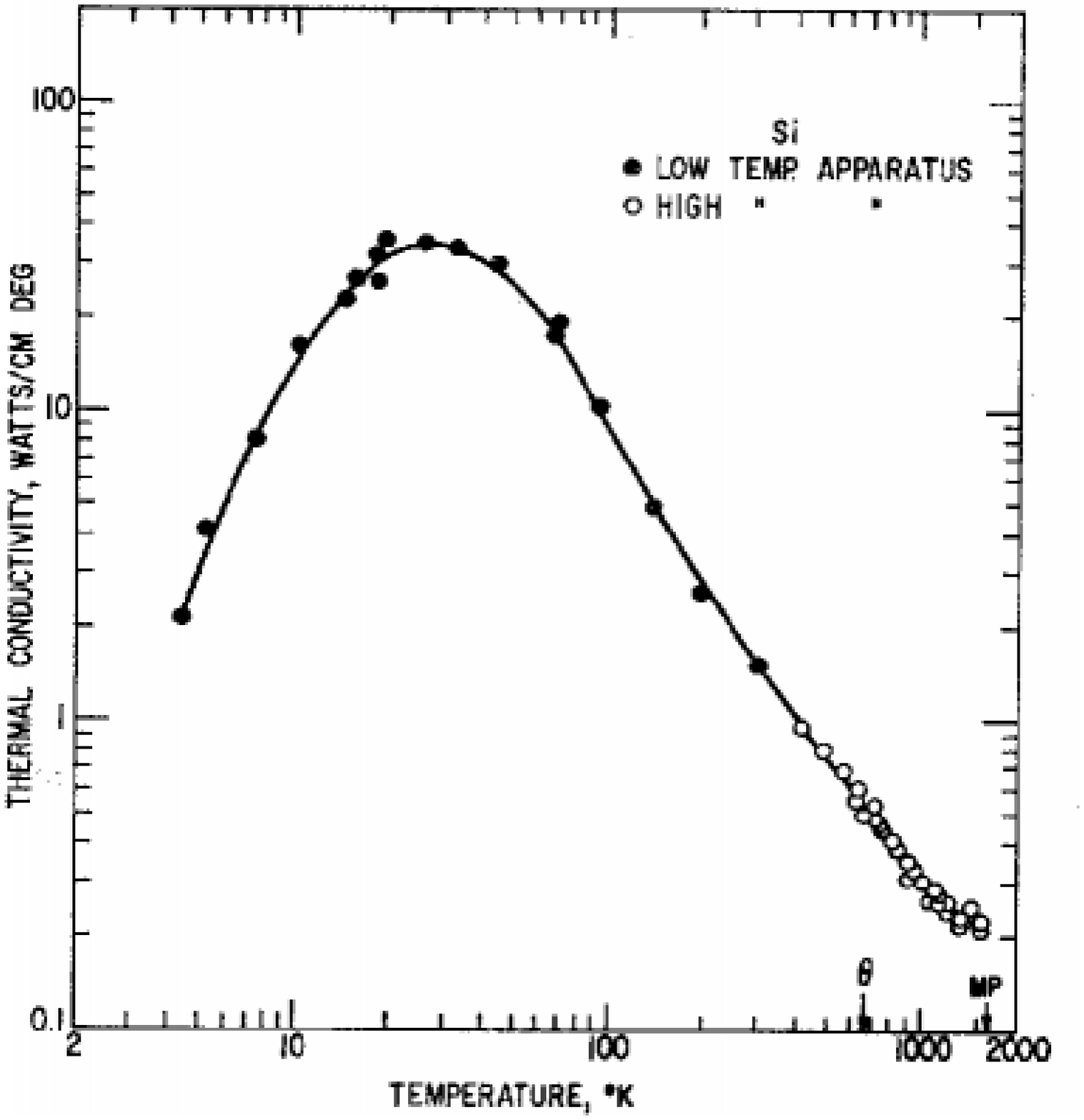}\hfill
 % \subfigure[]{
  \includegraphics[width=6.5cm,height=7cm]{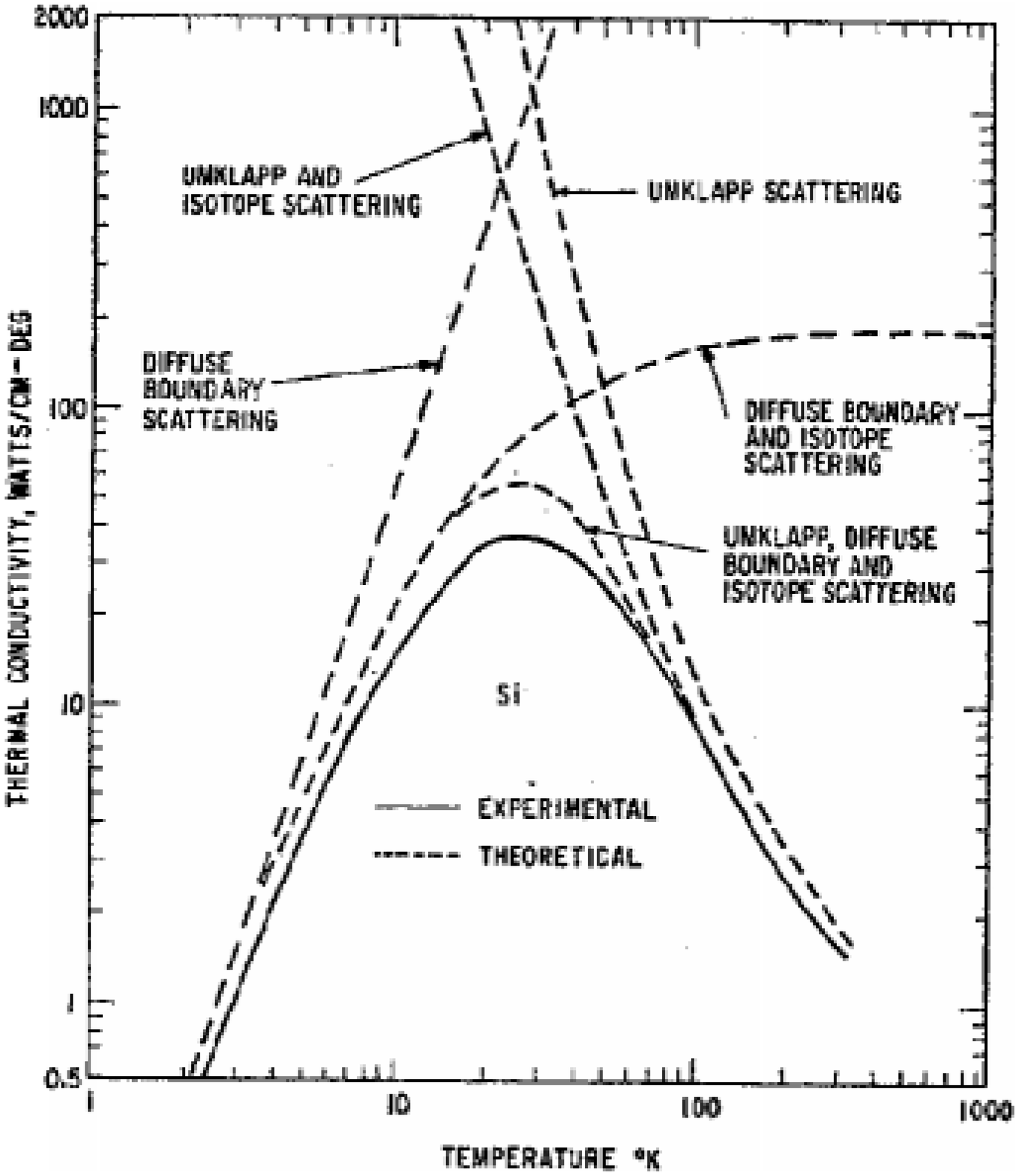}
  \caption{Thermal conductivity of Si (natural abundance)\cite{GlSl}.}\label{Fig1}
  \end{center}
\end{figure}

If the goal is to compute the thermal conductivity of real
crystals, the derivation of the Boltzmann equation is considered
as a minor issue, where the emphasis varies from author to author.
Much more relevant is to have reliable information on the lattice
structure, on the phonon dispersion law, and on the lowest order
anharmonic elastic constants. Furthermore, on the kinetic level
the conductivity is determined through the inverse of the
linearized collision operator, which cannot be computed by hand.
Hence suitable approximation schemes had to be developed. I will
have nothing to say on these topics.

On a qualitative level kinetic theory provides a rather simple
picture for the temperature dependence of the thermal
conductivity, $\kappa(T)$. At ``high'' temperatures a
semiclassical approximation suffices, which predicts
$\kappa(T)=\theta_\textrm{h}/T$ with some temperature independent
coefficient $\theta_\textrm{h}$. At ``low'' temperatures the
quantization of lattice vibrations must be taken into account. The
total number of phonons then equals $\int
d^3k(e^{\omega(k)/k_\textrm{B}T}-1)^{-1}$ which reflects the
freezing of the number of energy carriers as $T\to 0$. On the
other hand also momentum nonconserving collisions become rare,
resulting in a phonon mean free path which diverges as $T\to 0$.
This latter effect dominates and yields the prediction
$\kappa(T)=e^{\theta_\textrm{l}/T}$, $\theta_\textrm{l}>0$, as
$T\to 0$. Experimentally such a behavior is masked by the finite
size of the sample and only over a narrow temperature range the
exponential increase in $1/T$ can be seen. A crucial point in the
experiment is to manufacture a crystal which has no dislocations
and is free of impurities. Even then, isotope disorder provides an
additional mechanism for diffusive energy transport, which
persists in the harmonic approximation. E.g., for Si the natural
abundance is $^{28}$Si 92.23\%, $^{29}$Si 4.76\%, and  $^{30}$Si
3.01\%, which means that the deviation from the perfect constant
atomic mass crystal can be considered as small.

To provide an example we reproduce in Figure \ref{Fig1} the thermal conductivity
for chemically pure and dislocation free Si as measured by
Glassbrenner and Slack \cite{GlSl}. On the right hand side the
importance of the various scattering mechanisms is displayed.
Above 100$^\circ$K one notes the classical 1/T-behavior. Below
100$^\circ$K the quantization of phonons becomes relevant. Diffuse
boundary scattering reflects the size of the probe which is 2 cm
long times 0.44 cm as average diameter. The umklapp scattering
refers to momentum nonconserving collisions, see Section
\ref{sec.4}. The experimental findings are well reproduced by the
theory \cite{Sr}, which is based on the linearized Boltzmann
equation, as will be explained in Section \ref{sec.11}.

In the kinetic theory of gases the central object is the Boltzmann
distribution function $Nf(r,v,t)$, $N$ the total number of
particles, which counts the number of gas molecules in the volume
element $d^3r d^3v$ in the one-particle phase space close to $r,v$
at time $t$. Phonons are not such local objects. In fact, upon
specifying the complete displacement field, including its
velocities, it is not so clear how to extract from it the
positions and momenta of the particle-like objects called phonons.
Most likely, for a general displacement field no such procedure
can be devised. Still in the kinetic limit the mechanical picture
becomes precise. As has been recognized for some time
\cite{Fu,RyPa}, the link between a wave field and transport
equations allowing for a mechanical interpretation is provided by
the Wigner function. This approach will be followed also in these
notes, noting already now that the collision between phonons,
while they conserve energy and momentum, are otherwise unlike
collisions between mechanical point particles.

For the purpose of a better understanding of the validity of the
kinetic description, my guiding principle is to discard all
details and to devise the arguably simplest of all models, which
still displays the same physics. I will even go as far as to
ignore the obvious fact that atoms deviate in three-space from
their equilibrium position. Hence I will assume that the
displacement field is \textit{scalar}. The virtue, I hope, is to
make the derivation of the transport equation maximally
transparent.

We propose to ignore quantization in the first round. One reason
is the hope that for a classical model techniques different from a
hierarchy of correlation functions and Feynman diagrams might
become available. As a further bonus, we establish the link to
weakly anharmonic, in general multicomponent, wave equations,
which are applied in the wave dynamics of the upper ocean, in
acoustic turbulence, and in other areas \cite{Za}. In this context
the phenomenon of interest is a turbulent state maintained through
external forcing. Again, kinetic theory is the natural theoretical
tool to explain and predict properties of the steady state.\medskip\\
{\bf Acknowledgements}: I am most grateful to Jani Lukkarinen for many
instructive discussions and a first reading of the notes. I thank Carlo
Cercignani for help towards the H-theorem and Eric Carlen for discussions on the Brout-Prigogine equation. These notes were
first presented as lectures at the workshop ``Quantum Dynamics and
Quantum Transport'', Warwick, September 6 - 12, 2004. I am
grateful to Gero Friesecke for this opportunity.

%%%%%%%%%%%%%%%%%%%%%%%%%

\section{A real crystal simplified}\label{sec.2}
\setcounter{equation}{0}

We consider the simple cubic lattice $\mathbb{Z}^3$ as the lattice
of mechanical equilibrium positions of the crystal atoms. The
deviations from their equilibrium position are denoted by
\begin{equation}\label{2.1}
q_x \in \mathbb{R}\,, \quad x \in \mathbb{Z}^3\,,
\end{equation}
with the canonically conjugate momenta
\begin{equation}\label{2.2}
p_x \in \mathbb{R}\,, \quad x \in \mathbb{Z}^3\,.
\end{equation}
We will use units in which the mass $m$ of an atom equals one. For
small deviations from the equilibrium position we may use the
harmonic approximation in lowest order. The corresponding
potential energy then reads
\begin{equation}\label{2.2a}
U_\textrm{harm}(q)=  \frac{1}{2} \sum_{x,y \in \mathbb{Z}^3}
\alpha(x-y)q_x q_y\,.
\end{equation}
The elastic constants $\alpha(x)$ satisfy
\begin{equation}\label{2.2b}
\alpha(x)=\alpha(-x)\,,\quad
|\alpha(x)|\leq\alpha_0e^{-\alpha_1|x|}
\end{equation}
for suitable  $\alpha_0,\alpha_1>0$, and
\begin{equation}\label{2.2d}
\sum_{x\in \mathbb{Z}^3}\alpha(x)=0
\end{equation}
because of the invariance of the interaction between the crystal
atoms under the translation $q_x\rightsquigarrow q_x+a$. Mechanical
stability requires
\begin{equation}\label{2.2e}
\widehat{\alpha}(k)>0\quad\textrm{for}\quad k\neq 0
\end{equation}
for the Fourier transform $\widehat{\alpha}$ of $\alpha$.

The anharmonicity is assumed to reside only in the on-site
potential which we divide into a harmonic piece and the rest
\begin{equation}\label{2.2f}
U_\textrm{site}=  \sum_{x \in \mathbb{Z}^3}
\big(\frac{1}{2}\omega^2_0 q^2_x +V_\textrm{an}(q_x)\big)\,.
\end{equation}\smallskip\\
Physically, the on-site potential is artificial and it would be
more natural to assume that the atoms are coupled through a weakly
anharmonic pair potential. As we will argue below, in the kinetic
limit only the collision rate turns out to be modified. Thus, for
the purpose of deriving the kinetic equation, we might as well
stick to the somewhat simpler on-site potential.

The Hamiltonian of the anharmonic lattice system is written as the
sum
\begin{equation}\label{2.3}
H=H_0+V\,.
\end{equation}
$H_0$ is the harmonic piece given through
\begin{equation}\label{2.4}
H_0= \frac{1}{2} \sum_{x \in \mathbb{Z}^3} \Big( p_x^2+
\omega_0^2q_x^2\Big)
+\frac{1}{2}\sum_{x,y\in\mathbb{Z}^3}\alpha(x-y)q_xq_y \,,
\end{equation}
$\omega_0 > 0$. The lowest order type of anharmonicity reads
\begin{equation}\label{2.6}
V= \sum_{x \in \mathbb{Z}^3} V_\textrm{an}(q_x)\,\quad
V_\textrm{an}(q_x)= \lambda \frac{1}{3}q^3_x
\end{equation}
with $\lambda$ small. The potential energy
$U_\textrm{harm}+U_\textrm{site}$ is then not bounded from below,
which however will not be visible on the kinetic time scale. If
preferred, one could add to $V_\textrm{an}$ the quartic term
$\lambda' q_x^4$ with $\lambda' =\lambda^2/18\omega_0^2$. Then $H
\geq 0$ and the quartic term disappears in the kinetic scaling.
For reasons of readability we will set $\lambda' =0$.

We work in the physical space dimension. Whether the kinetic
approximation is valid in one and two dimensions remains debated.
On the other hand only for such low dimensional systems extensive
numerical results are available, to which we will turn in Section
\ref{sec.14}.

The equations of motion are
\begin{eqnarray}\label{2.7}
&&\frac{d}{dt}q_x (t) = p_x (t)\,,\nonumber\\
&&\frac{d}{dt}p_x (t) = -\sum_{y\in\mathbb{Z}^3}\alpha(y-x)q_y(t)
- \omega_0^2 q_x (t)- \lambda q_x (t)^2\,, \quad x \in
\mathbb{Z}^3\,.
\end{eqnarray}
We will consider only finite energy solutions. In particular, it
is assumed that $|p_x|\to 0$, $|q_x|\to 0$ sufficiently fast as
$|x|\to \infty$. In fact, later on there will be the need to
impose random initial data, which again are assumed to be
supported on finite energy configurations. As to be explained in
great detail, in the kinetic limit the average energy diverges
suitable linked to the nonlinearity $\lambda\to 0$.

We will mostly work in Fourier space and set up the notation. Let
$\mathbb{T}^3 = [-\frac{1}{2},\frac{1}{2}]^3$ be the first Brillouin zone of the dual lattice. For
$f:\mathbb{Z}^3\to \mathbb{R}$ we use the following convention
for the Fourier transform,
\begin{equation}\label{2.2g}
\widehat{f}(k)=\sum_{x\in\mathbb{Z}^3}e^{-i 2\pi k\cdot
x}f_x\,,\quad k\in \mathbb{T}^3\,.
\end{equation}
$\widehat{f}(k)$ extends to a $2\pi$-periodic function on $\mathbb{R}^3$. 
The inverse Fourier transform is given by
\begin{equation}\label{2.2h}
f_x= \int_{\mathbb{T}^3}dk e^{i 2\pi k\cdot x}\widehat{f}(k)\,,
\end{equation}
where $dk$ is the 3-dimensional Lebesgue measure. This convention
has the advantage of maximally avoiding prefactors of $2\pi$. The
dispersion relation for the harmonic part $H_0$ is easily computed
as
\begin{equation}\label{2.2i}
\omega(k)=\big(\omega^2_0 + \widehat{\alpha}(k)\big)^{1/2}\,.
\end{equation}
By mechanical stability $\omega(k)\geq\omega_0$. If $\omega_0>0$,
then $\omega$ is a real analytic function on $\mathbb{T}^3$. If
$\omega_0=0$, $\omega$ may still be real analytic, one example
being $\widehat{\alpha}(k)\simeq|k|^4$ for small $k$. In Fourier
space the equations of motion become
\begin{eqnarray}\label{2.10}
&&\hspace{-26pt}\frac{\partial}{\partial t}\widehat{q}(k,t) = \widehat{p}(k,t)\,,\nonumber\\
&&\hspace{-26pt}\frac{\partial}{\partial t}\widehat{p}(k,t) = -
\omega (k)^2
\widehat{q}(k,t)\nonumber\\
&&\hspace{36pt}-\lambda \int_{\mathbb{T}^6}dk_1 dk_2
\delta(k-k_1-k_2) \widehat{q}(k_1,t) \widehat{q}(k_2,t)
\end{eqnarray}
with $k\in \mathbb{T}^3$. Here $\delta$ is the $\delta$-function
on the unit torus, to say, $\delta (k')$ carries a point mass
whenever $k'\in \mathbb{Z}^3$.

It will be convenient to concatenate $q_x$ and $p_x$ into a single
complex-valued field. We set
\begin{equation}\label{2.11}
a(k)= \frac{1}{\sqrt{2}} \big(\sqrt{\omega(k)}\, \widehat{q}(k) + i
\frac{1}{\sqrt{\omega(k)}}\, \widehat{p}(k)\big)
\end{equation}
with the inverse
\begin{equation}\label{2.12}
\widehat{q}(k) = \frac{1}{\sqrt{2}}\frac{1}{\sqrt{\omega(k)}}\big(
a(k) + a(-k)^\ast\big)\,,\; \widehat{p}(k) = \frac{1}{\sqrt{2}} i
{\sqrt{\omega(k)}}\big(-a(k) + a(-k)^\ast \big)\,.
\end{equation}
The $a$-field evolves as
\begin{eqnarray}\label{2.13}
&&\hspace{-10pt}\frac{\partial}{\partial t} a(k,t) = -i \omega
(k)a(k,t)- i \lambda \int_{\mathbb{T}^6}d k_1
d k_2  \delta(k-k_1-k_2)\nonumber\\
&&\hspace{-10pt}(8 \omega(k) \omega(k_1) \omega(k_2))^{-1/2}
(a(k_1,t) + a(-k_1,t)^\ast) (a(k_2,t) + a(-k_2,t)^\ast)\,.
\end{eqnarray}
In particular for $\lambda =0$,
\begin{equation}\label{2.14}
a(k,t) = e^{-i \omega(k)t} a(k)\,.
\end{equation}

For real crystals the $a$-field would be vector-valued for two
reasons: the displacements are in $\mathbb{R}^3$ and the unit cell
contains usually more than one atom. Correspondingly $\omega$ then
becomes a $k$-dependent matrix. Furthermore by translation
invariance the potential energy of the crystal depends only on the
differences $q_y-q_x$. As long as the interest is merely in the
derivation of the Boltzmann equation such extra features can be
ignored.

If one simplifies anyhow, the reader may wonder why we do not
switch to the continuum wave equation. In our context a natural
option would be the Klein-Gordon equation with a weak quadratic
nonlinearity,
\begin{equation}\label{2.14a}
\frac{\partial^2}{\partial t^2}\phi (x,t)=  \Delta
\phi(x,t)-\omega^2_0 \phi(x,t) -\lambda\phi(x,t)^2\,,\quad x\in
\mathbb{R}^3\,.
\end{equation}
Another possibility would be the standard wave equation with a
cubic nonlinearity
\begin{equation}\label{2.14b}
\frac{\partial^2}{\partial t^2}\phi (x,t)= \Delta
\phi(x,t)-\lambda\phi(x,t)^3\,,\quad x\in \mathbb{R}^3\,.
\end{equation}
We will discuss continuum equations in Section \ref{sec.7a}, from
which it will become clear that the underlying lattice structure
plays a crucial role.

Having agreed upon the basic model (\ref{2.7}) of our enterprise,
we have reached a point of bifurcation. Physically we should
quantize (\ref{2.3}), together with (\ref{2.4}), (\ref{2.6}),
according to the standard rules and then investigate the effects
small anharmonicities. On the other hand it seems to be worthwhile
not to hurry so much and to explore the classical model, which has
an interesting structure of its own. In addition there could be
help from the theory of nonlinear wave equations, which would put
our claims on firmer ground. Thus in Sections \ref{sec.3} to
\ref{sec.5} we treat the derivation of the Boltzmann equation for
the classical model. The same program is repeated for the
quantized crystal in Sections \ref{sec.7} and \ref{sec.8} with the
approach through Feynman diagrams explained in Section
\ref{sec.5a}. In Section \ref{sec.7a} we address wave turbulence
which is concerned with continuum wave equations such as
(\ref{2.14a}) and (\ref{2.14b}). Sections \ref{sec.6} and
\ref{sec.9} study properties of Boltzmann equation, in particular
the H-theorem. The nonlinear part concludes with a discussion of
the thermal conductivity. In the final part of our notes we
investigate the harmonic crystal with random isotope substitution.

%%%%%%%%%%%%%%%%%%%%%%%%%%%%%%%%%%%%
\section{Local stationarity, Wigner function}\label{sec.3}
\setcounter{equation}{0}

The kinetic theory of dilute gases relies on the scale separation
between typical interatomic distances and the mean free path. As a
consequence locally, in regions of linear size much larger than
atomic distances and much smaller than the mean free path, the
statistics of particles is Poisson in a good approximation. If
$f(r,v,t)$ denotes the Boltzmann distribution function at time
$t$, then close to $r$ the particles are uniformly distributed
with density $\rho(r) = \int d^3 v f(r,v,t)$ and their velocities
are independent with common distribution $f(r,v,t)/ \rho(r)$. The
Poisson distribution is singled out from all other conceivable
distributions, because it is stationary in time with respect to
the free gas dynamics, translation invariant in space, and has a
strictly positive entropy per unit volume. In fact, there are no
other such probability measures \cite{EySp}.

To transcribe this kinetic picture to weakly interacting phonons,
as building blocks we need, on the phase space of microscopic
configurations $\{q_x,p_x,\; x\in \mathbb{Z}^3\}$, probability
measures which are invariant under the free dynamics generated by
$H_0$, stationary under lattice shifts, and have a strictly
positive entropy per unit volume. The obvious candidates are
Gaussian measures with zero mean. By translation invariance, their
covariance reads
\begin{equation}\label{3.1}
\langle q_x q_y \rangle = Q(x-y)\,, \quad \langle p_x p_y \rangle
= P(x-y)\,,\quad \langle q_x p_y \rangle = C(x-y)\,.
\end{equation}
Let $\widehat{Q}$, $\widehat{P}$, $ \widehat{C}$ denote the
corresponding Fourier transforms. Then $\widehat{Q}(k) \geq 0$,
$\widehat{Q}(k) = \widehat{Q}(-k)$, $\widehat{P}(k) \geq 0$,
$\widehat{P}(k) = \widehat{P}(-k)$, $\widehat{C}(k) =
\widehat{C}(-k)^\ast$, and $|\widehat{C}|^2 \leq
\widehat{Q}\widehat{P}$. Stationarity in time yields in addition
the relations
\begin{equation}\label{3.2}
\widehat{P}= \omega^2\widehat{Q}\,, \quad
C(x)=-C(-x)\,,\quad\textrm{i.e.}\; \widehat{C}(k) = -
\widehat{C}(-k)\,.
\end{equation}
Such properties are more concisely expressed through the
$a$-field. Stationarity in space-time is equivalent to
\begin{equation}\label{3.3}
\langle a(k)\rangle = 0\,,\quad\langle a(k)a(k')\rangle = 0\,,
\quad \langle a(k)^\ast a(k')\rangle = W(k) \delta (k-k')\,.
\end{equation}
$W(k) \geq 0$ and, by convention, $W(k)$ is a $2\pi$-periodic function on 
 $\mathbb{R}^3$. Inserting the definition (\ref{2.11}) and comparing with
(\ref{3.2}) results in
\begin{eqnarray}\label{3.4}
&&\frac{1}{2}\big( W(k)+W(-k)\big) = \frac{1}{2}\big(\omega
\widehat{Q}(k)+ \frac{1}{\omega} \widehat{P}(k)\big) =\omega
\widehat{Q}(k)
=\frac{1}{\omega} \widehat{P}(k)\,,\nonumber\\
&&\frac{1}{2}\big( W(k)- W(-k)\big) = i \widehat{C}(k)\,.
\end{eqnarray}

The meaning of the covariance $W$ is grasped better by considering
expectations of some physical quantities. Let us first study the
local energy $H_x$, for which we equally divide the potential
energy between the two elastically coupled sites. Then
\begin{eqnarray}\label{3.5}
&&H_x = \frac{1}{2} p_x^2 +\frac{1}{2}\omega^2_0 q^2_x +
\frac{1}{2}\sum_{y\in\mathbb{Z}^3}\alpha(x-y)q_x q_y
\end{eqnarray}
and
\begin{equation}\label{3.6}
H_0 = \sum_{x \in \mathbb{Z}^3} H_x\,.
\end{equation}
Clearly
\begin{equation}\label{3.7}
\langle H_x \rangle = \int_{\mathbb{T}^3} dk \omega(k) W(k)\,.
\end{equation}
To probe further, we study the flow of energy out of a big box
$\Lambda\subset\mathbb{Z}^3$. Setting
$H_\Lambda=\sum_{x\in\Lambda}H_x$ one finds
\begin{equation}\label{3.8}
\frac{d}{dt}H_\Lambda=\frac{1}{2}
\sum_{x\in\Lambda}\sum_{y\in\mathbb{Z}^3 \setminus \Lambda}
\alpha(x-y)(-q_xp_y+q_yp_x)\,.
\end{equation}
Since the coupling is not only nearest neighbor, the division into
local currents is somewhat arbitrary. To be specific, let us
choose as one face of $\Lambda$ the coordinate plane
$\{x,x^1=0\}$. Then, in the limit $\Lambda\to\infty$, the
one-component of the energy current becomes
\begin{equation}\label{3.9}
j^1_{\mathrm{e}}= \frac{1}{2}\sum_{x^1\leq0}\sum_{y^1\geq1}
\alpha(x-y)(-q_{(x^1,0,0)}p_y +q_y p_{(x^1,0,0)})\,.
\end{equation}
Upon averaging, using (\ref{3.4}) and (\ref{3.9}),
\begin{equation}\label{3.10a}
\langle j_{\mathrm{e}}\rangle = \frac{1}{4\pi} \int_{\mathbb{T}^3}
dk \nabla \widehat{\alpha}(k) W(k)=\frac{1}{2\pi}
\int_{\mathbb{T}^3} dk (\omega\nabla \omega)(k) W(k)\,.
\end{equation}
Thus it is natural to regard $W$ as number density in wave number
space. $\omega W$ is the energy density and $(2\pi)^{-1}\nabla
\omega (\omega W)$ is the energy current density. Note that if $W$
is even, the total energy current vanishes.

A further important quantity is the entropy per unit volume, which
on general grounds is defined as the logarithm of the phase space
volume at prescribed values of the ``macrovariables", see Appendix
\ref{sec.16c} for further discussion. Here we use an equivalent
short-cut and compute the Gibbs entropy of the Gaussian measure
with covariance given through $W$. To do so let us choose the
periodic box $[1,\ell]^3$, $\ell$ integer, and consider the finite
volume analogue of the Gaussian measure from (\ref{3.3}). Then $k$
takes the discrete values $k\in(\ell^{-1}[1,...,\ell])^3$. Let
$\rho_\textrm{G}$ be the corresponding probability density. As
usual, the entropy of $\rho_\textrm{G}$ is given through
\begin{equation}\label{3.10}
S_\ell= -\int_{\mathbb{R}^{\ell^3}}d^{\ell^3}q
d^{\ell^3}p\rho_\textrm{G}\log\rho_\textrm{G} =
\sum_{k\in(\ell^{-1}[1,...,\ell])^3} \big(\log W(k) +
\log\pi+1\big)
\end{equation}
and thus the entropy per unit volume by
\begin{equation}\label{3.11}
\lim_{\ell\to \infty} \ell^{-3} S_\ell= \int_{\mathbb{T}^3}d
k\big(\log W(k)+\log\pi+1\big)\,.
\end{equation}

The next step is to construct, out of the Gaussian measures
introduced in (\ref{3.3}), Gaussian measures which have a slow
variation in physical space $\mathbb{Z}^3$ and which are locally
stationary. For this purpose we give ourselves the local power
spectrum $W(r,k) \geq 0$, $r\in\mathbb{R}^3$, which vanishes
rapidly as $|r| \to \infty$, and introduce
 \begin{eqnarray}\label{3.12}
&& Q(r,x)=
\int_{\mathbb{T}^3}dk W(r,k) \omega(k)^{-1} \cos (2\pi k\cdot x)\,,\nonumber\\
&& P(r,x)=
\int_{\mathbb{T}^3} dk W(r,k) \omega(k) \cos (2\pi k\cdot x)\,,\nonumber\\
&&C(r,x) = \int_{\mathbb{T}^3} dk W(r,k) \sin(2\pi k\cdot x)\,,
\end{eqnarray}
$x \in \mathbb{Z}^3$, by which we define the family $\langle \cdot
\rangle^{\textrm{G},\varepsilon}$ of Gaussian measures through
 \begin{eqnarray}\label{3.12a}
&&\langle q_x \rangle^{\textrm{G},\varepsilon} = 0\,, \quad
\langle p_x \rangle^{\textrm{G},\varepsilon}=0\,,\nonumber\\
&&\langle q_xq_{x'} \rangle^{\textrm{G},\varepsilon} =
Q(\varepsilon(x +x')/2, x - x')+\mathcal{O}(\varepsilon)\,,\nonumber\\
&&\langle p_xp_{x'} \rangle^{\textrm{G},\varepsilon} =
P(\varepsilon(x +x')/2, x - x')+\mathcal{O}(\varepsilon)\,,\nonumber\\
&&\langle q_x p_{x'} \rangle^{\textrm{G},\varepsilon} =
C(\varepsilon(x +x')/2, x - x')+\mathcal{O}(\varepsilon)\,.
\end{eqnarray}
The error of order $\varepsilon$ has to be allowed so to ensure a
positive definite covariance matrix.

This family has two important properties.\smallskip\\
(i) Relative to the reference point $r/\varepsilon$, $r  \in
\mathbb{R}^3$, the measure becomes stationary in the limit
$\varepsilon \to 0$.
This is the property of {\it local stationarity}.\smallskip\\
(ii) For two distinct reference points $r$ and $r'$, $r \neq r'$,
the local distributions become \textit{independent} in the limit
$\varepsilon \to 0$ as can be inferred from
  \begin{eqnarray}\label{3.12b}
&&\hspace{-10pt}\lim_{\varepsilon \to 0}\big\{ \langle q_{\lfloor
r/\varepsilon\rfloor+x} q_{\lfloor
r/\varepsilon\rfloor+x'}q_{\lfloor r'/\varepsilon\rfloor+y}
q_{\lfloor r'/\varepsilon\rfloor+y'}\rangle^{\textrm{G},{\varepsilon}}\nonumber\\
&&\hspace{10pt} - \langle q_{\lfloor r/\varepsilon\rfloor+x}
q_{\lfloor
r/\varepsilon\rfloor+x'}\rangle^{\textrm{G},{\varepsilon}} \langle
q_{\lfloor r'/\varepsilon\rfloor+y} q_{\lfloor
r'/\varepsilon\rfloor+y'}\rangle^{\textrm{G},{\varepsilon}}\big\}
=0\,,
\end{eqnarray}
with $\lfloor\cdot\rfloor$ denoting integer part, since $Q(r,x) \to 0 $ as $|x| \to \infty$ . The analogous property holds for
 the remaining covariances. Thus under the Gaussian measure $\langle \cdot \rangle^{\textrm{G},\varepsilon}$
 two macroscopically far apart regions are statistically independent.\smallskip

The construction (\ref{3.12a}) is computationally not so flexible
and it is more convenient to invert the order. Thus the primary
object is a family $\langle \cdot \rangle^{\textrm{G},\varepsilon}$
of Gaussian measures (non-Gaussian measures to be discussed
further on). They have mean zero and a local covariance, which is
almost time stationary and slowly varying in space. These
conditions are most easily imposed through the lattice analogue of
the local power spectrum $W$ expressed in terms of $a$-field,
compare with (\ref{3.3}). Firstly we require
\begin{equation}\label{3.13}
\langle a(k)\rangle^{\textrm{G},\varepsilon} =0\,,\quad  \langle a(k) a(k')
\rangle^{\textrm{G},\varepsilon} =0\,.
\end{equation}
The local $a^\ast a$ spectrum is defined through
 \begin{equation}\label{3.14}
W^1(x,k) =2^{-3} \int_{(2\mathbb{T})^3} d \eta e^{i2\pi x \cdot
\eta}\langle a (k- \eta/2)^\ast a (k+
\eta/2)\rangle^{\textrm{G},\varepsilon}\,.
\end{equation}
$\langle a(k-\eta/2)^\ast
a(k+\eta/2)\rangle^{\textrm{G},\varepsilon}$ is
$\mathbb{T}^3$-periodic in $k$ and $(2\mathbb{T})^3$-periodic in
$\eta$. Therefore $W(x,k)$ as inverse Fourier transform with
respect to $\eta$ is $\mathbb{T}^3$-periodic in $k$ and lives on
the half-integer lattice $(\mathbb{Z}/2)^3$ with respect to $x$.

We rescale the lattice to have lattice spacing $\varepsilon$
through the substitution $x=\varepsilon^{-1}y$, $y\in
(\varepsilon\mathbb{Z}/2)^3$, and obtain the rescaled local power
spectrum
\begin{equation}\label{3.15}
W^{\varepsilon} (y,k) = (\varepsilon/2)^3
\int_{(2\mathbb{T}/\varepsilon)^3} d \eta e^{i2\pi
y\cdot\eta}\langle a(k-\varepsilon \eta/2)^\ast a(k+ \varepsilon
\eta/2)\rangle^{\textrm{G},\varepsilon} \,.
\end{equation}
Then, denoting $\lfloor\cdot\rfloor_\varepsilon$ as modulo
$\varepsilon$, one requires
\begin{equation}\label{3.16}
\lim_{\varepsilon \to 0} W^{\varepsilon}(\lfloor
r\rfloor_\varepsilon,k) = W(r,k)
\end{equation}
pointwise. If $\langle \cdot \rangle^{\textrm{G},\varepsilon}$ is
defined through (\ref{3.12a}), then $W$ of (\ref{3.16}) agrees
with the one in (\ref{3.12}). $W^{\varepsilon} (y,k)$ is
normalized as
\begin{equation}\label{3.17}
\sum_{y\in (\varepsilon \mathbb{Z}/2)^3} \int_{\mathbb{T}^3}dk
W^{\varepsilon}(y,k) = \int_{\mathbb{T}^3}dk \langle a(k)^\ast
a(k) \rangle^{\textrm{G},\varepsilon}\,.
\end{equation}
The condition that the limit in (\ref{3.16}) exists thus implies
that the average phonon number increases as $\varepsilon^{-3}$,
equivalently the average total energy increases as
\begin{equation}\label{3.18}
\int_{\mathbb{T}^3}dk \omega(k) \langle a(k)^\ast a(k)
\rangle^{\textrm{G},\varepsilon}= \langle
H_0\rangle^{\textrm{G},\epsilon} = \mathcal{O}(\varepsilon^{-3})\,.
\end{equation}

(\ref{3.15}) has a familiar touch. Recall that for a quantum wave
function $\psi$ on physical space $\mathbb{R}^3$ the
\textit{Wigner function} is defined by
\begin{equation}\label{3.18b}
W^{\varepsilon} (x,k) =  \int_{\mathbb{R}^3} d \eta
e^{ix\cdot\eta}\widehat{\psi}(k-\varepsilon \eta/2)^\ast
\widehat{\psi}(k+ \varepsilon \eta/2)
\end{equation}
with $x,k\in\mathbb{R}^3$ and $\widehat{\psi}$ the Fourier
transform of $\psi$. $\varepsilon$ is the semiclassical parameter,
$\varepsilon\to 0$ in the semiclassical limit. The main difference
to (\ref{3.15}) is that for the semiclassical limit usually one
considers a sequence $\psi^\varepsilon$ of wave functions, while
in (\ref{3.15}) one has a sequence of probability measures over
the wave field and its time derivative. Because of this obvious
analogy we call (\ref{3.15}) the Wigner function, more properly
the one-point Wigner function. The  $n$-point Wigner function is understood as the
$n$-th moment of $a^\ast a$.

For a family $\langle \cdot\rangle^{\varepsilon}$ of general
measures on phase space one defines the one-point Wigner function
\begin{equation}\label{3.18a}
W^{\varepsilon} (y,k) =  (\varepsilon/2)^3
\int_{(2\mathbb{T}/\varepsilon)^3} d \eta e^{i2\pi
y\cdot\eta}\langle a(k-\varepsilon \eta/2)^\ast a(k+ \varepsilon
\eta/2)\rangle^{\varepsilon}\,,
\end{equation}
i.e.~through (\ref{3.15}) with $\langle
\cdot\rangle^{\textrm{G},\varepsilon}$ replaced by $\langle
\cdot\rangle^{\varepsilon}$. The rescaled two-point Wigner function
becomes
\begin{eqnarray}\label{3.19}
W^{\varepsilon} (y_1,k_1,y_2,k_2) =  (\varepsilon/2)^6
\int_{(2\mathbb{T}/\varepsilon)^6} d \eta_1 d \eta_2 \exp [i2\pi
y_1\cdot\eta_1+i 2\pi y_2\cdot\eta_2]\nonumber\\
\langle a(k_1- \varepsilon \eta_1/2)^\ast a(k_1+ \varepsilon
\eta_1/2) a(k_2- \varepsilon \eta_2/2)^\ast a(k_2+ \varepsilon
\eta_2/2)\rangle^{\varepsilon}\,,
\end{eqnarray}
and similarly for higher-point Wigner functions. We require
(\ref{3.16}) and
\begin{equation}\label{3.19a}
\lim_{\varepsilon\to 0}\langle\prod^m_{j=1}a(k_j)^\ast
\prod^n_{i=1}a(k_i')\rangle^{\varepsilon}=0
\end{equation}
whenever $m\neq n$. The condition of statistical independence of
far apart regions then reads
\begin{equation}\label{3.20}
\lim_{\varepsilon\to 0} \{W^\varepsilon (\lfloor r_1
\rfloor_\varepsilon, k_1, \lfloor r_2 \rfloor_\varepsilon, k_2)-
W^\varepsilon (\lfloor r_1 \rfloor_\varepsilon, k_1)
W^\varepsilon( \lfloor r_2 \rfloor_\varepsilon, k_2)\}=0
\end{equation}
for $r_1\neq r_2$, which in the context of low density gases is
known as \textit{assumption of molecular chaos}. Since
(\ref{3.20}) is a law of large numbers, it implies that
\begin{equation}\label{3.21}
\lim_{\varepsilon\to 0} W^\varepsilon (\lfloor r_1
\rfloor_\varepsilon, k_1,\ldots,\lfloor r_n \rfloor_\varepsilon,
k_n)= \prod^n_{j=1} W (\lfloor r_j \rfloor_\varepsilon, k_j)
\end{equation}
whenever the family $\{r_1, \ldots, r_n\}$ is free of double
points.

There is no reason that $\langle \cdot\rangle^{\varepsilon}$ becomes
locally stationary as $\varepsilon\to 0$. Still the condition of
local stationarity can be expressed through the limiting behavior
of multi-point Wigner functions. For example, in the case of the
two-point function the condition would read
\begin{eqnarray}\label{3.22}
&&\lim_{\varepsilon\to 0}W^\varepsilon (\lfloor r
\rfloor_\varepsilon, k_1, \lfloor r \rfloor_\varepsilon, k_2)=
W (r,k_1) W (r,k_2)\nonumber\\
&&+ \delta(k_1+k_2) \int_{\mathbb{T}^3} d\eta W(r,k_1+\eta/2)
W(r,k_2+\eta/2)\,.
\end{eqnarray}
For a sequence $\langle \cdot \rangle^{\textrm{G},\varepsilon}$ of
Gaussian measures satisfying (\ref{3.16}) the identity
(\ref{3.22}) holds by construction.

%%%%%%%%%%%%%%%%%%%%%%%%%%%%%%%%%%%

\section{Kinetic limit}\label{sec.4}
\setcounter{equation}{0}

As initial measures for (\ref{2.7}) we adopt the scale of Gaussian
measures $\langle \cdot \rangle^{\textrm{G},\varepsilon}$ satisfying
(\ref{3.13}) - (\ref{3.16}). The time-evolved measure at time $t$
is denoted by $\langle\cdot\rangle_t$. Let us first consider the
harmonic lattice dynamics, $\lambda = 0$. Then by linearity,
$\langle\cdot\rangle_t$ is again Gaussian. Since the deviations
from stationarity are on the spatial scale $\varepsilon^{-1}$ and
since there is a finite speed of propagation, one has to wait for
times of order $\varepsilon^{-1}t$ to observe appreciable changes
of the Wigner function, which defines the kinetic time scale
$\varepsilon^{-1}t$. On that scale one has
\begin{eqnarray}\label{4.1}
&&\hspace{-24pt}\frac{\partial}{\partial t} \langle a(k- \varepsilon \eta/2)^\ast
a (k+ \varepsilon \eta/2) \rangle_{t/\varepsilon,}\nonumber\\
&&= -i \varepsilon^{-1} \big(\omega (k+ \varepsilon \eta/2) -
\omega (k- \varepsilon \eta/2)\big) \;\langle a(k- \varepsilon
\eta/2)^\ast a (k+ \varepsilon \eta/2) \rangle_{t/\varepsilon}\,.
\end{eqnarray}
Taking the limit $\varepsilon \to 0$ one obtains
\begin{equation}\label{4.1a}
\frac{\partial}{\partial t} \widehat{W}(\eta,k,t) = - i\nabla
\omega (k) \cdot \eta \widehat{W}(\eta,k,t)
\end{equation}
and, upon inverting the Fourier transform, the limit Wigner
function is the solution of the transport equation
\begin{equation}\label{4.2}
\frac{\partial}{\partial t} W(r,k,t) = - \frac{1}{2\pi}\nabla
\omega (k) \cdot \nabla_r W( r,k,t)\,.
\end{equation}
Thus in the kinetic limit, $\varepsilon \to 0$, we can think of
the phonon counting function $W$ as arising from a gas of
independent particles, the \textit{phonons}, with kinetic energy
$\omega(k)$. Detailed proofs for the validity of the free
transport equation (\ref{4.2}) are given by Mielke \cite{Mi}. He
allows for rather general deterministic initial data and for
harmonic lattice dynamics with vector displacements and a general
unit cell.

If one adjusts the strength of collisions in such a way as to have
an effect of the same order as the transport term, then kinetic
theory claims that the locally stationary state imposed at $t=0$
retains its structure in the course of time. Of course, the
time-evolved measure $\langle \cdot \rangle_{t/\varepsilon}$ is no
longer exactly Gaussian. But for small $\varepsilon$ and on a
local scale it does remain so in a good approximation. As crucial
difference to (\ref{4.2}) the evolution equation will contain a
collision term taking account of the anharmonicities. As to be
shown in the following section, the cubic term is of the right
strength if one substitutes
\begin{equation}\label{4.3}
\lambda \rightsquigarrow \sqrt{\varepsilon}\lambda
\end{equation}
with $\lambda$ fixed and independent of $\varepsilon$. Then the
stabilizing quartic term has the strength $\lambda' =
(\lambda^2/18 \omega_0^2)\varepsilon$, which is indeed small
compared to the cubic term. The Wigner function at the kinetic
time $t$ is given through
\begin{equation}\label{4.5}
W^\varepsilon(y,k,t) =  \varepsilon^3
\int_{(\mathbb{T}/\varepsilon)^3}d \eta e^{i2\pi y\cdot
\eta}\langle a(k-\varepsilon \eta/2)^\ast a (k+\varepsilon \eta/2)
\rangle_{t/\varepsilon}\,.
\end{equation}
It is expected that the limit $\varepsilon \to 0$ exists,
\begin{equation}\label{4.6}
\lim_{\varepsilon \to 0} W^{\varepsilon}(\lfloor
r\rfloor_\varepsilon,k,t) = W(r,k,t)\,,
\end{equation}
and the limit phonon counting function $W$ is the solution of a
Boltzmann-like equation. Its derivation will be explained in the
section to follow, but let us state the result already now,
\begin{eqnarray}\label{4.7}
&&\hspace{-16pt}\frac{\partial}{\partial t} W(r,k,t) +
\frac{1}{2\pi}\nabla
\omega (k) \cdot \nabla_r W( r,k,t)\\
&&= \gamma \int_{\mathbb{T}^6} d k_1 d k_2 (\omega(k)
\omega(k_1) \omega(k_2))^{-1}\big\{2\delta
(\omega(k)+\omega(k_1)-\omega(k_2))
\delta(k+k_1-k_2)\nonumber\\
&&\hspace{16pt}\big(W(r,k_1,t) W(r,k_2,t)+ W(r,k,t)W(r,k_2,t)-W(r,k,t)W(r,k_1,t)\big)
\hspace{20pt} \textrm{(I)}\nonumber\\
&&\hspace{16pt}+\; \delta (\omega(k)- \omega(k_1)-\omega(k_2))\delta(k-k_1-k_2)\nonumber\\
&&\hspace{16pt}\big(W(r,k_1,t)
W(r,k_2,t)-W(r,k,t)W(r,k_1,t)-W(r,k,t)W(r,k_2,t)\big)\big\}\quad
\textrm{(II)} \nonumber
\end{eqnarray}
with $\gamma$ the strength of the collision term,
\begin{equation}\label{4.7a}
\gamma= \frac{\pi}{2}\lambda^2\,.
\end{equation}
$\delta$ is the torus $\delta$-function, see the explanation below
Equation (\ref{2.10}). As shorthand the collision operator is
denoted by $\mathcal{C}(W)$.

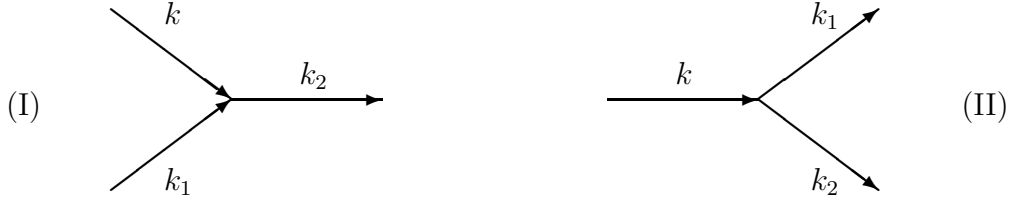
\begin{figure}
\setlength{\unitlength}{1cm}
\begin{picture}(10,4)(-2,-1.5)\thicklines
\put(2,0){\vector(1,0){2}} \put(0.4,1.2){\vector(4,-3){1.6}}
\put(0.4,-1.2){\vector(4,3){1.6}} \put(7,0){\vector(1,0){2}}
\put(9,0){\vector(4,3){1.6}} \put(9,0){\vector(4,-3){1.6}}

\put(2.9,0.2){$k_2$} \put(1.1,1.0){$k$} \put(1.1,-1.2){$k_1$}
\put(7.9,0.2){$k$} \put(9.7,1.0){$k_1$} \put(9.7,-1.2){$k_2$}

\put(-1,-0.2){(I)} \put(11.7,-0.2){(II)}
\end{picture}
\caption{Three phonon collisions.}\label{Fig2}
\end{figure}
Dynamically the terms (I) and (II) can be viewed as given in Figure \ref{Fig2}.
In (I) the phonon with wave vector $k$ collides with a phonon with
wave vector $k_1$ in order to merge into a phonon with wave vector
$k_2$. The loss term is the term proportional to $W(k)$, hence $2W(k)(W(k_2)-W(k_1))$,
and the gain term is the remainder, i.e. $2W(k_1)W(k_2)$. Note that the gain term has a definite sign
while the loss term takes both signs.  Correspondingly in (II) the phonon with
wave vector $k$ splits into two phonons with wave vector $k_1$ and
$k_2$. The gain term is again $W(k_1)W(k_2)$ and the loss term is
$-W(k)(W(k_1)+W(k_2))$. The precise way of how the phonon
distribution functions appear in (\ref{4.7}) does not seem to have
a mechanical interpretation in terms of colliding point particles.
As can be seen from the $\delta$-functions in the collision
operator, in both collision processes energy is conserved, while
momentum is conserved only modulo integers. E.g. for term (I) the
$\delta$-function yields the constraint
\begin{equation}\label{4.8}
k + k_1 = k_2 +  n\,, \quad n \in \mathbb{Z}^3\,, \quad k,
k_1, k_2 \in \mathbb{T}^3\,.
\end{equation}
In case $n=0$ one speaks of a \textit{normal} process while in
case $n \neq 0$ of an \textit{umklapp} process.

The rates appearing in (I) and (II) come out of the computation to
be presented in Section \ref{sec.5}. However, their relative
strength $1/2$ is required in order for energy to be locally
conserved.

Note that the Boltzmann equation preserves the positivity of $W$.
Obviously the free streaming term has this property. If $W$ first
hits 0 at some point $k$, $W(r,k,t)=0$, then $d
W(r,k,t)/d t >0$, $d/dt$ denoting the total time derivative, due to  the positive
gain term and vanishing loss term of the collision
operator. Hence at that point $W$ cannot turn negative.

This seems to be a good moment to return to the issue of a
potential energy which depends only on the differences in the
displacements, as would be the case for a real crystal. Then
$\omega_0=0$ and $V$ of (\ref{2.6}) is replaced by
\begin{equation}\label{4.9}
V_3 = \lambda \frac{1}{3}\sum_{x\in \mathbb{Z}^3}\sum_{\alpha=1}^3
(q_{x+e_\alpha}-q_x)^3\,,
\end{equation}
$e_1,e_2,e_3$ the standard basis of $\mathbb{Z}^3$, which
expressed in terms of the $a$-fields becomes
\begin{eqnarray}\label{4.10}
&&\hspace{-30pt}V_3 = \lambda \frac{1}{3}\sum_{\alpha=1}^3
\int_{\mathbb{T}^9}d k_1
d k_2 dk_3 \delta(k_1+k_2+k_3)\nonumber\\
&&\hspace{30pt} \prod^3_{j=1}\big\{(2 \omega(k_j))^{-1/2}
(\exp[i2\pi k^{\alpha}_j]-1)(a(k_j)+a(-k_j)^\ast)\big\}\,.
\end{eqnarray}
Compared to $V$ of (\ref{2.6}), only the weight in Fourier space
has changed. Thus the Boltzmann equation remains as in (\ref{4.7})
provided the collision rate
$(\omega(k)\omega(k_1)\omega(k_2))^{-1}$ is replaced by
\begin{equation}\label{4.11}
\prod^3_{j=1}\big|\sum_{\alpha=1}^3 \omega(k_j)^{-1/2} (\exp[i2\pi
k^{\alpha}_j]-1)\big|^2\,,\quad k_3=k\,.
\end{equation}
Close to the origin this collision rate is more singular  than the
one in (\ref{4.7}). But the general properties of the Boltzmann
equation, as to be discussed in Section \ref{sec.6}, remain in
force.

We hurried a little bit to write down the Boltzmann equation. So
the reader might wonder why we claim that on the kinetic time
scale local stationarity is maintained. The point is that the free
dynamics, generated by $H_0$, does not tolerate deviations from
local stationarity as long as the free dynamics is given some time
act. Such a property has been studied in considerable detail by
Dobrushin \textit{et al.} \cite{DoPe}, for recent improvements see
 \cite{DuSp}. Roughly speaking, they
consider initial measures $\langle\cdot\rangle^{\varepsilon}$ on
phase space for which $\langle q_x\rangle^{\varepsilon}=0=\langle
p_x\rangle^{\varepsilon}$ and for which the Wigner function
$W^\varepsilon$ of (\ref{3.18a}) has a limit as in (\ref{3.16}).
In addition they require that under
$\langle\cdot\rangle^{\varepsilon}$ spatial regions separated by a
distance $\ell$ with $1\ll \ell\ll\varepsilon^{-1}$ are in essence
statistically independent. $\langle\cdot\rangle^{\varepsilon}$  is
\textit{non}-Gaussian, in general. This initial state is evolved
under the dynamics generated by $H_0$. Then, for times $t$ where
$1\ll t\ll\varepsilon^{-1}$, the Wigner function does not change.
However locally the oscillators adjust such that the measure
becomes to a very good approximation Gaussian and satisfies the
conditions (\ref{3.13}) and (\ref{3.16}). Thus for times which are
short on the kinetic scale the harmonic lattice dynamics forces
local stationarity.

%%%%%%%%%%%%%%%%%%%%%%%%%%%%%%%%%%%%%%

\section{Conditions on the dispersion relation}\label{sec.4a}
\setcounter{equation}{0}

Our discussion seems to indicate that the kinetic description
holds independently of the particular form of the (short ranged)
harmonic interaction potential, in other words independently of
the (analytic) dispersion relation. As far as the convergence to
locally stationary Gaussian measures is concerned, this impression
is well supported \cite{DoPe}. However, for three-phonon collision
processes it cannot be taken for granted to have a non-vanishing
collision operator. If one sets
\begin{equation}\label{4a.1}
E_q(k)=\omega(k)+\omega(q)-\omega(k+q)\,,
\end{equation}
clearly conservation of energy can be satisfied only if
\begin{equation}\label{4a.2}
E_q(k)=0
\end{equation}
admits solutions when considered as a function on $\mathbb{T}^6$.
For nearest neighbor coupling only, to say $\alpha(e)=-1$ for
$|e|=1$, $\alpha(0)=6$, and $\alpha(x)=0$ otherwise, the
dispersion relation reads
\begin{equation}\label{4a.3}
\omega(k)=\Big(\omega^2_0+2\sum^3_{j=1}\big(1-\cos(2\pi
k^j)\big)\Big)^{1/2}\,,\quad k=(k^1,k^2,k^3)\,.
\end{equation}
As shown in Appendix \ref{sec.16a}, for this choice
$E_q(k)\geq\omega_0/2>0$. The physically most obvious model does
not admit three-phonon collisions.

From this perspective one might wonder whether (\ref{4a.2}) can be
satisfied at all. An example which can be checked still by hand is
given by
\begin{equation}\label{4a.4}
\omega(k)=\omega_0+2\sum^3_{j=1}\big(1-\cos(2\pi
k^j)\big)\,.
\end{equation}
It corresponds to the harmonic couplings
\begin{equation}\label{4a.5}
(6+\omega_0)(q_{(1,0,0)}-q_0)^2\,,\;-\frac{1}{2}
(q_{(2,0,0)}-q_0)^2\,,\;-(q_{(1,1,0)}-q_0)^2\,,
\end{equation}
all others determined by isotropy and translation invariance. Note
that the next nearest neighbor couplings are destabilizing.
Clearly $E_0(0)=\omega_0$ while for $q=(1/4)(1,1,1)$,
$k=(1/8)(1,1,1)$ one has $E_q(k)=\omega_0-6(\sqrt{2}-1)<0$
provided $\omega_0$ is not too large.

To have a nonvanishing collision operator we require
\begin{equation}\label{4a.6}
\int_{\mathbb{T}^6}dkdq
\delta(\omega(k)+\omega(q)-\omega(k+q))>0\,.
\end{equation}
There seems to be no simple sufficient criterion on $\omega$,
which would ensure (\ref{4a.6}). Numerically one plots $E_q(k)$
for random choices for $q$ to find out whether $E_q(k)$ takes
negative values which then implies (\ref{4a.6}).

Observe that $E_q(0)=\omega_0$ for all $q\in\mathbb{T}^3$. If
$\omega_0>0$, by continuity there is then a neighborhood
$\Lambda_0$ of 0 defined through $\Lambda_0=\{k\in\mathbb{T}^3$,
$E_q(k)>0$ for all $q\in\mathbb{T}^3\}$ and $0\in \Lambda_0$. If
$W(r,k)$ is supported in $\Lambda_0$ for every $r$, then
$\mathcal{C}(W)=0$. The free flow leaves this set of $W$'s
invariant and therefore such $W$'s evolve merely by free
streaming. In general, there will other components of
$\mathbb{T}^3$ where no collision partner is available. In
addition, there can be components
$\Lambda_1,\Lambda_2,\ldots,\Lambda_m$ such that if
$k\in\Lambda_j$ it will remain so under any sequence of
collisions. Then in each $\Lambda_j$ the system equilibrates in
the long time limit, but in general the equilibration temperature
will differ from component to component. For this reason we
introduce the notion that $k\in\mathbb{T}^3\setminus\{0\}$ is linked by a
collision to $q\in\mathbb{T}^3\setminus\{0\}$ if $E_q(k)=0$. Clearly,
linkage is symmetric.\medskip\\
{\bf Ergodicity Condition (E)}: For every
$k,k'\in\mathbb{T}^3\setminus\{0\}$ there is a finite sequence of
collisions such that $k$ is linked to $k'$.\medskip\\
In particular for every $k\neq 0$ there is at least one collision
partner $q\neq 0$ such that $E_q(k)=0$. A necessary condition for
ergodicity to hold is $\omega_0=0$. If in (\ref{4a.4}) we set
$\omega_0=0$, then ergodicity is satisfied with one intermediate
collision, as can be seen from an explicit computation.

There is a further condition related to the issue of existence of
solutions of the Boltzmann equation (\ref{4.7}). If $\|W\|_\infty$
denotes the sup-norm, the collision operator can be trivially
estimated as
\begin{equation}\label{4a.7}
\|\mathcal{C}(W)\|_\infty\leq c
\Big(\sup_{q\in\mathbb{T}^3}\int_{\mathbb{T}^3}dk\delta(E_q(k))\Big)
\|W\|_\infty^2
\end{equation}
provided $\omega_0>0$. By standard methods of kinetic theory, if
\begin{equation}\label{4a.8}
\int_{\mathbb{T}^3}dk\delta(E_q(k))\leq e_{\max}<\infty\,,
\end{equation}
then the Boltzmann equation (\ref{4.7}) has a unique bounded
solution for $0\leq t\leq t_0$ with suitable $t_0$. If
(\ref{4a.8}) does not hold, resp.~if $\omega_0=0$, to establish
the existence of solutions local in time would require more
efforts.

For the dispersion relation (\ref{4a.4}) the condition
(\ref{4a.8}) is satisfied. In general, for (\ref{4a.8}) to hold
$E_q(k)$ has to be a Morse function uniformly in $q$. 
To see why, assume that $E_q$ is not a Morse function, and
try to locate points $q$ where the integral in
(\ref{4a.8}) diverges. On the level set $\{k,E_q(k)=0\}$ one must
have
\begin{equation}\label{4a.9}
\nabla_k E_q(k)=0\,,
\end{equation}
which can be solved locally to yield $q=q(k)$. Thus
\begin{equation}\label{4a.10}
E_{q(k)}(k)=0
\end{equation}
must have solutions. Secondly the Hessian of $E_q(k)$ must have at
least one vanishing eigenvalue which leads to the condition
\begin{equation}\label{4a.11}
\det( \textrm{Hess}\,E_q(k))=0\quad \textrm{at}\;q=q(k)\,.
\end{equation}
The surfaces in $\mathbb{T}^3$ defined through the level zero sets
in (\ref{4a.10}) and (\ref{4a.11}) will generically intersect
along a curve. Thus we must be prepared that the integral in
(\ref{4a.8}) diverges along a curve in $\mathbb{T}^3$. Again, no
simple sufficient criterion is available to ensure (\ref{4a.8}).

%%%%%%%%%%%%%%%%%%%%%%%%%%%%%%%%

\section{Derivation of the phonon Boltzmann equation (classical model)}\label{sec.5}
\setcounter{equation}{0}

The textbook derivation of the Boltzmann equation starts from the
quantized theory as to be discussed in Section \ref{sec.7}, and
uses the Fermi golden rule to compute the transition rate, see
\cite{Gu} for a particularly lucid discussion. While such a
procedure yields the correct rates, it provides little theoretical
insight why the Fermi golden rule would be applicable in such a
field theoretical context. Of course, the best of all
possibilities would be to have a mathematically rigorous
derivation. We are far from such a goal at present. Instead we
offer in this section a derivation based on the concept of local
stationarity through which higher order correlations can be
suitably decoupled, see \cite{ErSaY} for a similar argument in
the case of a weakly interacting Fermi gas on the lattice. Physically, this seems to me the most
transparent procedure, admittedly with the disadvantage that the
approximate local stationarity cannot be checked directly. A more
systematic approach uses Feynman diagrams, as will be explained in
Section \ref{sec.5a}.

To properly argue for the validity of the Boltzmann equation
(\ref{4.7}), it is convenient to work in atomic units for a while.
We give ourselves the Wigner function $W(r,k)\geq 0$ and assume
that the initial measure, $\langle\cdot\rangle_0$, is Gaussian
satisfying (\ref{3.13}) and (\ref{3.16}). The average with respect
to the measure at time $t$ is denoted by $\langle \cdot
\rangle_t$. We introduce the shorthands
\begin{equation}\label{5a.6}
a(k,1)=a(k)^\ast\,,\quad a(k,-1)=a(k)\,,
\end{equation}
and
\begin{equation}\label{5a.3d}
\phi(k,k_1,k_2)=\lambda(8\omega(k)\omega(k_1)\omega(k_2))^{-1/2}\,.
\end{equation}
Then the equations of motion (\ref{2.13}) can be written in the more compact form
\begin{eqnarray}\label{5a.7}
&&\hspace{-20pt}\frac{d}{dt} a(k,\sigma) = i \sigma \omega(k)
a(k,\sigma)+ i \sqrt{\varepsilon}\sigma
\sum_{\sigma_1,\sigma_2=\pm1}\int_{\mathbb{T}^6} dk_1
dk_2\phi(k,k_1,k_2)\nonumber\\
&&\hspace{35pt} \times \delta(-\sigma k+\sigma_1k_1+\sigma_2k_2)
a(k_1,\sigma_1) a(k_2,\sigma_2)\,, \quad \sigma=\pm 1\,.
\end{eqnarray}

The two-point function satisfies
\begin{equation}\label{5.4}
\frac{d}{d t} \langle a(p)^\ast a(q)\rangle_t =
i(\omega(p) -\omega(q)) \langle a(p)^\ast a(q)\rangle_t +
\sqrt{\varepsilon}F(q,p,t)
\end{equation}
with
\begin{eqnarray}\label{5.5a}
&&\hspace{-40pt}F(q,p,t) = i \sum_{\sigma_1,\sigma_2=\pm1}\int_{\mathbb{T}^6} d
k_1 d k_2 \Big( \phi(p,k_1,k_2)
\delta (-p + \sigma_1 k_1+  \sigma_2
k_2) \langle  a(k_1,\sigma_1) \nonumber\\
&&\hspace{-30pt}\times a(k_2,\sigma_2)a(q)\rangle_t -  \phi(q,k_1,k_2)
\delta (q + \sigma_1 k_1+  \sigma_2)\langle  a(p)^\ast a(k_1,\sigma_1)a(k_2,\sigma_2)
\rangle_t\Big)\,.
\end{eqnarray}
We need a second iteration, which we write in integrated form as
\begin{equation}\label{5.6}
F(q,p,t) = F_{\mathrm{hom}}(q,p,t) + \sqrt{\varepsilon}\int^t_0 ds
G(q,p,t-s,s)\,.
\end{equation}
The homogeneous term in (\ref{5.6}) reads
\begin{eqnarray}\label{5.5}
&&\hspace{-20pt} F_{\mathrm{hom}}(q,p,t) = i \sum_{\sigma_1,\sigma_2=\pm1}\int_{\mathbb{T}^6} d
k_1 d k_2 e^{it(\sigma_1\omega(k_1) +\sigma_2\omega(k_2))}
\nonumber\\
&&\hspace{0pt}\times \Big( \phi(p,k_1,k_2)\delta (-p + \sigma_1 k_1+  \sigma_2
k_2) e^{-it\omega(q)}\langle  a(k_1,\sigma_1) a(k_2,\sigma_2)a(q)\rangle_0  \nonumber\\
&&\hspace{0pt} - \phi(q,k_1,k_2)
 \delta (q + \sigma_1 k_1+  \sigma_2)e^{it\omega(p)}\langle  a(p)^\ast a(k_1,\sigma_1)a(k_2,\sigma_2)
\rangle_0\Big)\nonumber\\
&&=0
\,,
\end{eqnarray}
since in the initial measure odd moments vanish.
We conclude that
 \begin{equation}\label{5.9}
\frac{d}{d t} \langle a(p)^\ast a(q)\rangle_t =
i(\omega(p) -\omega(q)) \langle a(p)^\ast a(q)\rangle_t +
\varepsilon \int^t_0 ds G(q,p,t-s,s)\,.
\end{equation}

Following (\ref{3.18a}) one switches to Wigner function variables and sets
 \begin{equation}\label{5.9a}
\widehat{W}^\varepsilon(\eta,k,t) = \varepsilon^3\langle a(k-\varepsilon \eta/2)^\ast a(k+ \varepsilon
\eta/2)\rangle_{t/\varepsilon}\,.
\end{equation}
Then
\begin{eqnarray}\label{5.11a}
&&\frac{\partial}{\partial t}\widehat{W}^\varepsilon(\eta,k,t) =
i\varepsilon^{-1}(\omega(k-\varepsilon\eta/2)-\omega(k+\varepsilon\eta/2))
\widehat{W}^\varepsilon(\eta,k,t)\nonumber\\
&&\hspace{82pt}+ \varepsilon^3 \int^{t/\varepsilon}_0 ds
G(k+\varepsilon\eta/2,k-\varepsilon\eta/2,\varepsilon^{-1}t-s,s)\,.
\end{eqnarray}
Assuming that $\widehat{W}^\varepsilon(\eta,k,t)$ converges to 
 $\widehat{W}(\eta,k,t)$ as $\varepsilon \to 0$, the remaining task is to establish that the inhomogeneous term on the right converges to the collision operator (\ref{4.7}) acting on 
 $\widehat{W}(\eta,k,t)$.\medskip\\
(1) {\it Local stationarity, Gaussian approximation}. The integrand of the inhomogeneous term in
(\ref{5.6}) is given by
\begin{eqnarray}\label{5.7}
&&\hspace{-16pt}G(q,p,t,s)=  \sum_{\sigma_1,\sigma_2=\pm1}\int_{\mathbb{T}^6} d
k_1 d k_2 \sum_{\tau_1,\tau_2=\pm1}\int_{\mathbb{T}^6} d
l_1 d l_2 \phi(p,k_1,k_2)\phi(q,l_1,l_2)\\
&&\hspace{6pt}\times \Big( \delta (-p + \sigma_1 k_1+  \sigma_2
k_2)\delta (q + \tau_1 l_1+  \tau_2 l_2)e^{-it\omega(q)} 
\nonumber\\
&&\hspace{36pt}+ \delta (q + \sigma_1 k_1+  \sigma_2
k_2)\delta (-p + \tau_1 l_1+  \tau_2 l_2)e^{it\omega(p)} \Big)\nonumber\\
&&\hspace{6pt}
\times e^{it(\sigma_1\omega(k_1)+\sigma_2\omega(k_2))}
\langle  a(k_1,\sigma_1)a(k_2,\sigma_2)a(l_1,\tau_1)a(l_2,\tau_2)
\rangle_s\nonumber\\
&&\hspace{6pt}
-2 \sum_{\sigma_1,\sigma_2=\pm1}\int_{\mathbb{T}^6} d
k_1 d k_2 \sum_{\tau_1,\tau_2=\pm1}\int_{\mathbb{T}^6} d
l_1 d l_2\nonumber\\
&&\hspace{6pt}
 \phi(k_1,l_1,l_2)\delta (-\sigma_1k_1 +\tau_1 l_1+ \tau_2 l_2)e^{it (\sigma_1\omega(k_1)+\sigma_2\omega(k_2))}\sigma_1\nonumber\\
&&\hspace{6pt} \times \Big(
\phi(p,k_1,k_2) \delta (-p + \sigma_1 k_1+  \sigma_2 k_2)e^{-it\omega(q)} 
\langle  a(q)a(k_2,\sigma_2)a(l_1,\tau_1)a(l_2,\tau_2)
\rangle_s \nonumber\\
&&\hspace{6pt}-
\phi(q,k_1,k_2) \delta (q + \sigma_1 k_1+  \sigma_2 k_2)e^{it\omega(p)} 
\langle  a(p)^\ast a(k_2,\sigma_2)a(l_1,\tau_1)a(l_2,\tau_2)
\rangle_s\Big) \,.\nonumber
\end{eqnarray}
As our basic assumption, in the kinetic scaling regime, the
average $\langle\cdot\rangle_s$ at the arguments in question is in
a good approximation a locally stationary measure. If so, the
averages appearing in (\ref{5.7}) can be substituted by Gaussian
pairings. Using the shorthand $k$ for $a(k,\sigma)$, the approximation amounts to 
\begin{eqnarray}\label{5.8}
&&\langle k_1k_2l_1l_2\rangle_s = 
\langle k_1l_1\rangle_s \langle k_2l_2\rangle_s +
\langle k_1l_2\rangle_s\langle k_2l_1\rangle_s + 
\langle k_1k_2\rangle_s \langle l_1l_2\rangle_s\,,\nonumber\\
&&\hspace{0pt}
\langle qk_2l_1l_2\rangle_s = 
\langle ql_1\rangle_s\langle k_2l_2\rangle_s +
\langle ql_2\rangle_s\langle k_2l_1\rangle_s
+ \langle qk_2\rangle_s \langle l_1l_2\rangle_s \,,
\end{eqnarray}
and correspondingly for $p$. By symmetry, upon inserting in (\ref{5.7}), the first two terms on the right are identical and will yield the gain and loss term, respectively. The third pairing is subleading and vanishes as  $\varepsilon \to 0$. Accordingly we set
 \begin{equation}\label{5.9c}
G = G_{\mathrm{gain}} + G_{\mathrm{loss}}+ G_{\mathrm{sub}}\,.\medskip
\end{equation}
{(2) {\it Gain and loss term}. In $G_{\mathrm{gain}}$ we change to Wigner fucntion variables as
\begin{eqnarray}\label{5.10}
&&k_1 = k' - \varepsilon\sigma_1\eta'/2\,,\quad l_1 = k' + \varepsilon\sigma_1\eta'/2\,,
\quad \tau_1 = - \sigma_1\,,\nonumber\\
&&\hspace{0pt}k_2 = k'' - \varepsilon\sigma_2\eta''/2\,,\quad l_2 = k'' + 
\varepsilon\sigma_2\eta''/2\,,\quad \tau_2 = - \sigma_2\,.
\end{eqnarray}
The $\varepsilon$-dependence of $\phi$ can be ignored and the $\eta$-integration 
is extended to $\mathbb{R}^3$, since by asumption $\widehat{W}^\varepsilon(\eta,k,t)$
has a good decay in $\eta$. The phases have to be expanded to first order in
 $\varepsilon$. Then 
 \begin{eqnarray}\label{5.11}
&&\hspace{-24pt}G_{\mathrm{gain}}(k+\varepsilon\eta/2,k-\varepsilon\eta/2,t,s) \nonumber\\
&&\hspace{-10pt}= 2 \varepsilon^6  \sum_{\sigma_1,\sigma_2=\pm1}\int_{\mathbb{T}^6} d
k' d k''\int_{\mathbb{R}^6}d\eta' d\eta'' \phi(k,k',k'')^2
 e^{it(\sigma_1\omega(k' - \varepsilon\sigma_1\eta'/2)+\sigma_2\omega(k'' - \varepsilon\sigma_2\eta''/2))}\nonumber\\
&&\hspace{2pt}\times\Big(
\delta (-k + \varepsilon(\eta/2) + \sigma_1 (k' - \varepsilon\sigma_1\eta'/2)+  \sigma_2
(k'' - \varepsilon\sigma_2\eta''/2))\nonumber\\
&&\hspace{2pt} \delta (k+ \varepsilon(\eta/2) - \sigma_1
(k' + \varepsilon\sigma_1\eta'/2)-
  \sigma_2 (k'' + 
\varepsilon\sigma_2\eta''/2))e^{-it\omega(k+ \varepsilon\eta/2)}\nonumber\\
&&\hspace{2pt} +
\delta (k+ \varepsilon(\eta/2)+ \sigma_1
(k' - \varepsilon\sigma_1\eta'/2)+
\sigma_2 (k'' - \varepsilon\sigma_2\eta''/2))
\nonumber\\
&&\hspace{2pt}\delta (-k + \varepsilon(\eta/2) - \sigma_1 (k' + \varepsilon\sigma_1\eta'/2)-  \sigma_2
(k'' + \varepsilon\sigma_2\eta''/2))e^{it\omega(k- \varepsilon\eta/2)}\Big)\nonumber\\
&&\hspace{2pt}
\langle  a(k' - \varepsilon\sigma_1\eta'/2,\sigma_1)
a(k' + \varepsilon\sigma_1\eta'/2,-\sigma_1)
\rangle_s \nonumber\\
&&\hspace{2pt}
\times  \langle a(k'' - \varepsilon\sigma_2\eta''/2,\sigma_2)a(k'' + 
\varepsilon\sigma_2\eta''/2,-\sigma_2)
\rangle_s\nonumber\\
&&\hspace{-10pt}
=  2 \varepsilon^{-3}  \sum_{\sigma_1,\sigma_2=\pm1}\int_{\mathbb{T}^6} d
k' d k''\int_{\mathbb{R}^6}d\eta' d\eta'' \phi(k,k',k'')^2
e^{it(\sigma_1\omega(k')+\sigma_2\omega(k''))}\nonumber\\
&&\hspace{2pt}
e^{-it \varepsilon(\nabla\omega(k)(\eta/2)+\nabla\omega(k')(\eta'/2)+\nabla\omega(k'')(\eta''/2)
)}
 \nonumber\\
&&\hspace{2pt}
\big(e^{-it\omega(k)}\delta(k-\sigma_1k' - \sigma_2k'') + 
e^{it\omega(k)}\delta(k+\sigma_1k' + \sigma_2k'')\big)\nonumber\\
&&\hspace{2pt}
\delta(\eta-\eta'-\eta'')\widehat{W}^\varepsilon(\eta',k',s)\widehat{W}^\varepsilon(\eta'',k'',s)
\nonumber\\
&&\hspace{-10pt}
= 2 \varepsilon^{-3}  \sum_{\sigma_1,\sigma_2=\pm1}\int_{\mathbb{T}^6} d
k' d k''\int_{\mathbb{R}^6}d\eta' d\eta'' \phi(k,k',k'')^2\big(
e^{it(-\omega(k) +\sigma_1\omega(k')+\sigma_2\omega(k''))} + c.c.\big)
\nonumber\\
&&\hspace{2pt}
\delta(k-\sigma_1k' - \sigma_2k'')e^{-it \varepsilon(\nabla\omega(k)(\eta/2)+\nabla\omega(k')(\eta'/2)+\nabla\omega(k'')(\eta''/2))}\nonumber\\
&&\hspace{2pt}
\delta(\eta-\eta'-\eta'')\widehat{W}^\varepsilon(\eta',k',s)\widehat{W}^\varepsilon(\eta'',k'',s)\,,
\end{eqnarray}
where in the last step the $c.c.$ term arises through replacing the sum over $\sigma_1,\sigma_2$ by the sum over $-\sigma_1,
-\sigma_2$.

In $G_{\mathrm{loss}}$ we change to Wigner function variables as
\begin{eqnarray}\label{5.12}
&&\int_{\mathbb{T}^3} dk_3\delta(q-k_3)  \langle k_3l_1\rangle_s\langle k_2l_2\rangle_s \,,\nonumber\\
&&\hspace{0pt}
l_1 = k' - \varepsilon\eta'/2\,,\quad k_3 = k' + \varepsilon\eta'/2\,,
\quad \tau_1 = 1\,,\nonumber\\
&&\hspace{0pt}k_2 = k'' - \varepsilon\sigma_2\eta''/2\,,\quad l_2 = k'' + 
\varepsilon\sigma_2\eta''/2\,,\quad \tau_2 = - \sigma_2\,,
\end{eqnarray}
and
\begin{eqnarray}\label{5.12a}
&&\int_{\mathbb{T}^3} dk_3\delta(p-k_3)  \langle k_3l_1\rangle_s\langle k_2l_2\rangle_s \,,\nonumber\\
&&\hspace{0pt}
k_3 = k' - \varepsilon\eta'/2\,,\quad l_1 = k' + \varepsilon\eta'/2\,,
\quad \tau_1 = -1\,,\nonumber\\
&&\hspace{0pt}k_2 = k'' - \varepsilon\sigma_2\eta''/2\,,\quad l_1 = k'' + 
\varepsilon\sigma_2\eta''/2\,,\quad \tau_2 = - \sigma_2\,.
\end{eqnarray}
Then
\begin{eqnarray}\label{5.13}
&&\hspace{-24pt}G_{\mathrm{loss}}(k+\varepsilon\eta/2,k-\varepsilon\eta/2,t,s) 
\nonumber\\
&&\hspace{-12pt}
= -4 \varepsilon^6  \sum_{\sigma_1,\sigma_2=\pm1}\int_{\mathbb{T}^9} dk_1d
k' d k''\int_{\mathbb{R}^6}d\eta' d\eta'' \phi(k_1,k',k'') \phi(k,k_1,k'')\sigma_1
\nonumber\\
&&\hspace{-2pt}
e^{it(\sigma_1\omega(k_1)+\sigma_2\omega(k''))}
e^{it \varepsilon(-\nabla\omega(k')(\eta'/2)-\nabla\omega(k'')(\eta''/2))}
\nonumber\\
&&\hspace{-2pt}
\Big(e^{-it\omega(k')}\delta(-\sigma_1k_1 + k'-  \varepsilon(\eta'/2) - \sigma_2k''-  \varepsilon(\eta''/2)) \nonumber\\
&&\hspace{-2pt}
\times \delta (-k + \varepsilon(\eta/2) + \sigma_1k_1+  \sigma_2 k'' - \varepsilon(\eta''/2))
\delta (k + \varepsilon(\eta/2) - k' - \varepsilon(\eta'/2))
\nonumber\\
&&\hspace{-2pt}- 
e^{it\omega(k')}\delta(-\sigma_1k_1 - k'-  \varepsilon(\eta'/2) - \sigma_2k''-  \varepsilon(\eta''/2)) \nonumber\\
&&\hspace{-2pt}
\times\delta (k + \varepsilon(\eta/2) + \sigma_1k_1 +  \sigma_2 k'' - \varepsilon(\eta''/2))
\delta (k - \varepsilon(\eta/2) - k' + \varepsilon(\eta'/2))\Big)\nonumber\\
&&\hspace{-2pt}
\widehat{W}^\varepsilon(\eta',k',s)\widehat{W}^\varepsilon(\eta'',k'',s)\,.
\end{eqnarray}
We integrate over $k'$ and neglect the shift of order $\varepsilon$ in the $\eta$-argument of 
$\widehat{W}^\varepsilon$. In the second summand $\sigma_1,\sigma_2$ is substituted by 
$-\sigma_1,-\sigma_2$ with the result
\begin{eqnarray}\label{5.14}
&&\hspace{-24pt}G_{\mathrm{loss}}(k+\varepsilon\eta/2,k-\varepsilon\eta/2,t,s)
\nonumber\\
&&\hspace{-12pt}
 = -4 \varepsilon^{-3}  \sum_{\sigma_1,\sigma_2=\pm1}\int_{\mathbb{T}^6} dk_1
d k''\int_{\mathbb{R}^6}d\eta' d\eta'' \phi(k,k_1,k'')^2\big(e^{it(\omega(k) - \sigma_1\omega(k_1)-\sigma_2\omega(k''))}  + c.c.\big)
\nonumber\\
&&\hspace{-2pt}
\times\sigma_1\delta(k-\sigma_1k_1 - \sigma_2k'')
e^{-it \varepsilon(\nabla\omega(k)(\eta/2)+\nabla\omega(k'')(\eta''/2))}\nonumber\\
&&\hspace{-2pt}
\times\delta(\eta-\eta'-\eta'')\widehat{W}^\varepsilon(\eta',k,s)\widehat{W}^\varepsilon(\eta'',k'',s)\,.
\end{eqnarray}

By assumption the Wigner function is varying on the kinetic scale. Thus the remaining time integration for $G_{\mathrm{gain}}$ and $G_{\mathrm{loss}}$ is of the generic form
\begin{eqnarray}\label{5.15}
&&\lim_{\varepsilon \to 0} \int d\omega g(\omega) \int _0^{t/\varepsilon}ds (e^{i\omega(t-s)} +
e^{-i\omega(t-s)} )f(\varepsilon s, \varepsilon(\varepsilon^{-1}t -s))
\nonumber\\
&&\hspace{0pt}
\lim_{\varepsilon \to 0} \int d\omega g(\omega) \varepsilon^{-1}
2\int _0^{t}ds\cos(\omega s/\varepsilon) f(t-s,s)
\nonumber\\
&&\hspace{10pt}
= 2\pi  \int d\omega g(\omega) \delta(\omega)f(t,0)\,,
\end{eqnarray}
where
$g(\omega)$ is some smooth  test function of rapid decay.

Combining (\ref{5.11}), (\ref{5.14}), (\ref{5.15}) and upon noting that the convolution 
becomes multiplication in position space, one concludes that
\begin{eqnarray}\label{5.16}
&&\hspace{-30pt}\lim_{\varepsilon \to 0} \int _0^{t/\varepsilon}ds
(\varepsilon/2)^3
\int_{(2\mathbb{T}/\varepsilon)^3} d \eta e^{ i 2\pi \lfloor
r\rfloor_\varepsilon \cdot \eta}
\Big( G^\varepsilon_{\mathrm{gain}}(k+\varepsilon\eta/2,k-\varepsilon\eta/2,
(\varepsilon^{-1}t-s),s) 
\nonumber\\
&&\hspace{100pt}+ G^\varepsilon_{\mathrm{loss}}(k+\varepsilon\eta/2,k-\varepsilon\eta/2,
(\varepsilon^{-1}t-s),s)\Big)
\nonumber\\
&&\hspace{-20pt}
= \lambda^2\frac{\pi}{2}\sum_{\sigma_1,\sigma_2=\pm1}
\int_{\mathbb{T}^6}dk_1dk_2(\omega(k)\omega(k_1)\omega(k_2))^{-1}
\delta(\omega-\sigma_1\omega_1-\sigma_2\omega_2)
\nonumber\\
&&\hspace{-20pt}\delta( k-\sigma_1k_1-\sigma_2k_2)\big(W(r,k_1,t)W(r,k_2,t)-
2\sigma_1W(r,k,t)
W(r,k_2,t)\big)\,,
\end{eqnarray}
which agrees with the collision term (\ref{4.7}).\medskip\\
(3) {\it Subleading terms}. There are two subleading terms from (\ref{5.8}), denoted here by
$G_{\mathrm{sub}} = G_{\mathrm{sub1}} + G_{\mathrm{sub2}}$. For $G_{\mathrm{sub1}}$
we change to Wigner function variables as
\begin{eqnarray}\label{5.17}
&&k_1 = k' - \varepsilon\sigma_1\eta'/2\,,\quad k_2 = k' + \varepsilon\sigma_1\eta'/2\,,
\quad \sigma_2 = - \sigma_1\,,\nonumber\\
&&\hspace{0pt}l_1 = k'' - \varepsilon\tau_1\eta''/2\,,\quad l_2 = k'' + 
\varepsilon\tau_1\eta''/2\,,\quad \tau_2 = - \tau_1\,.
\end{eqnarray}
Then
\begin{eqnarray}\label{5.18}
&&\hspace{-24pt}G_{\mathrm{sub1}}(k+\varepsilon\eta/2,k-\varepsilon\eta/2,t,s) \nonumber\\
&&\hspace{-10pt}
=     \sum_{\sigma_1,\tau_1=\pm1}\int_{\mathbb{T}^6} d
k' d k''\int_{\mathbb{R}^6}d\eta' d\eta'' \phi(k,k',k') \phi(k,k'',k'')e^{-it \varepsilon(\nabla\omega(k')\eta'+\nabla\omega(k)(\eta/2))}\nonumber\\
&&\hspace{2pt}\Big(
 e^{-it\omega(k)}
\delta (-k + \varepsilon(\eta/2)  - \varepsilon\eta')
\delta (k + \varepsilon(\eta/2)  - \varepsilon\eta'')
\nonumber\\
&&\hspace{2pt}
+ e^{it\omega(k)}\delta (k + \varepsilon(\eta/2)  - \varepsilon\eta')
\delta (k - \varepsilon(\eta/2)  + \varepsilon\eta'')\Big)
\widehat{W}^\varepsilon(\eta',k',s)\widehat{W}^\varepsilon(\eta'',k'',s)
\nonumber\\
&&\hspace{-10pt}
= 4 \varepsilon^{-3} \int_{\mathbb{T}^6} d
k' d k''\int_{\mathbb{R}^6}d\eta' d\eta'' \phi(k,k',k') \phi(k,k'',k'')\big(
e^{-it\omega(k)} + e^{it\omega(k)} \big)\nonumber\\
&&\hspace{2pt}
 \delta(k)e^{-it \varepsilon(\nabla\omega(k')\eta'+\nabla\omega(k)(\eta/2))}\delta(\eta-\eta'-\eta'')\widehat{W}^\varepsilon(\eta',k',s)\widehat{W}^\varepsilon(\eta'',k'',s)\,.
\end{eqnarray}
The remaining time integration is of the generic form
\begin{eqnarray}\label{5.18a}
&&\hspace{0pt} \int _0^{t/\varepsilon}ds \cos(\omega(0)( \varepsilon^{-1}t-s))
f(\varepsilon s) =   \varepsilon^{-1}
\int _0^{t}ds \cos (\omega(0)s/ \varepsilon)
f(t- s)
\nonumber\\
&&\hspace{0pt}
= \omega(0)^{-1}\Big( \sin (\omega(0)t/ \varepsilon)f(0) + \int _0^{t}ds  \sin(\omega(0)s/\varepsilon) 
f'(t- s)\Big)\,.
\end{eqnarray}
The second summand is of order $\varepsilon$. The first summand oscillates fastly around
zero average and thus vanishes by one further integration in time.

Our argument indicates that  $\omega(0) > 0$ is required. If  $\omega(0) = 0$, then the product
$\delta(k)\omega(k)^{-1}$ is not defined. Whether this is an artifact of the
derivation or signals a limit in the validity of the kinetic description remains to be understood.

For $G_{\mathrm{sub2}}$
we change to Wigner function variables as
\begin{eqnarray}\label{5.18b}
&&\int_{\mathbb{T}^3} dk_3\delta(q-k_3)  \langle k_3k_2\rangle_s\langle l_1l_2\rangle_s \,,\nonumber\\
&&k_2 = k' - \varepsilon\eta'/2\,,\quad k_3 = k' + \varepsilon\eta'/2\,,
\quad \sigma_2 = 1\,,\nonumber\\
&&\hspace{0pt}l_1 = k'' - \varepsilon\tau_1\eta''/2\,,\quad l_2 = k'' + 
\varepsilon\tau_1\eta''/2\,,\quad \tau_2 = - \tau_1\,,
\end{eqnarray}
and
\begin{eqnarray}\label{5.19}
&&\int_{\mathbb{T}^3} dk_3\delta(p-k_3)  \langle k_3k_2\rangle_s\langle l_1l_2\rangle_s \,,\nonumber\\
&&k_3 = k' - \varepsilon\eta'/2\,,\quad k_2 = k' + \varepsilon\eta'/2\,,
\quad \sigma_2 = -1\,,\nonumber\\
&&\hspace{0pt}l_1 = k'' - \varepsilon\tau_1\eta''/2\,,\quad l_2 = k'' + 
\varepsilon\tau_1\eta''/2\,,\quad \tau_2 = - \tau_1\,.
\end{eqnarray}
Then
\begin{eqnarray}\label{5.20}
&&\hspace{-24pt}G_{\mathrm{sub2}}(k+\varepsilon\eta/2,k-\varepsilon\eta/2,t,s) \nonumber\\
&&\hspace{0pt}
=    -2 \sum_{\sigma_1,\tau_1=\pm1}\int_{\mathbb{T}^9} dk_1d
k' d k''\int_{\mathbb{R}^6}d\eta' d\eta'' \phi(k,k_1,k')\phi(k_1,k'',k'')\nonumber\\
&&\hspace{0pt} 
e^{-it \varepsilon(\nabla\omega(k')(\eta'/2)+\nabla\omega(k)(\eta/2))}
\delta(-\sigma_1k_1 - 
 \varepsilon \eta'')\Big(
 e^{it(\sigma_1\omega(k_1) + \omega(k') -\omega(k) )}
 \nonumber\\
&&\hspace{0pt}\times\delta (k + \varepsilon(\eta/2) -k' - \varepsilon(\eta'/2))
\delta (-k + \varepsilon(\eta/2) +\sigma_1k_1 + k' - \varepsilon(\eta'/2))
 \nonumber\\
&&\hspace{0pt}
 - e^{it(\sigma_1\omega(k_1) - \omega(k') +\omega(k) )}
\delta (k - \varepsilon(\eta/2) -k' + \varepsilon(\eta'/2))
\nonumber\\
&&\hspace{0pt}\times\delta (k + \varepsilon(\eta/2) +\sigma_1k_1 - k' - \varepsilon(\eta'/2))\Big)
\sigma_1
\widehat{W}^\varepsilon(\eta',k',s)\widehat{W}^\varepsilon(\eta'',k'',s)
\nonumber\\
&&\hspace{0pt}
= -4 \sum_{\sigma_1=\pm1}\int_{\mathbb{T}^9} dk_1d
k' d k''\int_{\mathbb{R}^6}d\eta' d\eta'' \phi(k_1,k,k)\phi(k_1,k'',k'') 
\nonumber\\
&&\hspace{0pt}
 e^{it\sigma_1\omega(k_1)}
\Big(\delta (k  -k' + \varepsilon(\eta/2)- \varepsilon(\eta'/2))
- \delta (k  -k' - \varepsilon(\eta/2) + \varepsilon(\eta'/2))  \Big)
\nonumber\\
&&\hspace{0pt}
\delta(\sigma_1k_1  + \varepsilon \eta'')e^{-it \varepsilon\nabla\omega(k')\eta'}\delta(\eta-\eta'-\eta'')\sigma_1\widehat{W}^\varepsilon(\eta',k',s)\widehat{W}^\varepsilon(\eta'',k'',s)\,.
\end{eqnarray}
Integrating over $k_1$ yields the phase $\omega(\varepsilon \eta'')$. If $\omega(0) > 0$, the 
remaining time integration is of order 1. The difference of $\delta$-functions in the large round bracket is of order $\varepsilon$, when integrated against  $\widehat{W}^\varepsilon$. Therefore the second subleading term vanishes as $\varepsilon \to 0$.

%%%%%%%%%%%%%%%%%%%%%%%%%%%%%%%%
%%%%%%%%%%%%%%%%%%%%%%%%%%%%%%%%%%%%

\section{Some properties of the classical phonon Boltzmann
equation}\label{sec.6} \setcounter{equation}{0}

\textit{(i) Energy}. The energy at position $r$ and time $t$ on
the kinetic scale is defined through
\begin{equation}\label{6.1}
e(r,t) =  \int_{\mathbb{T}^3} d k \omega(k)
W(r,k,t)\,.
\end{equation}
It satisfies the local conservation law
\begin{equation}\label{6.2}
\frac{\partial}{\partial t}e(r,t) + \nabla \cdot j_\textrm{e}(r,t)
=0\,.
\end{equation}
From the transport term one concludes that the energy current is
given by
\begin{equation}\label{6.3}
j_\textrm{e}(r,t) = (2\pi)^{-1} \int_{\mathbb{T}^3} d k
(\nabla\omega(k))\omega(k) W(r,k,t)\,.
\end{equation}
The vanishing of the contribution from the collision term can be
seen from
\begin{eqnarray}\label{6.4}
&&\hspace{-12pt}\int_{\mathbb{T}^9} d k_1 d k_2 dk_3
(\omega(k_1)\omega(k_2)\omega(k_3))^{-1}
\big\{2\delta(\omega(k_1)+\omega(k_2)-\omega(k_3))\delta(k_1+k_2-k_3)\nonumber\\
&&\omega(k_1)\big(W(k_2)W(k_3)+W(k_1)W(k_3)-W(k_1)W(k_2)\big)+
\delta(\omega(k_1)-\omega(k_2)-\omega(k_3))\nonumber\\
&&\delta(k_1-k_2-k_3)\omega(k_1)\big(W(k_2)W(k_3)-W(k_1)W(k_2)-W(k_1)W(k_3)\big)\big\}\nonumber\\
&&\hspace{-12pt}= \int_{\mathbb{T}^9} d k_1 d k_2 dk_3
(\omega(k_1)\omega(k_2)\omega(k_3))^{-1}
\delta(\omega(k_1)+\omega(k_2)-\omega(k_3))\delta(k_1+k_2-k_3)\nonumber\\
&&\hspace{26pt}\omega(k_3)\big(W(k_2)W(k_3)+W(k_1)W(k_3)-W(k_1)W(k_2)\nonumber\\
&&\hspace{26pt}+W(k_1)W(k_2)-W(k_3)W(k_1)-W(k_3)W(k_2)\big)=0\,.
\end{eqnarray}
We used here the symmetrization of $2\omega(k_1)$ to $\omega(k_1)
+ \omega(k_2)$, the energy conservation
$\omega(k_1)+\omega(k_2)=\omega(k_3)$ in term (I), and the cyclic
substitution $k_1 \to k_3$, $k_3 \to k_2$, $k_2 \to k_1$ in term
(II).

If the ergodicity condition (E) holds, energy is the only conservation law, see the discussion at
the end of Section \ref{sec.9}.\medskip\\
\textit{(ii) Entropy}. Following (\ref{3.11}), up to a constant,
the local entropy at position $r$ and time $t$ on the kinetic
scale is defined through
\begin{equation}\label{6.5}
s(r,t) =  \int_{\mathbb{T}^3}dk \log W(r,k,t)\,.
\end{equation}
It satisfies the semi-conservation law
\begin{equation}\label{6.6}
\frac{\partial}{\partial t}s(r,t) + \nabla \cdot j_\textrm{s}(r,t)
= \sigma(r,t)
\end{equation}
with the entropy flow
\begin{equation}\label{6.7}
j_\textrm{s}(r,t) = (2\pi)^{-1} \int_{\mathbb{T}^3}dk
\nabla\omega(k) \log W(r,k,t)
\end{equation}
and the entropy production
\begin{eqnarray}\label{6.8}
&&\sigma(r,t) = \gamma \int_{\mathbb{T}^9} d k_1 d k_2 dk_3
(\omega(k_1)\omega(k_2)\omega(k_3))^{-1}
\delta(\omega(k_1)+\omega(k_2)-\omega(k_3))\notag\\
&&\hspace{96pt}\delta(k_1+k_2-k_3)W(r,k_1,t)W(r,k_2,t)W(r,k_3,t)\notag\\
&&\hspace{96pt}\big(W(r,k_1,t)^{-1}+W(r,k_2,t)^{-1}-W(r,k_3,t)^{-1}\big)^2\,.
\end{eqnarray}
Clearly $\sigma \geq 0$. To derive the expression (\ref{6.8}) one
uses the same identities as for the energy,
\begin{eqnarray}\label{6.9}
&&\hspace{-8pt}\gamma \int_{\mathbb{T}^9} d k_1 d k_2 dk_3
(\omega(k_1)\omega(k_2)\omega(k_3))^{-1}W(k_1)^{-1}
\big\{2\delta(\omega(k_1)+\omega(k_2)-\omega(k_3))\nonumber\\
&&\hspace{26pt}\delta(k_1+k_2-k_3)\big(W(k_2)W(k_3)+W(k_1)W(k_3)-W(k_1)W(k_2)\big)\nonumber\\
&&\hspace{26pt}+\delta(\omega(k_1)-\omega(k_2)-\omega(k_3))\delta(k_1-k_2-k_3)\nonumber\\
&&\hspace{26pt}\big(W(k_2)W(k_3)-W(k_1)W(k_2)-W(k_1)W(k_3)\big)\big\}\nonumber\\
&&\hspace{-8pt}= \gamma \int_{\mathbb{T}^9} d k_1 d k_2 dk_3
(\omega(k_1)\omega(k_2)\omega(k_3))^{-1}
\delta(\omega(k_1)+\omega(k_2)-\omega(k_3))\delta(k_1+k_2-k_3)\nonumber\\
&&\hspace{26pt}W(k_1)W(k_2)W(k_3) \big(W(k_1)^{-2}+W(k_2)^{-2} +
2W(k_1)^{-1}W(k_2)^{-1}\nonumber\\
&&\hspace{26pt}-2W(k_1)^{-1}W(k_3)^{-1}+W(k_3)^{-2}-W(k_3)^{-1}
W(k_2)^{-1}-W(k_3)^{-1}W(k_2)^{-1}\big)\nonumber\\
&&\hspace{-8pt}= \sigma\,.
\end{eqnarray}

The entropy production vanishes if and only if
\begin{equation}\label{6.10}
W(k_1)^{-1}+W(k_2)^{-1}-W(k_1+k_2)^{-1}=0
\end{equation}
on the set $\{(k_1,k_2)\in \mathbb{R}^6\,|\,
\omega(k_1)+\omega(k_2)=\omega(k_1+k_2)\}$. As will be discussed in
Section \ref{sec.9}, if the ergodicity condition (E) holds, the
only solution to (\ref{6.10}) is
\begin{equation}\label{6.11}
W_\beta(k)=\frac{1}{\beta\omega(k)}\,.
\end{equation}
$\beta>0$ is a free parameter. Physically, $\beta$ is the inverse
temperature, $\beta=(k_\textrm{B}T)^{-1}$. We will use temperature
units such that $k_\textrm{B}=1$.\medskip\\
\textit{(iii) Stationary solutions}. For the spatially homogeneous
Boltzmann equation under the ergodicity condition (E)
the only stationary solutions are of the form (\ref{6.11}). If
there would be another stationary solution, its entropy production
has to vanish, which means (\ref{6.10}) has to be satisfied, in contradiction to
(\ref{6.11}) being the only solution of (\ref{6.10}).\medskip

(\ref{6.11}) is in accordance with equilibrium statistical
mechanics. In thermal equilibrium, the nonlinearity can be ignored
in the kinetic limit
and the Gibbs distribution is $Z^{-1}\exp[-\beta H_0]$. This is a
Gaussian measure with Wigner function
$W_\beta(k)=(\beta\omega(k))^{-1}$.

To have the one-parameter family (\ref{6.11}) as the only
stationary solutions is a remarkable prediction of the phonon
Boltzmann equation. It means that the weak nonlinearity
thermalizes the gas of phonons. For example, one could set up an
initial state with nonvanishing phonon current,
$j_\textrm{n}(0)=(2\pi)^{-1}$ $\int_{\mathbb{T}^3} dk
\nabla\omega(k)W(k,t=0)\neq 0$. Through umklapp processes this
current degrades in the course of time and $\lim_{t\to
\infty}j_\textrm{n}(t)=0$. If there are no umklapp processes, as for the continuum 
wave equation below, on the kinetic level there are stationary states which maintain 
a constant phonon current.

%%%%%%%%%%%%%%%%%%%%%%%%%%%%%%%%%%%%%

\section{Wave turbulence}\label{sec.7a}
\setcounter{equation}{0}

Wave turbulence has become a generic term for, possibly
multicomponent, wave equations with weak nonlinearity. Examples
are listed in \cite{Za} and include waves on liquid surfaces,
acoustic turbulence, and the nonlinear Schr\"{o}dinger equation
for dispersive media. The link to our discussion comes from the
fact that apparently kinetic theory is the most powerful method
available to handle the nonlinearities, a concrete field of
application being the dynamics of ocean waves \cite{Jan}. The
underlying physical space is $\mathbb{R}^3$, possibly $
\mathbb{R}^2$, which means that we briefly return to the continuum
setting from the end of Section \ref{sec.2}. For wave turbulence,
typically one is interested in a stationary nonequilibrium state
which is sustained by pumping in energy at large scales and
dissipating it at small scales. Thus the focus is on stationary
solutions of the spatially homogeneous equation with the
appropriate source terms added. Here we only discuss the
derivation of the kinetic equation from the Klein-Gordon equation
(\ref{2.14a}).

(\ref{2.14a}) has the dispersion relation
$\omega(k)=(\omega_0^2+k^2)^{1/2}$, $\omega_0\geq 0$. For
three-wave interactions the resonance condition reads
\begin{equation}\label{7a.1}
\omega(k_1)+\omega(k_2)= \omega(k_1+k_2)\,,
\end{equation}
where momentum conservation, $k_1+k_2=k_3$, has been used already.
If $\omega_0>0$, then
\begin{equation}\label{7a.2}
\omega(k_1+k_2)<\omega(k_1)+\omega(k_2)
\end{equation}
and (\ref{7a.1}) cannot be satisfied. If $\omega_0=0$, the vectors
must be collinear which again yields a vanishing collision term.
Thus on the kinetic time scale we have to turn to four-wave
interactions in order to have the nonlinearity still in effect.

The ``simplest'' example is
\begin{equation}\label{7a.3}
 \frac{\partial^2}{\partial t^2}\phi=
 \Delta\phi-\omega_0^2\phi -\sqrt{\varepsilon}\lambda\phi^3\,.
\end{equation}
As in (\ref{7a.2}) one concludes that the merging of three phonons
into one and the splitting of one phonon into three are forbidden
processes on the kinetic scale. The only remaining possibility are
pair collisions, see Figure \ref{Fig3}.
For the formal derivation of the kinetic equation one proceeds as
in Section \ref{sec.5} with the result
\begin{eqnarray}\label{7a.4}
&&\hspace{-10pt}\frac{\partial}{\partial t}W(k)
+ \nabla\omega(k)\cdot\nabla_r W(k)=\\
&&\hspace{15pt} \frac{9\pi}{4}\lambda^2 (2\pi)^{-3} \int d^3k_1
d^3k_2 d^3k_3 \big(\omega(k)\omega(k_1)\omega(k_2)\omega(k_3)\big)^{-1}\nonumber\\
&&\hspace{15pt}\delta(\omega(k)+\omega(k_1)-\omega(k_2)-\omega(k_3))
\delta(k+k_1-k_2-k_3)\nonumber\\
&&\hspace{15pt}\big[W(k_1)W(k_2)W(k_3)+W(k)(W(k_2)W(k_3)-
W(k_1)W(k_3)-W(k_1)W(k_2))\big]\,.\nonumber
\end{eqnarray}
Here we use the standard convention for Fourier transformation in
$\mathbb{R}^3$ and $\int d^3k$ is understood as the integration
over all of $\mathbb{R}^3$.

In a recent series of studies Nazarenko and coworkers reconsider
the derivation of (\ref{7a.4}) from a different perspective. We
explain the method in Appendix \ref{sec.16d}.

\begin{figure}
\setlength{\unitlength}{1cm}
\begin{picture}(10,2)(-2,-1)\thicklines
\put(3.4,1.2){\vector(4,-3){1.6}}
\put(3.2,-0.45){\vector(4,1){1.8}}
\put(5,0){\vector(4,0){2.0}}
\put(5,0){\vector(2,-1){1.7}}
\put(6.3,0.2){$k_2$} \put(5.7,-0.9){$k_3$}
\put(4.0,0.2){$k$} \put(3.4,-0.9){$k_1$}
\end{picture}
\caption{A four phonon collision with number conservation.}\label{Fig3}
\end{figure}
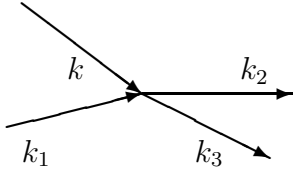

The kinetic equation (\ref{7a.4}) preserves number, momentum, and
energy of phonons. This is also reflected by the formally
stationary solutions
\begin{equation}\label{7a.5}
W_{\alpha\beta\gamma}(k)= (\beta\omega + \alpha\cdot k
+\gamma)^{-1}
\end{equation}
with $\beta>|\alpha|$ and
\begin{equation}\label{7a.6}
\gamma+(\beta^2-\alpha^2)^{-1/2}(\beta^2+\alpha^2)>0\,,
\end{equation}
so to have $W_{\alpha\beta\gamma}(k)\geq0$. The solutions
(\ref{7a.5}) have infinite energy because of the divergence at
large $k$. Such states have not been included in our set-up. In
particular, starting from finite energy initial data, the system
cannot properly reach thermal equilibrium.

The  additional conservation laws are also reflected in the size
dependence of the thermal conductance. At the high temperature
side of the sample on the average more phonons are created than at
the low temperature side. The collisions conserve momentum. Thus
there is a laminar flow of phonons which transports energy
independently of the size of the sample. In distinction, a real
fluid has diffusive energy transport, since no particles are
created, resp. destroyed, at the boundary.

To turn to the issue of wave turbulence, one considers a spatially
homogeneous situation and augments the kinetic equation
(\ref{7a.4}) phenomenologically with a driving term as
\begin{equation}\label{7a.7}
\frac{\partial}{\partial t}W(k,t) = \mathcal{C}(W(t))(k) +
\Gamma(k)W(k,t)\,,
\end{equation}
where as a shorthand the collision operator is denoted by
$\mathcal{C}(W)$. One is interested in the steady state
$W_\textrm{s}$, for which $\partial W_\textrm{s}/\partial t=0$.
From the $H$-theorem we know that $\int d^3k
W_\textrm{s}(k)^{-1}\mathcal{C}(W_\textrm{s})(k)>0$. Therefore
\begin{equation}\label{7a.8}
\int d^3k \Gamma (k)<0\,.
\end{equation}
In addition, to have energy and phonon number conservation in the
steady state it must hold that

\begin{equation}\label{7a.9}
\int d^3k \omega(k) W_\textrm{s}(k)\Gamma(k)=0\,,\quad \int d^3k
W_\textrm{s}(k)\Gamma(k)=0\,.
\end{equation}
We imagine to have a narrow band source of energy at small $k$ and
a sink at large $k$, compare with (\ref{7a.7}). In the
intermediate regime one has to solve then
\begin{equation}\label{7a.10}
\mathcal{C}(W_\textrm{s})=0\,.
\end{equation}

To be specific let us consider the wave equation with
$\omega(k)=|k|$, i.e.~$\omega_0=0$. Since $\omega$ is homogeneous,
and so are the collision rates, it is natural to look for a
self-similar solution of (\ref{7a.10}) of the form
$W_\textrm{s}(k)=|k|^{-\sigma}$. Indeed, besides the equilibrium
values $\sigma=0,1$, one obtains the solutions
\begin{equation}\label{7a.11}
W_\textrm{s}^{(\textrm{e})}(k)=|k|^{-5/3}\,,\quad
W_\textrm{s}^{(\textrm{n})}(k)=|k|^{-4/3}\,,\quad
\textrm{dimension}\; d=3\,.
\end{equation}
As their equilibrium counterpart, the solutions (\ref{7a.11}) have
infinite energy because of ultraviolet divergence. The true steady
state for (\ref{7a.7}) has the power law of (\ref{7a.11}) only in
some intermediate regime and the at large $|k|$ negative $\Gamma$
supposedly ensures that $\int d^3k W_\textrm{s} (k)<\infty$.

The physical meaning of the steady states in (\ref{7a.11}) can be
understood through studying the flux in energy space \cite{Za}. It
turns out that $W_\textrm{s}^{(\textrm{e})}$ supports a constant
energy flux directed from small $\omega$ to large $\omega$, while
$W_\textrm{s}^{(\textrm{n})}$ supports a constant phonon number
flux directed from large $\omega$ to small $\omega$.

\section{Quantizing phonons, locally quasifree states}\label{sec.7}
\setcounter{equation}{0}

The basic Hamiltonian (\ref{2.3}) is readily quantized by
regarding $q_x$ as multiplication operator and substituting
$-i\partial/\partial q_x$ for $p_x$ as acting on the Hilbert space
$L^2(\mathbb{R},dq_x)$ attached to the site $x\in \mathbb{Z}^3$.
To derive the phonon Boltzmann equation it is convenient to switch
immediately to the notation of second quantization and to work in
Fourier space rather than with the spatial lattice. The
one-particle Hilbert space is then
\begin{equation}\label{7.1}
\mathfrak{h} =L^2(\mathbb{T}^3,dk)
\end{equation}
out of which we construct the bosonic Fock space through
\begin{equation}\label{7.2}
\mathcal{F}=\bigoplus_{n=0}^\infty(\mathfrak{h}^{\otimes
n})_{\textrm{symm}}\,.
\end{equation}
Here $(\mathfrak{h}^{\otimes n})_\textrm{symm}$ is the $n$-fold
tensor product restricted to wave functions symmetric under
permutation of labels. On $\mathcal{F}$ we define a scalar Bose
field with creation/annihilation operators $a(k),a(k)^\ast$, which
satisfy the canonical commutation relations
\begin{equation}\label{7.3}
[a(k),a(k')]=0=[a(k)^\ast,a(k')^\ast]\,,\quad[a(k),a(k')^\ast]=
\delta(k-k')\,.
\end{equation}
Properly speaking, one has to smear $a(k)$ to
$a(f)=\int_{\mathbb{T}^3} dk f(k)a(k)$ with $f\in\mathfrak{h}$ to
have a well-defined operator on Fock space.

In terms of the Bose field $a(k)$ the quantization of $H$ from
(\ref{2.3}) results in
\begin{equation}\label{7.4}
H=H_0+V+V_4
\end{equation}
with
\begin{eqnarray}\label{7.5}
&&\hspace{-10pt}H_0 = \int_{\mathbb{T}^3}dk \omega(k)a(k)^\ast a(k)\,,\nonumber\\
&&\hspace{-10pt}V=\frac{1}{3}\lambda\int_{\mathbb{T}^9}dk_1 dk_2
dk_3 \delta(k_1+k_2+k_3)\prod_{j=1}^3(2\omega(k_j))^{-1/2}
\big(a(k_j)+a(-k_j)^\ast \big)\,,\nonumber\\
&&\hspace{-10pt}V_4 =\lambda'\int_{\mathbb{T}^{12}}dk_1 dk_2 dk_3
dk_4\delta(k_1+k_2+k_3+k_4)\nonumber\\
&&\hspace{20pt}\prod_{j=1}^4(2\omega(k_j))^{-1/2}\big(a(k_j)+a(-k_j)^\ast\big)\,,\;
\lambda'=\lambda^2/18\omega^2_0\,.
\end{eqnarray}
We added explicitly the stabilizing quartic term. Then $H\geq 0$
and we may take the Friedrich extension to make out of $H$ a
self-adjoint operator acting on Fock space.
Physically, the Fock
vacuum $\Omega$ corresponds to the ground state of $H_0$ for the
infinitely extended lattice. States $\psi\in \mathcal{F}$ thus
describe local excitations away from the ground state. In
particular, far away from the origin the particles are in their
state of lowest energy. $H_0$ is normalized to have ground state
energy zero. Thus states of finite mean $H_0$-energy, $\langle
\psi,H_0 \psi \rangle_{\mathcal{F}}<\infty$, are the finite energy
excitations of the harmonic lattice.
As discussed already, for the purpose of the kinetic
limit this set-up is of sufficient generality.

The role of the Gaussian measures in the classical model is taken
over by the quasifree states. They can be defined through their
moments
\begin{eqnarray}\label{7.6}
&&\hspace{40pt}\langle a(k)^\ast a(k')\rangle^\textrm{Q}= R(k,k')\,,\nonumber\\
&&\langle \prod_{j=1}^m a(k_j)^\ast \prod_{j=1}^n
a(k'_j)\rangle^\textrm{Q} =\delta_{mn}
\textrm{perm}\{R(k_i,k'_j)\}_{1\leq i,j\leq n}
\end{eqnarray}
with perm denoting the permanent of a matrix. Clearly, the
positivity of the state $\langle\cdot\rangle^\textrm{Q}$ is
ensured only if $R\geq 0$ as a quadratic form.

If $R$ is a projection, then the state
$\langle\cdot\rangle^\textrm{Q}$ is pure (i.e.~given by a vector
in $\mathcal{F}$) and is a coherent state in the usual
terminology. Let $N$ denote the number of phonons,
\begin{equation}\label{7.7a}
N= \int_{\mathbb{T}^3} dk a(k)^\ast a(k)\,.
\end{equation}
Then
\begin{equation}\label{7.7}
\textrm{tr} R=\int_{\mathbb{T}^3} dk R(k,k) = \langle N
\rangle^\textrm{Q}\,.
\end{equation}
Thus $R$ to be of trace class is a sufficient condition for the
state $\langle\cdot\rangle^\textrm{Q}$ to live on Fock space.

In kinetic theory the building blocks are states which are locally
translation invariant, stationary under the dynamics generated by
$H_0$, and have a strictly positive entropy per unit volume. The
obvious candidates are quasifree states characterized by the
covariance
\begin{equation}\label{7.8}
\langle a(k)^\ast a(k')\rangle^\textrm{Q}=W(k)\delta(k-k')\,,\quad
W(k)\geq 0\,.
\end{equation}
Such a state has infinite energy and is thus outside of Fock
space. The required mathematical framework is well studied
\cite{BrRo}, but will not be needed here. Instead we will consider
a scale of states in Fock space labelled by $\varepsilon$ such
that locally a state of the form (\ref{7.8}) is approximated in
the limit $\varepsilon\to 0$.

We still need to compute the entropy per unit volume of a
quasifree state, compare with (\ref{3.10}), (\ref{3.11}). We
choose the periodic box $[1,\ell]^3$ and a quasifree state of the
form (\ref{7.6}), (\ref{7.8}) with discrete $k$,
$k\in(\ell^{-1}[1,\ldots,\ell])^3$. The corresponding density
matrix is denoted by $\rho_\ell$ and has the entropy
\begin{equation}\label{7.8a}
S_\ell= -\textrm{tr}\rho_\ell \log \rho_\ell\,,
\end{equation}
trace over Fock space. $\rho_\ell$ is of the form
$Z^{-1}\exp[-\sum_k \lambda(k)a^\ast(k)a(k)]$ with
$\lambda=\log\big((1+W)/W\big)$. Therefore
\begin{equation}\label{7.8b}
S_\ell= \sum_{k\in(\ell^{-1}[1,\ldots,\ell])^3}
\big((1+W(k))\log(1+W(k))-W(k)\log W(k)\big)
\end{equation}
which becomes
\begin{equation}\label{7.8c}
\lim_{\ell\to\infty} \ell^{-3} S_\ell=
\int_{\mathbb{T}^3}dk \big((1+W(k))\log(1+W(k))-W(k)\log
W(k)\big)\,.
\end{equation}

We now follow the classical intuition and give ourselves a phonon
distribution function $W(r,k)$. Then
$\{\langle\cdot\rangle^{\textrm{Q},\varepsilon}\,,\;\varepsilon>0\}$
is a family of quasifree states with the property that, defining
\begin{equation}\label{7.9}
W^\varepsilon(y,k)= (\varepsilon/2)^3
\int_{(2\mathbb{T}/\varepsilon)^3}d\eta e^{i2\pi
y\cdot\eta}\langle a(k-\varepsilon\eta/2)^\ast
a(k+\varepsilon\eta/2 \rangle^{\textrm{Q},\varepsilon}
\end{equation}
for $y\in(\varepsilon\mathbb{Z})^3$, one has
\begin{equation}\label{7.10}
\lim_{\varepsilon\to 0}W^\varepsilon(\lfloor r\rfloor_\varepsilon,k)=W(r,k)\,.
\end{equation}
Note that
\begin{equation}\label{7.11}
\sum_{y\in(\varepsilon\mathbb{Z}/2)^3} \int_{\mathbb{T}^3} dk
W^\varepsilon(y,k)= \langle N\rangle^{\textrm{Q},\varepsilon}\,,
\end{equation}
which means that
\begin{equation}\label{7.12}
\langle N\rangle^{\textrm{Q},\varepsilon} \cong \varepsilon^{-3}
\end{equation}
under the condition (\ref{7.10}). We impose
$\langle\cdot\rangle^{\textrm{Q},\varepsilon}$ as the scale of
initial states. Adopting the central assumption of kinetic theory,  the state 
$\langle \cdot\rangle_{t/\varepsilon}$ at the
long time $\varepsilon^{-1}t$ is well approximated by a
locally quasifree state, which,  as to be argued in more detail in the following section,
results in the phonon Boltzmann equation for the quantized
lattice vibrations.

%%%%%%%%%%%%%%%%%%%%%%%%%%%%%%%%%%%%%%

\section{Derivation of the phonon Boltzmann equation (quantized model)}\label{sec.8}
\setcounter{equation}{0}

We follow the scheme of Section \ref{sec.5} and use atomic units. The evolution equations are still given by (\ref{5a.7}), now interpreted as Heisenberg equations for the quantized field.
As major difference to Section \ref{sec.5}, the order of the field operators must be respected.
Let $\langle \cdot\rangle_t$ be the state at time $t$ under the dynamics
$e^{-iHt}$ with initial quasifree state as in (\ref{7.6}),
(\ref{7.9}). The two-point function still satisfies (\ref{5.4}). 
However, in the expression  (\ref{5.7}) for $G(q,p,t,s)$ we used the commutativity of the fields to lump terms together, which 
 has to be undone in the quantum context. Also, the Gaussian factorization
(\ref{5.8}) is to be replaced by the expectation over the locally quasifree state  $\langle \cdot\rangle_s$, which amounts to, for example,
\begin{eqnarray}\label{8.4}
&&\hspace{-20pt}\langle a(k_1)  a(k_2)^\ast a(k_3)^\ast a(k_4)\rangle_s = \langle
a(k_1) a(k_2)^\ast\rangle_s \langle a(k_3)^\ast a(k_4)\rangle_s\nonumber\\
&&\hspace{20pt}+\langle a(k_1) a(k_3)^\ast\rangle_s \langle
a(k_2)^\ast a(k_4)\rangle_s +\langle a(k_1) a(k_4)\rangle_s \langle
a(k_2)^\ast a(k_3)^\ast\rangle_s\,.
\end{eqnarray}
Note that  the last term on the right vanishes by assumption.  Transferred to the Wigner function 
the ordering results in 
\begin{eqnarray}\label{8.5}
&&\varepsilon^3\langle a(k-\varepsilon\eta/2)^\ast a(k+\varepsilon\eta/2)\rangle_{s/\varepsilon} = \widehat{W}^\varepsilon(\eta,k,s)\,,
\nonumber\\
&&
\varepsilon^3\langle  a(k+\varepsilon\eta/2) a(k-\varepsilon\eta/2)^\ast\rangle_{s/\varepsilon}
=\delta(\eta)+\widehat{W}^\varepsilon(\eta,k,s)\,.
\end{eqnarray}

Otherwise the computation of Section \ref{sec.5} can be repeated {\it verbatim}. Perhaps somewhat unexpected at first glance, the collision term is modified only through a linear term and becomes
\begin{eqnarray}\label{8.7}
&&\gamma \int_{\mathbb{T}^6} dk_1 dk_2
(\omega(k)\omega(k_1)\omega(k_2))^{-1}\Big(2\delta
(\omega(k)+\omega(k_1)-\omega(k_2))\nonumber\\
&&\delta(k+k_1-k_2) \big(\tilde{W}(r,k_1)W(r,k_2)
+W(r,k)W(r,k_2)-W(r,k)W(r,k_1)\big)\nonumber\\
&&+\delta(\omega(k)-\omega(k_1)-\omega(k_2))\delta(k-k_1-k_2)
\big(W(r,k_1)W(r,k_2)-W(r,k)
\Tilde{W}(r,k_1)\nonumber\\
&&- W(r,k)W(r,k_2)\big)\Big)\,,
\end{eqnarray}
where
\begin{equation}\label{8.6a}
\tilde{W}(r,k)=1+W(r,k)\,.
\end{equation}

Properties of the Boltzmann equation are more readily seen by
writing the collision operator with an apparent cubic
nonlinearity. This results in the conventional form of the phonon
Boltzmann equation,
\begin{eqnarray}\label{8.16}
&&\hspace{-16pt}\frac{\partial}{\partial t}W(r,k,t)+
\frac{1}{2\pi}\nabla\omega(k)\cdot\nabla_rW(r,k,t)\\
&&=\gamma \int_{\mathbb{T}^6} dk_1 dk_2
(\omega(k)\omega(k_1)\omega(k_2))^{-1}
\Big\{2\delta(\omega(k)+\omega(k_1)-\omega(k_2))
\delta(k+k_1-k_2)\nonumber\\
&&\hspace{28pt}\big(\Tilde{W}(r,k,t)\Tilde{W}(r,k_1,t)
W(r,k_2,t) - W(r,k,t)W(r,k_1,t)\Tilde{W}(r,k_2,t)\big)\hspace{32pt} \textrm{(I)}\nonumber\\
&&\hspace{28pt}+
\delta(\omega(k)-\omega(k_1)-\omega(k_2))\delta(k-k_1-k_2)\nonumber\\
&&\hspace{28pt}\big(\Tilde{W}(r,k,t)W(r,k_1,t) W(r,k_2,t) -
W(r,k,t)\Tilde{W}(r,k_1,t)\Tilde{W}(r,k_2,t)\big)\Big\}\,.\hspace{15pt}
\textrm{(II)}\nonumber
\end{eqnarray}
The Boltzmann equation (\ref{8.16}) is one of our central results.
It reduces to the classical phonon equation (\ref{4.7}) through
omitting the tilde in (\ref{8.7}).

%%%%%%%%%%%%%%%%5

\section{Feynman diagrams}\label{sec.5a} \setcounter{equation}{0}

The iteration leading to Equation (\ref{5.6}) suggests to develop
more systematically the time-dependent perturbation theory. In
this section we will do the first step in a program which needs to
be completed. For simplicity let us assume an initial state which
is translation invariant and quasifree with covariance
\begin{eqnarray}\label{5a.1}
&&\hspace{-10pt}\langle a(k)\rangle^\textrm{Q} =0\,,\;\langle
a(k)a(k')\rangle^\textrm{Q} =0\,,\nonumber\\
&&\hspace{-10pt}\langle a(k)^\ast a(k')\rangle^\textrm{Q}
=\delta(k-k')W(k)\,,
\end{eqnarray}
compare with (\ref{7.6}). By the magic of Wigner functions, 
a slowly varying initial measure would require small
modifications only. Since there is no spatial variation, kinetic
scaling amounts to merely consider the long times $\varepsilon^{-1}t$. By
translation invariance
\begin{equation}\label{5a.2}
\langle a(k)^\ast a(k')\rangle_{t/\varepsilon}
=\delta(k-k')W^\varepsilon(k,t)\,.
\end{equation}
As discussed already, one expects that
\begin{equation}\label{5a.3}
\lim_{\varepsilon\to 0}W^\varepsilon(k,t)=W(k,t)
\end{equation}
and $W(k,t)$ to satisfy the spatially homogeneous version of the
phonon Boltzmann equation (\ref{8.16}). Let us set
\begin{equation}\label{5a.3a}
W(k,1)=1+W(k)\,,\quad W(k,-1)=W(k)
\end{equation}
and, as before,
\begin{equation}\label{5a.3b}
\phi(k,k_1,k_2)=\lambda(8\omega(k)\omega(k_1)\omega(k_2))^{-1/2}\,.
\end{equation}
Then the phonon Boltzmann equation  (\ref{8.16}) is written more concisely
as
\begin{eqnarray}\label{5a.3c}
&&\hspace{-20pt}\frac{\partial}{\partial
t}W(k,\sigma,t)=4\pi\sum_{\sigma_1,\sigma_2=\pm1}
\int_{\mathbb{T}^6}dk_1dk_2\phi(k,k_1,k_2)^2\nonumber\\
&&\hspace{20pt}\delta(\sigma\omega+\sigma_1\omega_1+\sigma_2\omega_2)
\delta(\sigma k+\sigma_1k_1+\sigma_2k_2)\nonumber\\
&&\hspace{20pt}\big(W(k_1,\sigma_1,t)W(k_2,\sigma_2,t)+\sigma\sigma_1W(k,\sigma,t)
\sum_{{\widetilde{\sigma}}=\pm1}W(k_2,{\widetilde{\sigma}},t)\big)
\end{eqnarray}
with initial conditions from (\ref{5a.1}). Here we use as
shorthand $\omega=\omega(k)$, $\omega(k_1)=\omega_1$,
$\omega(k_2)=\omega_2$.
Note that the term with $\sigma,\sigma_1,\sigma_2=1$ vanishes,
since $\omega(k)+\omega(k_1)+\omega(k_2)\geq 0$. 

To derive (\ref{5a.3c}) from the microscopic dynamics, we take
 the time-dependent perturbation theory as starting point. The Heisenberg
equations for the quantum field are given by (\ref{5a.7}) with the shorthand (\ref{5a.6}).
Inserting them in time-integrated form yields 
the identity
\begin{eqnarray}\label{5a.8}
&&\hspace{-60pt} \langle\prod^m_{j=1} a(k_j,\sigma_j)\rangle_t =
\exp\big[it\big(\sum^m_{j=1} \sigma_j\omega(k_j)\big)\big]\langle
\prod^m_{j=1}
a(k_j,\sigma_j)\rangle^\textrm{Q}\nonumber\\
&&\hspace{26pt} + i\sqrt{\varepsilon}\int^t_0 ds
\exp\big[i(t-s)\big(\sum^m_{j=1} \sigma_j\omega(k_j)\big)\big]\nonumber\\
&&\hspace{26pt}\Big(\sum^m_{\ell=1}
\sum_{\sigma',\sigma''=\pm1}\sigma_\ell
\int_{\mathbb{T}^6} dk'dk''\phi(k_\ell,k',k'')\delta(-\sigma k_\ell+\sigma'k'+\sigma''k'')\nonumber\\
&&\hspace{26pt} \langle (\prod^{\ell-1}_{j=1} a(k_j,\sigma_j))
a(k',\sigma')a(k'',\sigma'')\prod^m_{j'=\ell+1}a(k_{j'},\sigma_{j'}\rangle_s\Big)\,.
\end{eqnarray}
Note that the operator ordering is properly maintained. To generate the perturbation series for $W^\varepsilon(k,t)$ one
starts with $m=2$. Then on the right there is a product of 3 $a$'s
for which one substitutes (\ref{5a.8}) with $m=3$, etc. Finally
one averages explicitly over the initial quasifree state $
\langle\cdot\rangle^\textrm{Q}$. This yields
\begin{equation}\label{5a.9}
\langle a(q,\sigma_q)^\ast a(p,\sigma_p)\rangle_{t/\varepsilon}=
\delta(\sigma_q,-\sigma_p)
\delta(q-p)\big(W(q,\sigma_q)+\sum^\infty_{n=1}W^\varepsilon_n(q,\sigma_q,t)\big)\,.
\end{equation}
Let us postpone the issue of the convergence of the sum over $n$
to the end of this section and first discuss $W^\varepsilon_n$ for
each $n$ separately.

$\delta(\sigma_{q'},-\sigma_p)\delta(q-p)W^\varepsilon_n(q,\sigma_1,t)$
is a sum of oscillating integrals. The summation comes from three
sources
\smallskip\\
-- the sum over $\sigma',\sigma''$ in (\ref{5a.8})\\
-- the sum over $\ell$ in (\ref{5a.8})\\
-- the sum over all oriented pairings due to the average in the
initial quasifree state,
\begin{equation}\label{5a.10}
\langle\prod^{2n}_{j=1}a(k_j,\sigma_j)\rangle^\textrm{Q}=
\sum_{\mathrm{pairings}\;\pi,\pi'}\prod^n_{i=1}\langle
a(k_{\pi(i)},\sigma_{\pi(i)})a(k_{\pi'(i)},\sigma_{\pi'(i)})\rangle^\textrm{Q}\,.
\end{equation}
Oriented means that in $\langle a(k_{\pi(i)},
\sigma_{\pi(i)})a(k_{\pi'(i)},
\sigma_{\pi'(i)})\rangle^\textrm{Q}$ on the right hand side the
operators appear in the same order as on the left hand side. Since
the integrals have a rather complicated structure, it is
convenient to visualize them as \textit{Feynman diagrams}.

A Feynman diagram is an oriented graph with labels. We first
construct the graph and then the labelling. The graph uses as
``backbone" $2n+2$ equidistant horizontal level lines labelled
from 0 to $2n+1$. The graph itself consists of two binary downward
trees. The roots are two vertical bonds from line $2n+1$ to $2n$.
These bonds are continued downwards. At level $m$
there is \textit{exactly one} branch point with two branches. Branches
do not cross, see Figure \ref{Fig4}. Thus at level 0 there are $2n+2$ vertical
bonds (branches). They are connected according to the pairing rule
resulting in $n+1$ pairs. Thereby the graph consists of internal
lines and two roots (external legs). The Feynman graph is
oriented, with lines pointing either up $(\sigma=+1)$ or down
$(\sigma=-1)$. If there is no branching the orientation is
inherited from the continuing vertical bond in the level
above. At a pairing the orientation must be maintained. Thus at
level 0 a branch with an up arrow can be paired only with a branch
with a down arrow. If the pairing is pointing to the left, it
corresponds to the order $\langle a^\ast a\rangle^\textrm{Q}$ in
(\ref{5a.10}), while a pairing pointing to the right corresponds
to $\langle a a^\ast\rangle^\textrm{Q}$. By construction, each
internal line has two orientations and starts and ends at a branch
point.
\begin{figure}
\setlength{\unitlength}{1cm}
\begin{picture}
(10,6.5)(-3.5,0) \put(-1,-0.1){$0$} \put(-1,0.9){$1$}
\put(-1,1.9){$2$} \put(-1,2.9){$3$} \put(-1,3.9){$4$}
\put(-1,4.9){$5$} \put(8,-0.1){$0$} \put(8,0.9){$t_1$}
\put(8,1.9){$t_2$} \put(8,2.9){$t_3$} \put(8,3.9){$t_4$}
\put(8,4.9){$t$}

\put(1.3,4.1){$1$}\put(3.5,4.1){$2$}\put(2.9,3.1){$3$}\put(4,3.1){$4$}
\put(4.9,2.1){$5$}\put(6,2.1){$6$}\put(0.9,1.1){$7$}\put(2,1.1){$8$}
\put(2.4,5.2){$q$}\put(5.4,5.2){$p$} \linethickness{0.1pt}
\put(0,0){\line(1,0){7}} \put(0,1){\line(1,0){7}}
\put(0,2){\line(1,0){7}} \put(0,3){\line(1,0){7}}
\put(0,4){\line(1,0){7}} \put(0,5){\line(1,0){7}} \thicklines
\put(1,-0.3){\line(0,1){1.3}} \put(2,-0.15){\line(0,1){1.15}}
\put(3,-0.15){\line(0,1){3.15}} \put(4,-0.15){\line(0,1){3.15}}
\put(5,-0.15){\line(0,1){2.15}} \put(6,-0.3){\line(0,1){2.3}}
\put(1.5,1){\line(0,1){3}} \put(2.5,4){\line(0,1){1}}
\put(3.5,3){\line(0,1){1}} \put(5.5,2){\line(0,1){3}}
\put(1,1){\line(1,0){1}} \put(3,3){\line(1,0){1}}
\put(1.5,4){\line(1,0){2}} \put(5,2){\line(1,0){1}}
\put(1,-0.3){\line(1,0){5}} \put(2,-0.15){\line(1,0){1}}
\put(4,-0.15){\line(1,0){1}} \put(1.5,3.5){\vector(0,1){0.15}}
\put(1,0.5){\vector(0,1){0.15}} \put(2,0.5){\vector(0,-1){0.15}}
\put(2.5,4.5){\vector(0,-1){0.15}} \put(3,2.5){\vector(0,1){0.15}}
\put(4,2.5){\vector(0,1){0.15}} \put(5,1.5){\vector(0,-1){0.15}}
\put(6,1.5){\vector(0,-1){0.15}} \put(5.5,4.5){\vector(0,1){0.15}}
\put(3.5,3.5){\vector(0,-1){0.15}}
\end{picture}\\
\caption{Example of a Feynman diagram at order $n=2$.}\label{Fig4}
\end{figure}
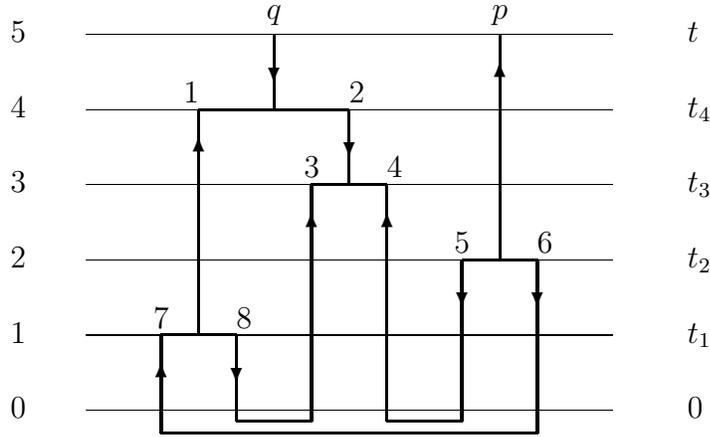

Next we insert the labels. The level lines 0 to $2n+1$ are
labelled by times $0<t_1\ldots<t_{2n}<t$. The left root carries
the label $q$ while the right root carries the label $p$. Each
internal line is labelled with a wave number $k$.

To each Feynman diagram one associates an integral through the
following steps.\smallskip\\
(i) The time integration is over the simplex $0\leq
t_1\ldots\leq t_{2n}\leq t$ as $dt_1\ldots dt_{2n}$.\smallskip\\
(ii) The wave number integration is over all internal lines as
$\int dk_1 \ldots \int dk_\kappa$, where $\kappa=3n-1$ is the
number of internal
lines.\smallskip\\
(iii) One sums over all orientations of the internal lines.\smallskip\\
The integrand is a product of 3 factors.\smallskip\\
(iv) There is a product over all branch points.

At each branchpoint there are a root, say wave vector $k_1$ and
orientation $\sigma_1$, and two branches, say wave vectors $k_2$,
$k_3$ and orientations $\sigma_2$, $\sigma_3$. Then each branch
point carries the weight
\begin{equation}\label{5a.11}
\delta(-\sigma_1k_1+\sigma_2k_2+\sigma_3k_3)\sigma_1\phi(k_1,k_2,k_3)\,.
\end{equation}
If one regards the wave vector $k$ as a current with orientation
$\sigma$, then (\ref{4.6}) expresses Kirchhoff's rule for
conservation of the current.\smallskip\\
(v) By construction each bond carries a time difference
$t_{m+1}-t_m$, a wave vector $k$, and an orientation $\sigma$.
Then to this bond one associates the phase factor
\begin{equation}\label{5a.12a}
\exp[i(t_{m+1}-t_m)\sigma\omega(k)/\varepsilon]\,.
\end{equation}
The second factor is the product of such phase factors over all bonds.\smallskip\\
(vi) The third factor of the integrand is
\begin{equation}\label{5a.12}
\prod^{n+1}_{j=1} W(k_j,\sigma_j)\,,
\end{equation}
where $k_1,\ldots,k_{n+1}$ are the labels of the branches between
level 0 and level 1. $\sigma_j=1$ if the pairing line is oriented
to the
right and $\sigma_j=-1$ if oriented to the left. \smallskip\\
(vii) Finally there is the prefactor $(-1)^n\varepsilon^{-n}$.\smallskip

$\delta(\sigma_q,-\sigma_p)\delta(q-p)W^\varepsilon_n(q,t)$ is the
sum over all integrals corresponding to Feynman graphs with $2n+2$
horizontal time slices, given the external legs $q,p$ with
orientations $\sigma_q,\sigma_p$.

To illustrate the method let us consider the case $n=1$. There are
then $(2\cdot3)\cdot3\cdot4=72$ diagrams.  There is a group of 24 diagrams 
for which each of the two trees
branches once.  Among them there are 8 
diagrams with the two trees disconnected.  They yield the term 
$\delta(q)\delta(p)\mathcal{O}(\varepsilon)$ provided $\omega(0) > 0$. 
According to the rules listed,
the remaining 16 diagrams sum up to the oscillating integral
\begin{eqnarray}\label{5a.14}
&&\hspace{-40pt}I^+_\varepsilon=\varepsilon^{-1}2\delta(\sigma_q,-\sigma_p)
\int^t_0 dt_2 \int^{t_2}_0 dt_1 \sum_{\sigma_1,\sigma_2=\pm1}\int
dk_1dk_2 \phi(q,k_1,k_2)
\phi(p,k_1,k_2)\nonumber\\
&&\hspace{-30pt}\big(\exp\big[i(t_2-t_1)
(\sigma_q\omega(q)-\sigma_1\omega(k_1)-\sigma_2\omega(k_2))/\varepsilon\big]+c.c.\big)\nonumber\\
&&\hspace{-30pt}\times\delta(-\sigma_q q+\sigma_1k_1+\sigma_2k_2)
\delta(-\sigma_p p-\sigma_1k_1-\sigma_2k_2)W(k_1,-\sigma_1) W(k_2,-\sigma_2)\,.
\end{eqnarray}
The limit $\varepsilon\to 0$ is covered by the argument from
(\ref{5.15}) and
\begin{eqnarray}\label{5a.15}
&&\hspace{-20pt}\lim_{\varepsilon\to 0} I^+_\varepsilon=
\delta(\sigma_q,-\sigma_p) \delta(q-p) 4\pi \int^t_0 dt_2
\sum_{\sigma_1,\sigma_2=\pm1}
\int dk_1dk_2\phi(q,k_1,k_2)^2\nonumber\\
&&\hspace{28pt}\delta(\sigma_q\omega(q)+\sigma_1\omega(k_1)+\sigma_2\omega(k_2))
\delta(\sigma_q q+\sigma_1k_1+\sigma_2k_2)
\nonumber\\
&&\hspace{28pt} W(k_1,\sigma_1)W(k_2,\sigma_2)\,.
\end{eqnarray}
 
Secondly there is a group of 48 diagrams for which one of the two trees does
not branch. Among them there are 16 
diagrams which have an internal line with $k=0$. They cancel amongst
each other by symmetry.
The remaining 32 diagrams sum up to $I^-_\varepsilon$.
Its oscillating integrals are handled as for $I^+_\varepsilon$. Thereby one obtains
\begin{eqnarray}\label{5a.17}
&&\hspace{-25pt}\lim_{\varepsilon\to 0}
I^-_\varepsilon=\delta(\sigma_q,-\sigma_p) \delta(q-p) 4\pi
\int^t_0 dt_2 \sum_{\sigma_1,\sigma_2=\pm1}
\int dk_1dk_2\phi(q,k_1,k_2)^2\nonumber\\
&&\hspace{23pt}\delta(\sigma_q\omega(q)+\sigma_1\omega(k_1)+\sigma_2\omega(k_2))
\delta(\sigma_q q+\sigma_1k_1+\sigma_2k_2)
\nonumber\\
&&\hspace{23pt}\sigma_q\sigma_1
W(q,\sigma_q)\big(W(k_2,1)+W(k_2,-1)\big)\,.
\end{eqnarray}

We note the analogy with the discussion in Section \ref{sec.5}. The computation there is more lengthy, since spatial variation is included. $G_{\mathrm{gain}}$ corresponds to $I^+_\varepsilon$,
$G_{\mathrm{loss}}$ to $I^-_\varepsilon$,  $G_{\mathrm{sub1}}$ to the 8 diagrams with both trees branched,
and  $G_{\mathrm{sub2}}$ to the 16 diagrams with only one tree branched. 

Kinetic theory claims that at any order diagrams divide into leading and
subleading. The subleading diagrams vanish in the limit
$\varepsilon\to 0$ while the leading ones have a finite limit. In
fact the leading diagrams can be characterized very concisely.\bigskip\\
\textbf{Kinetic Conjecture}: \textit{In a leading Feynman diagram
the Kirchhoff rule never forces an internal wave number 0 i.e.~a
factor of the form $\delta(k_j)$ with some wave vector $k_j$. In
addition, the sum of the $2(n-m+1)$ phases from the bonds between
level lines $2m$ and $2m+1$ vanishes for every choice of internal
wave numbers. This cancellation must hold for
$m=0,\ldots,n$.}\bigskip

By a tricky combinatorial argument \cite{HS} it can be shown that the sum of
all leading, according to the Kinetic Conjecture, diagrams satisfy
a set of differential equations, which in analogy to the kinetic
theory of gases is called Boltzmann hierarchy. Let
$(f_1,f_2,\ldots)$ be a vector of functions where
$f_n(k_1,\sigma_1,\ldots,k_n,\sigma_n)$ is symmetric in its
arguments. We define the collision operator $\mathcal{C}_{n,n+1}$
through
\begin{eqnarray}\label{5a.18}
&&\hspace{-30pt}(\mathcal{C}_{n,n+1}f_{n+1})(k_1,\sigma_1,\ldots,k_n,\sigma_n)
= 4\pi\sum^n_{\ell=1} \sum_{\sigma',\sigma''=\pm
1}\int_{\mathbb{T}^6} d k' d k''
\phi(k_\ell,k',k'')^2\nonumber\\
&&\hspace{75pt}
\delta(\sigma_\ell\omega_\ell+\sigma'\omega'+\sigma''\omega'')
\delta(\sigma_\ell k_\ell+\sigma'k'+\sigma''k'')\nonumber\\
&&\hspace{75pt} [f_{n+1}(k_1,\sigma_1,
\ldots,k_{\ell-1},\sigma_{\ell-1},k',\sigma',\ldots,k'',\sigma'')\nonumber\\
&&\hspace{75pt} +\sigma_\ell \sigma'\sum_{\widetilde{\sigma}=\pm1}
f_{n+1}(k_1,\sigma_1,\ldots,k_n,\sigma_n,k'',\widetilde{\sigma})]\,.
\end{eqnarray}
Then the Boltzmann hierarchy reads
\begin{equation}\label{5a.19}
\frac{d}{dt}f_n(t)= \mathcal{C}_{n,n+1}f_{n+1}(t)\,.
\end{equation}
Note that the result from (\ref{5a.15}), (\ref{5a.17}) can be
stated as
\begin{equation}\label{5a.20}
\lim_{\varepsilon\to
0}(I^+_\varepsilon+I^-_\varepsilon)=t(\mathcal{C}_{1,2}f_2)(k_1,\sigma_1)
\end{equation}
provided one sets
\begin{equation}\label{5a.21}
f_2(k_1,\sigma_1,k_2,\sigma_2)=W(k_1,\sigma_1)W(k_2,\sigma_2)\,.
\end{equation}

The Boltzmann hierarchy has the property that initial
factorization of $f_n$ is maintained in time,
\begin{equation}\label{5a.21a}
f_n(k_1,\sigma_1,\ldots,k_n,\sigma_n,t)= \prod^n_{j=1}
f(k_j,\sigma_j,t)
\end{equation}
and each factor satisfies the Boltzmann equation
\begin{eqnarray}\label{5a.22}
&&\hspace{-16pt}\frac{\partial}{\partial t}f(k,\sigma,t)
=4\pi\lambda^2\sum_{\sigma',\sigma''=\pm 1}
\int_{\mathbb{T}^6}dk'dk''(8\omega\omega'\omega'')^{-1}\\
&&\hspace{16pt}\delta(\sigma\omega+\sigma'\omega'+\sigma''\omega'')
\delta(\sigma k+\sigma'k'+\sigma''k'')\nonumber\\
&&\hspace{16pt}[f(k',\sigma',t)f(k'',\sigma'',t)+
\sigma\sigma'f(k,\sigma,t)(f(k'',1,t)+f(k'',-1,t))]\,.\nonumber
\end{eqnarray}
For the particular choice
\begin{equation}\label{5a.23}
f(k,1)= 1+W(k)\,,\quad f(k,-1)= W(k)
\end{equation}
Equation (\ref{5a.22}) agrees with the phonon Boltzmann equation
(\ref{5a.3c}). Thereby the Kinetic Conjecture amounts to the
assertion
\begin{equation}\label{5a.24}
\lim_{\varepsilon\to 0}W^\varepsilon_n(k,\sigma,t)=
\frac{1}{n!}t^n(\mathcal{C}_{1,2}\ldots
\mathcal{C}_{n,n+1}f_{n+1})(k,\sigma)
\end{equation}
with the initial $f_n$ factorized as in (\ref{5a.21}) and single
factor (\ref{5a.23}).

The difference between the quantized theory, discussed so far, and
the classical theory is surprisingly minor when viewed on the
level of Feynman diagrams. The classical $a$-field commutes, which
however does not simplify the structure of the diagram. Only in
the average of the initial state quasifree is replaced by Gaussian
which according to (\ref{5a.10}) induces the modification $\langle
aa^\ast\rangle^\textrm{G}=\langle a^\ast a\rangle^\textrm{G}$.
Thus the classical phonon Boltzmann equation is obtained by
setting the initial conditions for the hierarchy as
\begin{equation}\label{5a.25}
f_n(k_1,\sigma_1,\ldots,k_n,\sigma_n)= \prod^n_{j=1}W(k_j)\,.
\end{equation}
One checks that (\ref{5a.22}) indeed coincides with (\ref{4.7}).

So far we avoided the issue of the convergence of the series in
(\ref{5a.9}). At order $n$ there are $(2n-1)!(2n)!2^n/n!$ Feynman
diagrams. If $W$ is bounded, then a single Feynman diagram is of
order $c^n t^{2n}/(2n)!$ with some suitable constant $c$. Thus,
unless cancellations are used, even at finite $\varepsilon$ the
sum over $n$ does not converge. The situation improves in the
kinetic level. At order $n$ there are only $(48)^n n!$ leading
diagrams. If $e_{\textrm{max}}$ of (\ref{4a.8}) is bounded, then
each diagram is of order $c^n(e_\textrm{max})^n t^n/n!$. Therefore
in the limit the sum over $n$ in (\ref{5a.24}) converges provided
$t$ is sufficiently small.

We conclude that the most
immediate project is to establish (\ref{5a.24}), which means  that all subleading diagrams
vanish in the limit $\varepsilon\to 0$. This would be a step
further when compared to the investigation \cite{BeCa},
see also \cite{Hu,HoLa}.
Of course a complete proof must deal with the uniform convergence
of the series in (\ref{5a.9}).\bigskip\\

%%%%%%%%%%%%%%%%%%%%%%%%%%%%%%%%%%%%%%%%%

\section{Properties of the quantum phonon Boltzmann equation}\label{sec.9}
\setcounter{equation}{0}

The Boltzmann equation (\ref{8.16}) for the quantized phonons
differs somewhat from its classical cousin (\ref{4.7}). But the
basic properties remain unaltered. As before, energy is locally
conserved. The entropy functional has to be modified, but results
again in a positive entropy production. Most significantly the
stationary distribution functions are now the one-parameter family
of Bose-Einstein distributions at zero chemical potential,
\begin{equation}\label{9.1}
W_\beta(k)=(e^{\beta\omega(k)}-1)^{-1}\,.
\end{equation}
In the high temperature limit, $\beta \to 0$, they reduce to
$(\beta\omega(k))^{-1}$, which are the stationary solutions of the
phonon Boltzmann equation for classical lattice dynamics.

In the sequel we discuss each item separately.\medskip\\
\textit{(i) Energy}. The properties (\ref{6.1}) to (\ref{6.3})
remain intact. One only has to show that $\int_{\mathbb{T}^6}
dk\omega(k)\mathcal{C}(W)(k) =0$. Inserting the collision operator
from (\ref{8.16}) one obtains
\begin{eqnarray}\label{9.2}
 &&\hspace{-19pt}\int_{\mathbb{T}^9} dk_1 dk_2 dk_3
(\omega(k_1)\omega(k_2)\omega(k_3))^{-1}\{2\delta
(\omega(k_1)+\omega(k_2)-\omega(k_3))\delta(k_1+k_2-k_3)\nonumber\\
&&\omega(k_1)\big( \Tilde{W}(k_1)\Tilde{W}(k_2)W(k_3)- W(k_1)
W(k_2)\Tilde{W}(k_3)\big)+ \delta
(\omega(k_1)-\omega(k_2)-\omega(k_3))\nonumber\\
&&\delta(k_1-k_2-k_3)\omega(k_1) \big(\Tilde{W}(k_1)W(k_2)W(k_3)-
W(k_1)\Tilde{W}(k_2)\Tilde{W}(k_3)\big)\}\nonumber\\
&&\hspace{-8pt}= \int_{\mathbb{T}^9} dk_1 dk_2 dk_3
(\omega(k_1)\omega(k_2)\omega(k_3))^{-1}\delta
(\omega(k_1)+\omega(k_2)-\omega(k_3))\delta(k_1+k_2-k_3)\nonumber\\
&&\omega(k_3)\big(\Tilde{W}(k_1)\Tilde{W}(k_2)W(k_3)-
W(k_1)W(k_2)\Tilde{W}(k_3)+\Tilde{W}(k_3)W(k_1)W(k_2)\nonumber\\
&&-W(k_3)\Tilde{W}(k_1)\Tilde{W}(k_2)\big) = 0\,,
\end{eqnarray}
where as in (\ref{6.4}) in the first summand we symmetrized
$2\omega(k_1)$ to $\omega(k_1)+\omega(k_2)$ and used energy
conservation, while in the second summand we employed the cyclic
substitution $k_1 \to
k_3$, $k_3 \to k_2$, $k_2 \to k_1$.\medskip\\
\textit{(ii) Entropy}. As can be seen from (\ref{7.8c}), the local
entropy is defined through
\begin{equation}\label{9.3}
s(r,t)= \int_{\mathbb{T}^3} dk
\big(\Tilde{W}(r,k,t)\log \Tilde{W}(r,k,t)-W(r,k,t)\log
W(r,k,t)\big)\,.
\end{equation}
It satisfies the semi-conservation law
\begin{equation}\label{9.4}
\frac{\partial}{\partial t}s(r,t) +\nabla \cdot
j_\textrm{s}(r,t)=\sigma(r,t)
\end{equation}
with a positive entropy production $\sigma$. The entropy flow is
easily deduced to
\begin{equation}\label{9.5}
j_\textrm{s}(r,t)= (2\pi)^{-1}\int_{\mathbb{T}^3} dk
\nabla\omega(k) \big(\Tilde{W}(r,k,t)\log
\Tilde{W}(r,k,t)-W(r,k,t)\log W(r,k,t)\big)\,.
\end{equation}
To compute the entropy production we symmetrize as in the case of
the energy, the role of $\omega(k_1)$ being taken over by
$\log\big(\Tilde{W}(r,k_1,t)/W(r,k_1,t)\big)$. Then
\begin{eqnarray}\label{9.6}
&&\hspace{-16pt}\sigma(r,t)= \gamma \int_{\mathbb{T}^9} dk_1 dk_2 dk_3
(\omega(k_1)\omega(k_2)\omega(k_3))^{-1}\delta
(\omega(k_1)+\omega(k_2)-\omega(k_3))\\
&&\hspace{-8pt}\delta(k_1+k_2-k_3)f \big(
\Tilde{W}(r,k_1,t)\Tilde{W}(r,k_2,t)W(r,k_3,t),
W(r,k_1,t)W(r,k_2,t)\Tilde{W}(r,k_3,t)\big)\nonumber
\end{eqnarray}
with
\begin{equation}\label{9.7}
f(x,y)=(x-y)\log(x/y)\,.
\end{equation}
Clearly, $\sigma(r,t)>0$ unless $x=y$ in (\ref{9.7}), i.e. unless
\begin{equation}\label{9.8}
\Tilde{W}(k_1)\Tilde{W}(k_2)W(k_3)= W(k_1)W(k_2)\Tilde{W}(k_3)
\end{equation}
on the set of $(k_1,k_2,k_3)$'s satisfying the constraints
\begin{equation}\label{9.9}
\omega(k_1)+\omega(k_2) =\omega(k_3)\,,\quad k_1+k_2=k_3\,,
\end{equation}
where we regard $W$ and $\omega$ as continued periodically to all
of $\mathbb{R}^3$. Thus $\sigma=0$ if and only if (\ref{9.8})
holds for all $(k_1,k_2,k_3)\in \mathbb{R}^9$ on the set defined
by (\ref{9.9}).

The total entropy is
\begin{equation}\label{9.10}
S(t)= \int d^3r s(r,t)\,.
\end{equation}
From (\ref{9.4}) it follows that
\begin{equation}\label{9.11}
\frac{d}{dt}S(t)\geq 0\,,
\end{equation}
which is the analogue of Boltzmann's H-theorem.\medskip\\
\textit{(iii) Stationary solutions}. We consider a spatially
homogeneous system for which the Boltzmann equation
(\ref{8.16}) reduces to
\begin{equation}\label{9.10a}
\frac{\partial}{\partial t}W= \mathcal{C}(W)\,.
\end{equation}
By definition a stationary solution has to satisfy $\mathcal{C}(W)
= 0$. This equality looks rather unapproachable and a better
strategy is to use that for a solution to be stationary its
entropy production has to vanish. To make the resulting functional
equations (\ref{9.8}) and (\ref{9.9}) more tractable we introduce
as auxiliary quantity
\begin{equation}\label{9.12}
\psi=\log(W/\Tilde{W})\,.
\end{equation}
Then (\ref{9.8}) becomes additive as
\begin{equation}\label{9.13}
\psi(k_1)+\psi(k_2)=\psi(k_3)\,.
\end{equation}

\begin{proposition}\label{9.prop1}
Let the ergodicity condition (E) be satisfied and let
$\det(\mathrm{Hess}\,\omega)=0$ at most on a set of codimension 1.
Let $\psi:\mathbb{T}^3\to \mathbb{R}$ be twice continuously
differentiable and satisfy the functional equation
\begin{equation}\label{9.14}
\psi(k_1)+\psi(k_2)=\psi(k_1+k_2)
\end{equation}
on the set $\Lambda_\omega = \{(k_1,k_2)\in
\mathbb{T}^6\,|\;\omega(k_1)+\omega(k_2)=\omega(k_1+k_2)\}$. Then
necessarily
\begin{equation}\label{9.15}
\psi(k) = a \omega(k)\,,\quad a\in \mathbb{R}\,.
\end{equation}
\end{proposition}
{\textit{Proof:}} In spirit we follow Cercignani and Kremer
\cite{CeKr}. We set $k_1=k\,,\;k_2=k'$. The collisional invariant
$\psi(k)+\psi(k')$ is constant on the set $\{(k,k')\in
\mathbb{T}^6\,|\;\omega(k)+\omega(k')= \textrm{const}\,,\;k+k'=
\textrm{const}\}$. Therefore there exists a function
$\phi:\,\mathbb{R}\times \mathbb{T}^3 \to \mathbb{R}$ such that
\begin{equation}\label{9.16}
\psi(k)+\psi(k')=\phi(\omega(k)+\omega(k')\,,\,k+k')\,.
\end{equation}
We set $k=(k^1,k^2,k^3)$, $\omega=\omega(k)$,
$\omega'=\omega(k')$,
\begin{equation}\label{9.17}
\partial_\omega\phi(\omega,k)=\partial\phi(\omega,k)/\partial\omega\,,\quad
\partial_\alpha\phi(\omega,k)=\partial\phi(\omega,k)/\partial
k^\alpha\,,
\end{equation}
$\alpha=1,2,3$, and differentiate (\ref{9.16}) with respect to
$k,k'$. Then
\begin{eqnarray}\label{9.18}
&&\partial_\alpha\psi(k)=\partial_\omega\phi(\omega+\omega',k+k')\partial_\alpha\omega
+\partial_\alpha\phi(\omega+\omega',k+k')\,,\nonumber\\
&&\partial_\alpha\psi(k')=\partial_\omega\phi(\omega+\omega',k+k')\partial_\alpha\omega'
+\partial_\alpha\phi(\omega+\omega',k+k')\,.
\end{eqnarray}
Subtracting and symmetrizing yields
\begin{eqnarray}\label{9.19}
&&\big(\partial_\alpha\psi(k)-\partial_\alpha
\psi(k')\big)\big(\partial_\beta\omega(k)-\partial_\beta\omega(k')\big)\nonumber\\
&&=\big(\partial_\beta\psi(k)-\partial_\beta
\psi(k')\big)\big(\partial_\alpha\omega(k)-\partial_\alpha\omega(k')\big)\,.
\end{eqnarray}
Differentiating with respect to $k$,
\begin{eqnarray}\label{9.20}
&&\partial_\alpha \partial_\gamma \psi(k)
\big(\partial_\beta\omega(k)-\partial_\beta\omega(k')\big) +
\big(\partial_\alpha\psi(k)-\partial_\alpha \psi(k')\big)
\partial_\beta \partial_\gamma \omega(k)\nonumber\\
&&= \partial_\beta \partial_\gamma \psi(k)
\big(\partial_\alpha\omega(k)-\partial_\alpha\omega(k')\big) +
\big(\partial_\beta\psi(k)-\partial_\beta \psi(k')\big)
\partial_\alpha \partial_\gamma \omega(k)\,,
\end{eqnarray}
and once more differentiating with respect to $k'$,
\begin{eqnarray}\label{9.21}
&&\partial_\alpha \partial_\gamma \psi(k)
\partial_\beta\partial_\delta\omega(k')+ \partial_\alpha \partial_\delta \psi(k')
\partial_\beta\partial_\gamma\omega(k)\nonumber\\
&&= \partial_\beta \partial_\gamma \psi(k)
\partial_\alpha\partial_\delta\omega(k')+ \partial_\beta \partial_\delta \psi(k')
\partial_\alpha\partial_\gamma\omega(k)\,,
\end{eqnarray}
which holds on $\Lambda_\omega$.

Let $\Lambda=\{k\in\mathbb{T}^3, \det
(\textrm{Hess}\,\omega(k))\neq0\}$. As proven in Appendix \ref{sec.16b}, if $k,k'\in \Lambda$, then the
only solution to (\ref{9.21}) reads
\begin{equation}\label{9.22}
\partial_\alpha \partial_\beta \psi(k)=a(k)
\partial_\alpha\partial_\beta\omega(k)\,,\; \partial_\alpha \partial_\beta
\psi(k')=a(k)
\partial_\alpha\partial_\beta\omega(k')
\end{equation}
with some constant $a(k)$ independent of $k'$. We choose now
$k''\in\Lambda$ linked through a collision to $k'$ and conclude
that also
\begin{equation}\label{9.23}
\partial_\alpha \partial_\beta \psi(k'')=a(k)
\partial_\alpha\partial_\beta\omega(k'')\,.
\end{equation}
By the ergodicity condition (E) the relation (\ref{9.23}) extends
to
\begin{equation}\label{9.23a}
\partial_\alpha \partial_\beta \psi(k)=a
\partial_\alpha\partial_\beta\omega(k)
\end{equation}
on $\Lambda$ with some constant $a$. By continuity (\ref{9.23a})
extends to all of $\mathbb{T}^3$. Integrating (\ref{9.23a}) yields
$\psi(k)=a\omega(k)+b\cdot k+c$. $b=0$ by continuity of $\psi$ and
$c=0$ by (\ref{9.14}). $\Box$\medskip\\
\textit{Remarks:} (i) Presumably our result holds under weaker
smoothness assumptions on $\psi$. The difficulty is that in
(\ref{9.21}) $k$ and $k'$ are constrained variables.\smallskip\\
(ii) The ergodicity condition may fail because at given $k$ no
collision is admitted by energy conservation. But there are more
subtle cases. For example $\mathbb{Z}^3$ could be partitioned into
two sublattices which are dynamically disconnected, i.e.~the
elastic constants $\alpha(x)$ couple only within each sublattice.
Then, at best, each sublattice thermalizes by itself and
ergodicity is violated. \smallskip\\
(iii) Assume that there is some function, $\psi(k)$, such that 
$ \int_{\mathbb{T}^3} d k \psi(k)
W(r,k,t)$ satisfies a local conservation law in the form (\ref{6.2}). Then the 
corresponding current is necessarily
\begin{equation}\label{9.25}
j_\psi(r,t) = (2\pi)^{-1} \int_{\mathbb{T}^3} d k
(\nabla\omega(k))\psi(k) W(r,k,t)
\end{equation}
and it must hold that
\begin{equation}\label{9.26}
\int_{\mathbb{T}^6}
dk\psi(k)\mathcal{C}(W)(k) =0
\end{equation}
for all $W$. Repeating the computation in (\ref{9.2}) leads to
\begin{eqnarray}\label{9.27}
&&\hspace{-24pt} \int_{\mathbb{T}^9} dk_1 dk_2 dk_3
(\omega_1\omega_2\omega_3)^{-1} \big(\psi(k_1)+\psi(k_2)-\psi(k_3)\big)\delta
(\omega_1+\omega_2-\omega_3)\nonumber\\
&&\hspace{-16pt}\delta(k_1+k_2-k_3)
\big(\Tilde{W}(k_1)\Tilde{W}(k_2)W(k_3)-
W(k_1)W(k_2)\Tilde{W}(k_3)\big) = 0\,.\end{eqnarray}
Hence $\psi$ is a collisional invariant in the sense of Proposition \ref{9.prop1}.
Under the assumptions stated there, it follows that $\psi(k) = a\omega(k)$
and energy is the only local conservation law.\smallskip\\

For the case at hand, $\omega(k)\geq 0$ and $W(k)\geq 0$, which
implies $a<0$. Thus we have shown that under the ergodicity
condition (E) the {\textit{only}} stationary solutions of the
spatially homogeneous Boltzmann equation are
\begin{equation}\label{9.24}
W_\beta(k) = \big(e^{\beta\omega(k)}-1\big)^{-1}\,,\quad
\beta>0\,.
\end{equation}
$\beta$ is fixed through the initial condition as
$\int_{\mathbb{T}^3} dk\omega(k)W(k,t=0)=\int_{\mathbb{T}^3}
dk\omega(k)W_\beta(k)$ by conservation of energy.
Thermodynamically $\beta$ is the inverse temperature and the
entropy of $W_\beta$ according to (\ref{9.3}) is the equilibrium
entropy of an ideal Bose gas at zero chemical potential.

%%%%%%%%%%%%%%%%%%%%%%%%%%%%%%%%%

\section{The linearized collision operator}\label{sec.10}
\setcounter{equation}{0}

The thermal conductivity, in the kinetic limit, is determined
through the inverse of the linearized collision operator. We will
explain the standard argument in the following section. Here we
merely study the linearized collision operator as a linear
operator in $L^2(\mathbb{T}^3,dk)$.

We consider the spatially homogeneous Boltzmann equation
(\ref{8.16}), which we write as
\begin{equation}\label{10.1}
\frac{\partial}{\partial t}W= \mathcal{C}(W)\,.
\end{equation}
Under the ergodicity condition (E) the only stationary solutions
of (\ref{10.1}) are the thermal $W_\beta$. We fix $\beta$ and
linearize at $W_\beta$, where the convenient way of writing the
perturbation is
\begin{equation}\label{10.2}
W=W_\beta+W_\beta \tilde W_\beta f\,.
\end{equation}
To linear order in $f$, (\ref{10.1}) then becomes
\begin{equation}\label{10.3}
W_\beta  \tilde W_\beta \frac{\partial}{\partial t}f= -L f
\end{equation}
with the linearized collision operator
\begin{eqnarray}\label{10.4}
&&\hspace{-20pt}L f(k)= -\gamma \int_{\mathbb{T}^6} dk_1 dk_2
(\omega(k)\omega(k_1)\omega(k_2))^{-1}\\
&&\hspace{-20pt}\Big(2\delta(\omega(k)+\omega(k_1)-\omega(k_2))\delta(k+k_1-k_2)
\big((W_\beta(k_2)-W_\beta(k_1))W_\beta(k)\tilde W_\beta(k)f(k)\nonumber\\
&&\hspace{-20pt}+(W_\beta(k_2)-W_\beta(k))W_\beta(k_1)\tilde W_\beta(k_1)f(k_1)
+(W_\beta(k)+ \tilde W_\beta(k_1))W_\beta(k_2)\tilde
W_\beta(k_2)f(k_2)\big)\nonumber\\
&&\hspace{-20pt}+ \delta(\omega(k)-\omega(k_1)-\omega(k_2))\delta(k-k_1-k_2)
\big(-(W_\beta(k_1)+ \tilde W_\beta(k_2))W_\beta(k)\tilde
W_\beta(k)f(k) \nonumber\\
&&\hspace{-20pt}+ (W_\beta(k_2)-W_\beta(k))W_\beta(k_1)\tilde W_\beta(k_1)f(k_1)
+(W_\beta(k_1)-W_\beta(k))W_\beta(k_2)\tilde
W_\beta(k_2)f(k_2)\big)\Big).\nonumber
\end{eqnarray}
In each term we use the $\delta$-constraint which leads to
identities of the type
\begin{equation}\label{10.5}
W_\beta(k_1)W_\beta(k_2)\tilde W_\beta(k_3) = \tilde
W_\beta(k_1)\tilde W_\beta(k_2)W_\beta(k_3)\;\textrm{on}\;
\omega(k_1)+\omega(k_2)=\omega(k_3)\,.
\end{equation}
Then (\ref{10.4}) simplifies to
\begin{eqnarray}\label{10.6}
&&\hspace{-18pt}L f(k)= \gamma \int_{\mathbb{T}^6} dk_1 dk_2
(\omega(k)\omega(k_1)\omega(k_2))^{-1}\Big(2\delta
(\omega(k)+\omega(k_1)-\omega(k_2))\delta(k+k_1-k_2)\nonumber\\
&&\hspace{36pt}\tilde W_\beta(k)\tilde W_\beta(k_1)
W_\beta(k_2)\big(f(k)+f(k_1)-f(k_2)\big)+ \delta
(\omega(k)-\omega(k_1)-\omega(k_2))\nonumber\\
&&\hspace{36pt}\delta(k-k_1-k_2)\tilde W_\beta(k)W_\beta(k_1) W_\beta(k_2)
\big(f(k)-f(k_1)-f(k_2)\big)\Big)\,.
\end{eqnarray}

Let $\langle\cdot,\cdot\rangle$ denote the inner product in
$L^2(\mathbb{T}^3,dk)$. Using once more (\ref{10.5}), the
quadratic form for $L$ is given by
\begin{eqnarray}\label{10.7}
&&\hspace{-44pt}\langle g,L f \rangle = \gamma \int_{\mathbb{T}^9} dk_1 dk_2
dk_3 (\omega(k_1)\omega(k_2)\omega(k_3))^{-1}\nonumber \\
&&\hspace{16pt}\delta
(\omega(k_1)+\omega(k_2)-\omega(k_3))
\delta(k_1+k_2-k_3)W_\beta(k_1)W_\beta(k_2) \tilde
W_\beta(k_3)\nonumber\\
&&\hspace{16pt}\big(g(k_1)+g(k_2)-g(k_3)\big)\big(f(k_1)+f(k_2)-f(k_3)\big)\,.
\end{eqnarray}
Thus it is evident that
\begin{equation}\label{10.8}
L^\ast =L\,,\quad L\geq 0\,,\quad L \omega =0\,.
\end{equation}
Note that any zero eigenvector of $L$, $\langle f,L f \rangle = 0$, must be a collisional invariant in the sense of 
(\ref{9.14}). Hence, under the stated assumptions, the eigenvalue zero is nondegenerate.
In the classical limit, $\beta \to 0$, $W_\beta$ and $\tilde
W_\beta$ are to be replaced by $(\beta\omega)^{-1}$. Then $L$
equals the linearization of (\ref{4.7}).

The spectral properties of $L$ have not been studied, to our
knowledge. But they seem to fall into the standard folklore
picture of kinetic theory. $L$ can be written as
\begin{equation}\label{10.9}
L f(k)= -\int_{\mathbb{T}^3} dk' A(k,k')f(k')+V(k)f(k)\,.
\end{equation}
The ``potential" follows from (\ref{10.6}) as
\begin{eqnarray}\label{10.10}
&&\hspace{-16pt}V(k) = \gamma \tilde W_\beta(k)\omega(k)^{-1} \int_{\mathbb{T}^3}
dk_1 (\omega(k_1)\omega(k+k_1))^{-1}\Big( 2\delta
(\omega(k)+\omega(k_1)-\omega(k+k_1))\nonumber\\
&&\hspace{-16pt}\tilde W_\beta(k_1)W_\beta(k+k_1)+ \delta
(\omega(k)-\omega(k_1)-\omega(k+k_1))
W_\beta(k_1)W_\beta(k+k_1)\Big)\,,
\end{eqnarray}
while the integral kernel $A$ has the form
\begin{eqnarray}\label{10.11}
&&\hspace{-22pt}A(k,k')=\\
&&\hspace{-16pt} 2\gamma
\big\{-(\omega(k)\omega(k')\omega(k+k'))^{-1} \tilde W_\beta(k)
\tilde W_\beta(k')W_\beta(k+k')
 \delta
(\omega(k)+\omega(k')-\omega(k+k'))\nonumber\\
&&\hspace{-16pt}+(\omega(k)\omega(k')\omega(k-k'))^{-1} \tilde
W_\beta(k) W_\beta(k')\tilde W_\beta(k-k') \delta
(\omega(k)-\omega(k')+\omega(k-k'))\nonumber\\
&&\hspace{-16pt}+(\omega(k)\omega(k')\omega(k-k'))^{-1} \tilde
W_\beta(k) W_\beta(k')W_\beta(k-k') \delta
(\omega(k)-\omega(k')-\omega(k-k'))\big\}\,. \nonumber
\end{eqnarray}
Under our assumptions on $\omega$ the potential is bounded away
from zero, $0<c_-\leq V(k)$ but not bounded sine $\omega_0=0$. For
$k$ fixed, $A(k,k')$ is concentrated on a set of codimension 1.
The kernel of $A^2$ is a function, but $A^2(k,k')$ has singular
points, in particular $A^2(k,k)=\infty$. We conjecture that
$\textrm{tr}A^4 <\infty$. If so, the bottom of the continuous
spectrum of $L$ is $c_-$. $L$ has a spectral gap and the continuous
spectrum extends to $\infty$. On the linearized level the
homogeneous system relaxes exponentially fast to equilibrium.

%%%%%%%%%%%%%%%%%%%%%%%%%%%%%

\section{Thermal conductivity}\label{sec.11}
\setcounter{equation}{0}

We look for a stationary solution of the Boltzmann equation
(\ref{8.16}), to say
\begin{equation}\label{11.1}
(2\pi)^{-1}\nabla\omega\cdot\nabla_r W= \mathcal{C}(W)\,,
\end{equation}
which has approximately a linear temperature profile
$T(r)=\beta^{-1}+\nabla T\cdot r$ with $|\nabla T|\ll 1$. Of
course, the ergodicity condition (E) has to be imposed. On the
left hand side in (\ref{11.1}) we assume local equilibrium in the
form $(e^{\omega(k)/T(r)}-1)^{-1}$ while on the right hand side we
expand $W=W_\beta+W_\beta\tilde{W}_\beta f$. Then (\ref{11.1})
becomes
\begin{equation}\label{11.2}
(\nabla\omega\cdot\nabla T)\omega W_\beta\tilde{W}_\beta \beta^2
=-L f\,.
\end{equation}
Since,  as argued before, the zero eigenvalue of $L$ is nondegenerate and the corresponding
eigenvector
 $\omega$ is orthogonal
to  $(\nabla\omega)\omega W_\beta\tilde{W}_\beta$ $\omega$, 
$L$ can be inverted and
\begin{equation}\label{11.3}
f= -(2\pi)^{-1}\beta^2 
L^{-1}W_\beta\tilde{W}_\beta\omega( \nabla T\cdot\nabla\omega)\,.
\end{equation}
The steady state heat (=energy) flux is then
\begin{eqnarray}\label{11.4}
&&\hspace{-12pt}j_\textrm{e}= (2\pi)^{-1}\int_{\mathbb{T}^3} dk
(W_\beta+W_\beta\tilde{W}_\beta f)\omega\nabla\omega\nonumber\\
&&= (2\pi)^{-2}\beta^2 \langle\omega\nabla\omega
W_\beta\tilde{W}_\beta,L^{-1}
W_\beta\tilde{W}_\beta\omega\nabla\omega\cdot\nabla T\rangle\,.
\end{eqnarray}
The thermal conductivity $\kappa$ is defined through Fourier's law
$j_\textrm{e}=-\kappa\nabla T$, hence
\begin{equation}\label{11.5}
\kappa_{\alpha\alpha'} (T)=\beta^{2} (2\pi)^{-2} \langle
W_\beta\tilde{W}_\beta\omega\nabla_\alpha\omega,L^{-1}
W_\beta\tilde{W}_\beta\omega\nabla_{\alpha'}\omega\rangle\,, \quad \beta= 1/T\,.
\end{equation}
For the case at hand, $\kappa$ is diagonal,
$\kappa_{\alpha\alpha'} = \delta_{\alpha\alpha'}\kappa$, and
\begin{equation}\label{11.6}
\kappa(T)= \frac{1}{3} T^{-2} (2\pi)^{-2} \langle
W_\beta\tilde{W}_\beta\omega\nabla\omega\,,\;L^{-1}
W_\beta\tilde{W}_\beta\omega\cdot\nabla\omega\rangle\,.
\end{equation}

Inserting Fourier's law into the local conservation of energy
(\ref{6.2}) yields a nonlinear diffusion equation for the energy
transport,
\begin{equation}\label{11.7}
\frac{\partial}{\partial t}e(r,t) =
\nabla\cdot\Big(\kappa(T(e))\frac{dT(e)}{de}\nabla e(r,t)\Big)
\end{equation}
with the thermodynamic relation
\begin{equation}\label{11.8}
e(T)= \int_{\mathbb{T}^3} dk \omega W_\beta\,,\quad
\beta=1/T\,.
\end{equation}
It would be of interest to establish (\ref{11.7}) as the
hydrodynamic limit of the Boltzmann equation (\ref{8.16}).

We discuss the qualitative temperature dependence of the thermal
conductivity. At high temperatures, $W_\beta$ and
$\tilde{W}_\beta$ are replaced by $(\beta\omega)^{-1}$. Then
\begin{equation}\label{11.9}
\kappa(T)=\frac{1}{3} T^{-1}(2\pi)^{-2} \langle
\frac{1}{\omega}\nabla\omega,(L_{\textrm{cl}})^{-1}\frac{1}{\omega}\cdot\nabla\omega\rangle
\end{equation}
with the classical linearized collision operator
\begin{eqnarray}\label{11.10}
&&\hspace{-20pt}L_{\textrm{cl}}f(k) = \gamma \int_{\mathbb{T}^6} dk_1 dk_2
dk_3 (\omega(k)\omega(k_1)\omega(k_2))^{-2}\nonumber\\
&&\Big( 2\delta
(\omega(k)+\omega(k_1)-\omega(k_2))\delta
(k+k_1-k_2)
\big(f(k)+f(k_1)-f(k_2)\big) \nonumber\\
&&+\delta (\omega(k)-\omega(k_1)-\omega(k_2))\delta(k-k_1-k_2)
\big(f(k)-f(k_1)-f(k_2)\big)\Big)\,.
\end{eqnarray}
Therefore the temperature dependence is multiplicative and
\begin{equation}\label{11.11}
\kappa(T)=\frac{\theta_\textrm{h}}{T}\,,\quad T
\;\textrm{large}\,,
\end{equation}
with $\theta_\textrm{h}$ determined by (\ref{11.9}).

At low temperatures the temperature dependence of $\kappa$ is not
so easily accessible. For $\beta\to \infty$
\begin{equation}\label{11.12}
W_\beta (k) \cong e^{-\beta\omega(k)}\,,
\end{equation}
which means that the number of energy carrying phonons is greatly
reduced. On the other hand, normal processes conserve momentum and
thus do not degrade the phonon current. Only in umklapp processes,
momentum is transferred to the lattice. But umklapp becomes rare
at low temperatures. It is argued in \cite{Gu}, Chapter 2.2, that the latter effect dominates
resulting in the exponential increase
\begin{equation}\label{11.13}
\kappa(T)\simeq e^{\theta_\textrm{l}/T}\,,\quad \theta_\textrm{l} > 0\,,\quad T \;\textrm{small}\,.
\end{equation}
It would be of interest to have bounds based directly on (\ref{11.5}) which confirm such a low temperature behavior.

For real materials the dependence (\ref{11.13}) is not so easily
resolved, since the conductivity is dominated by scattering from
isotope mass disorder, as will be discussed in the next section.
For mass purified samples the conductance is limited by the
size of the probe.

%%%%%%%%%%%%%%%%%%%%%%%%%%%%%%%%

\section{Isotope disorder}\label{sec.12}
\setcounter{equation}{0}

At low temperatures the thermal conductivity is limited by
impurities. Even for a chemically pure crystal, in their natural
abundance the crystal atoms come as a random mixture of isotopes.
Artifically enriched, resp. purified, samples have also been
manufactured so to provide a test of the predictions by the
theory. In the kinetic limit, the effects of  impurities and small
anharmonicities are additive. Therefore we study here random
isotope substitution in the harmonic approximation. If $m_x$
denotes the mass of the atom at site $x$, in the frame of our toy
model the equations of motion read
\begin{eqnarray}\label{12.1}
&&\frac{d}{dt}q_x(t) = \frac{1}{m_x}p_x(t)\nonumber\\
&&\frac{d}{dt}p_x(t) = -\sum_{y\in\mathbb{Z}^3}\alpha(y-x)q_y(t)
-\omega_0^2 q_x(t)\,,\quad x\in \mathbb{Z}^3\,,
\end{eqnarray}
compare with (\ref{2.7}). For isotope disorder the mass ratio is
of order $10^{-2}$. Therefore, in a good approximation we may set
\begin{equation}\label{12.2}
\frac{1}{m_x}= (1+\sqrt{\varepsilon}\xi_x)^2\,,\quad
\varepsilon\ll 1\,,
\end{equation}
with $\{\xi_x\,,\; x\in \mathbb{Z}^3\}$ a collection of
independent, identically distributed random variables. Let us denote by
$\mathbb{E}$ the expectation with respect to the
$\xi_x$'s, i.e.~the disorder average. We assume
$\mathbb{E}(\xi_x)=0$ and $|\xi_x|\leq c_0$ so that  $m_x>0$
for sufficiently small $\varepsilon$ as required for
mechanical stability.

To derive the kinetic equation we first follow the scheme devised
for the weak nonlinearity, also to emphasize that the structure is
in parallel. To mathematically justify the decoupling step a
distinct strategy is required, however, see Section \ref{sec.13}.

We rewrite the equations of motion (\ref{12.1}) in terms of the
$a$-field as defined in (\ref{2.11}), which means in terms of the
homogeneous system. Since the evolution is linear, there is no
difference between the classical and quantum model, possibly
except for the choice of the initial state and thus the initial
Wigner function. One obtains
\begin{eqnarray}\label{12.3}
&&\hspace{-36pt}\frac{d}{d t}a(k,\sigma,t) = i \sigma
\omega(k)a(k,t) - i\sqrt{\varepsilon}\sigma \sum_{\sigma_1 = \pm 1}\int_{\mathbb{T}^6}
dk_1 dk_2(\omega(k)\omega(k_1))^{1/2}\nonumber\\
&&\hspace{24pt} \times\delta(-\sigma k + \sigma_1k_1-k_2)\sigma_1a(k_1,\sigma_1,t)
 \widehat{\xi}(k_2)+\mathcal{O}(\varepsilon)\,,
\end{eqnarray}
where we take the quantum framework, to be definite. Compared to
(\ref{5a.7}) in essence one of the $a$-factors has been replaced
be $ \widehat{\xi}$. Since $a(k,\sigma,t)$ depends on the disorder, the
equations of motion are, so to speak, nonlinear in the couple $(a,
\widehat{\xi}\,)$.

The object of interest is the Wigner function
\begin{equation}\label{12.4}
\widehat{W}{^\varepsilon}(\eta,k,t)=\varepsilon^3
\mathbb{E}(\langle a(k-\varepsilon\eta/2)^\ast
a(k+\varepsilon\eta/2) \rangle_{t/\varepsilon})
\end{equation}
on the kinetic time scale $\varepsilon^{-1}t$. In (\ref{12.4})
there are two averages, one over the disorder, $\mathbb{E}$, and
one over the initial state. To be physically consistent we think
of a scale of initial states as explained in Sections
\ref{sec.3} and \ref{sec.7}. Since the equations of motion are
linear, there is however a much wider choice. In the classical
model the initial configuration could be deterministic. Quantum
mechanically the initial wave function could be in the
one-particle space $\mathfrak{h}$. All what is required is that
the Wigner function (\ref{12.4}) (possibly substituting
$\varepsilon^3$ by some other prefactor) has a limit at the
initial time $t=0$. The disorder average is taken only to avoid
extra difficulties in the derivation. Physically one expects $\langle
a(k-\varepsilon\eta/2)^\ast
a(k+\varepsilon\eta/2)\rangle_{t/\varepsilon}$ to be
self-averaging in the limit $\varepsilon\to 0$. More precisely,
the random variable $\int d\eta dk f(\eta,k)\langle
a(k-\varepsilon\eta/2)^\ast a(k+\varepsilon\eta/2)
\rangle_{t/\varepsilon}$, $f$ a smooth $\varepsilon$-independent
test function, will tend with probability one to a deterministic
limit as $\varepsilon\to 0$. No disorder average should be needed, in
fact.

Let us see how the arguments from Sections \ref{sec.5} and
\ref{sec.8} transcribe to the present situation. As before 
the atomic scale is used. Then
\begin{eqnarray}\label{12.5}
&&\hspace{-130pt}\frac{d}{d t}\mathbb{E}(\langle a(p)^\ast
a(q)\rangle_t)= i(\omega(p)-\omega(q)) \mathbb{E}(\langle
a(p)^\ast a(q)\rangle_t)\nonumber\\
\hspace{84pt}&&+  \varepsilon \int^t_0 ds G(q,p,t-s,s)
\end{eqnarray}
with
\begin{eqnarray}\label{12.6}
&&\hspace{-26pt}G(q,p,t,s)=-\mathbb{E}\Big[\sum_{\sigma_1 = \pm 1}
\int_{\mathbb{T}^6}
dk_1 dk_2 \sum_{\tau_1 = \pm 1}\int_{\mathbb{T}^6} dl_1 dl_2 (\omega(k_1)\omega(\sigma_1k_1-k_2))^{1/2}\nonumber
\\
&&\hspace{-10pt}(\omega(l_1)\omega(\tau_1l_1-l_2))^{1/2}\widehat{\xi}(k_2) \widehat{\xi}(l_2)\Big( e^{it(-\omega (q)+\sigma_1
\omega(k_1))}\delta(-p+\sigma_1k_1-k_2) 
\nonumber\\
&&\hspace{-10pt} \times \big(\delta(q + \tau_1l_1-l_2)\sigma_1\tau_1\langle
a(k_1,\sigma_1)a(l_1,\tau_1)\rangle_s +\delta(-\sigma_1k_1+\tau_1l_1-l_2)\nonumber\\
&&\hspace{-10pt}
\times \tau_1 \langle a(l_1,\tau_1)
a(q)\rangle_s\big)+e^{it(\omega (p)+\sigma_1 \omega(k_1))}\delta(q+\sigma_1k_1-k_2)
\big(\delta(-p+\tau_1l_1-l_2)\nonumber\\
&&\hspace{-10pt}
\times \sigma_1\tau_1\langle a(l_1,\tau_1)a(k_1,\sigma_1)\rangle_s 
+ \delta(-\sigma_1k_1+\tau_1l_1-l_2)\tau_1\langle a(p)^\ast
a(l_1,\tau_1)\rangle_s\big) \Big)\,
\end{eqnarray}
The homogeneous term vanishes, since $\mathbb{E} (\xi_x)=0$.

There is no need to carry the argument any further, since we have
seen it already. The analogue of the assumption of local
stationarity is to factorize, on the kinetic scale,  the disorder average as
\begin{equation}\label{12.7}
\mathbb{E}\big(\widehat{\xi}(k_1)\widehat{\xi}(k_2)\langle
a(l_1)^\ast a(q)\rangle_s\big) \cong \mathbb{E}\big(\widehat{\xi}(k_1)
\widehat{\xi}(k_2)\big) \mathbb{E}\big(\langle
a(l_1)^\ast a(q)\rangle_s\big)\,,
\end{equation}
for example. The rapidly oscillating time integral generates the
$\delta$-function for the energy conservation and makes terms as
$\langle aa\rangle$ and $\langle a^\ast a^\ast \rangle$ to vanish.
After these steps only four terms are left which combine into the
linear Boltzmann equation for the limit Wigner function $W$,
\begin{eqnarray}\label{12.8}
&&\hspace{-20pt}\frac{\partial}{\partial t}W(r,k,t) +
\frac{1}{2\pi}\nabla\omega(k)\cdot \nabla_r W(r,k,t)\nonumber\\
&&\hspace{2pt}= 2\pi\mathbb{E}(\xi^2_0)\omega(k)^2
\int_{\mathbb{T}^3} dk_1 \delta(\omega(k)-\omega(k_1))
\big(W(r,k_1,t)- W(r,k,t)\big)\,.
\end{eqnarray}

If one wants to compute the thermal conductivity including the
isotope disorder, one merely has to add to $L$ in (\ref{11.2}) the
impurity scattering in the form
\begin{equation}\label{12.9}
L_\textrm{i} f(k) =2\pi\mathbb{E}(\xi_0)^2 \omega^2 W_\beta
\tilde{W}_\beta \int_{\mathbb{T}^3} dk_1
\delta(\omega(k)-\omega(k_1)) \big(f(k)-f(k_1)\big)\,.
\end{equation}
By energy conservation $L_\textrm{i}$ randomizes on each energy
shell. Thus the zero eigenvectors of $L_\textrm{i}$ are of the
form $h(\omega(k))$ with arbitrary $h$. The Planck distribution is
singled out by the anharmonicities and can be thought of as a
specific initial condition in the current context. Following the
arguments in Section \ref{sec.11}, the thermal conductivity is
given through
\begin{equation}\label{12.10}
\kappa(T) =\frac{1}{3}\beta^2(2\pi)^{-2}\langle W_\beta
\tilde{W}_\beta \omega \nabla\omega,(L+L_\textrm{i})^{-1}
W_\beta \tilde{W}_\beta \omega\cdot\nabla\omega\rangle\,.
\end{equation}

In the limit of vanishing anharmonicity, $\gamma\to 0$, $\kappa$
can be computed more explicitly, since, by symmetry
$k\rightsquigarrow -k$, $\nabla\omega$ is an eigenfunction of
$L_\textrm{i}$. Then
\begin{equation}\label{12.11}
\kappa_\textrm{i} (T) =\frac{1}{3}\beta^2(2\pi)^{-2}(\frac{\pi}{2}
\mathbb{E}(\xi_0^2))^{-1} \int_{\mathbb{T}^3} dk W_\beta
\tilde{W}_\beta (\nabla\omega)^2 \frac{1}{\tau(\omega)}
\end{equation}
with
\begin{equation}\label{12.12}
\tau(\omega)=\int_{\mathbb{T}^3} dk_1
\delta(\omega-\omega(k_1))\,.
\end{equation}
If $\omega_0>0$, $\kappa_\textrm{i}$ vanishes exponentially as
$e^{-\beta\omega_0}$. If $\omega_0=0$, the competition between the
divergence of $\tau(\omega)$ for small $\omega$ and the
suppression of phonons results in a dependence as $
\kappa_\textrm{i}(T)\simeq T^{-1}$ for $T\to 0$.\bigskip\\

%%%%%%%%%%%%%%%%%%%%%%%%%%%%%%%%%%%%

\section{Mapping to a Schr\"{o}dinger-like equation with a weak random
potential}\label{sec.13}
\setcounter{equation}{0}

The linear evolution equation (\ref{12.1}) suggests to use
time-dependent perturbation theory. Let us set
\begin{equation}\label{13.1}
A=
\begin{pmatrix} 0 & 1 \\ \Delta-\omega_0^2 & 0
\end{pmatrix}\,,
\quad V=
\begin{pmatrix} 0 & \xi_x \\ 0 & 0
\end{pmatrix}\,.
\end{equation}
Then
\begin{equation}\label{13.2}
\frac{d}{dt}
\begin{pmatrix} q \\ p \end{pmatrix}= (A+\sqrt{\varepsilon}V)
\begin{pmatrix} q \\ p \end{pmatrix}
\end{equation}
and
\begin{equation}\label{13.3}
e^{(A+\sqrt{\varepsilon}V)t}=e^{At} + \sum^\infty_{n=1}
\varepsilon^{n/2}\int_{0\leq t_1\leq \ldots \leq t_n\leq t} d t_n
\ldots d t_1 e^{A(t-t_n)} Ve^{A(t_n-t_{n-1})} \ldots Ve^{At_1}\,.
\end{equation}
We insert the propagator (\ref{13.3}) into the definition of the
Wigner function and average over disorder. The leading term is
$\exp[-t(1/2\pi)\nabla\omega\cdot \nabla_r]$ on the kinetic scale.
The term of order $\sqrt{\varepsilon}$ vanishes because
$\mathbb{E}(\xi_x)=0$ and the term of order $\varepsilon$ yields,
when kinetically scaled and taking the limit $\varepsilon\to 0$,
\begin{equation}\label{13.4}
\int^t_0 ds e^{-(t-s)(1/2\pi)\nabla\omega\cdot\nabla_r} L_i
e^{-s(1/2\pi)\nabla\omega\cdot\nabla_r} W
\end{equation}
with $W$ the initial Wigner function and $L_i$ the linear
collision operator of (\ref{12.8}). Thus we only have to study
systematically the higher orders of the perturbation series and to
convince ourselves that they yield the corresponding
time-dependent perturbation series for (\ref{12.8}).
Unfortunately, while the principle is correct, it will never lead
to a proof, since there are too many terms in the perturbation
series. Even if we postulate that the $\{\xi_x\}$ are independent
Gaussians, the number of pairings is $n!/2^{n/2} (n/2)!$ which are
balanced by a factor $a^n t^{n/2}/(n/2)!$ from the time
integrations. Thus the series converges only for $|t|\leq t_0$
with a suitable $t_0$ on the kinetic time scale.

Erd\"{o}s and Yau \cite{ErY} study the one-particle
Schr\"{o}dinger equation with a random potential which has a
mathematical structure comparable to (\ref{13.2}). Thus the problem
of an exploding number of terms in the perturbation series also
arises. To circumvent this blockage, they expand only up to
$N=N(\varepsilon)$ and estimate the remainder by using the
unitarity of the unexpanded Schr\"{o}dinger evolution. To copy
their method we have to exploit that the energy
\begin{equation}\label{13.5}
H= \frac{1}{2}\sum_{x\in\mathbb{Z}^3}
\Big((1+\sqrt{\varepsilon}\xi_x)^2
p_x^2+\sum_{y\in\mathbb{Z}^3}\alpha(y-x)q_y q_x\Big)
\end{equation}
is conserved for each realization of the disorder. The energy
depends on $\xi$. This is rather inconvenient and we transform to
new fields such that the flat $\ell_2$-norm is conserved. Let us
regard
\begin{equation}\label{13.6}
\Omega_{xy}=\int_{\mathbb{T}^3} dk e^{i2\pi k\cdot(x-y)}\omega(k)
\end{equation}
as a linear operator in $\ell_2=\ell_2(\mathbb{Z}^3)$. Under our
assumption $\Omega$ has an exponential decay in $|x-y|$, which
possibly worsens as $\omega_0 \to 0$. We define
\begin{equation}\label{13.7}
\psi=(\psi^+,\psi^-)
\end{equation}
with components
\begin{equation}\label{13.8}
\psi^\pm = \frac{1}{\sqrt{2}}\big(\Omega q_x
\pm i(1+\sqrt{\varepsilon}\xi_x)p_x\big)\,.
\end{equation}
Note that $\|\psi^+\|^2=\|\psi^-\|^2=H$. Thus the
$\ell_2$-norm of $\psi$ is conserved in time. The $\psi$-field
evolves according
\begin{equation}\label{13.9}
i \frac{\partial}{\partial t}\begin{pmatrix} \psi^+
\\ \psi^-\end{pmatrix}=
\begin{pmatrix} \Omega & 0 \\ 0 & -\Omega\end{pmatrix}
\begin{pmatrix} \psi^+ \\ \psi^-\end{pmatrix}+ \sqrt{\varepsilon}
\begin{pmatrix} \xi\Omega+\Omega\xi & \xi\Omega-\Omega\xi \\
-\xi\Omega+\Omega\xi & -\xi\Omega-\Omega\xi\end{pmatrix}
\begin{pmatrix} \psi^+ \\ \psi^-\end{pmatrix} \,,
\end{equation}
where $\xi_x$ is regarded as a multiplication operator, $(\xi
f)_x=\xi_x f_x$. We use the short hand
\begin{equation}\label{13.10}
i \frac{\partial}{\partial
t}\psi=(H_0+\sqrt{\varepsilon}V)\psi\,,\quad
H_\varepsilon=H_0+\sqrt{\varepsilon}V
\end{equation}
and regard (\ref{13.9}) as an evolution equation in $\ell_2 \oplus
\ell_2$. Clearly, $H_\varepsilon$ is bounded and
$H_\varepsilon=H^\ast_\varepsilon$. Thus $e^{-iH_\varepsilon t}$
is unitary. Physical initial data are constrained to satisfy
$(\psi^+)^\ast=\psi^-$, but this will be imposed only at the very
end.

Since $\psi$ is a 2-spinor, the Wigner function becomes a $2\times
2$ matrix. Inserting the kinetic scaling, one has
\begin{equation}\label{13.11}
W^\varepsilon_{\sigma\sigma'}(y,k,t)=
2^{-3}\int_{(2\mathbb{T}/\varepsilon)^3}d\eta e^{i2\pi y\cdot\eta}
 \widehat{\psi}^\sigma(k-\varepsilon\eta/2,t/\varepsilon)^\ast
\widehat{\psi}^{\sigma'}(k+\varepsilon\eta/2,t/\varepsilon)\,,
\end{equation}
$\sigma = \pm,\sigma' = \pm$, with $k\in \mathbb{T}^3$ and $y\in
(\varepsilon\mathbb{Z})^3$. Note that, because of the definition
(\ref{13.8}), we deviated slightly from previous conventions. In
particular $\int_{\mathbb{T}^3}dk W_{++}^\varepsilon\\
(y,k,t)$ now acquires the meaning of an energy density at kinetic
time $t$. The off-diagonal element, $W^\varepsilon_{+-}(t)$, picks
up the fastly oscillating phase factor $\exp[\pm
2i\omega(k)t/\varepsilon]$. Hence it vanishes upon time averaging.
For example, $W^\varepsilon_{+-}(t)$ determines the difference
between kinetic and potential energy, which is indeed a fast
variable. By symmetry $W^\varepsilon_{--}(t)$ is obtained from
$W_{++}(t)$ by substituting $\omega$ by $-\omega$. Thus we only
have to deal with $W^\varepsilon_{++}(t)$.

The Wigner function at fixed $t$ typically oscillates on small
scales and only upon integrating against a smooth test function
one expects to have a limit. Thus let
$J:\mathbb{R}^3\times\mathbb{T}^3\to\mathbb{R}$ be a smooth,
rapidly decreasing function with its Fourier transform with
respect to the spatial argument denoted by $\widehat{J}$. The
Wigner function integrated against $J$ becomes then
\begin{equation}\label{13.12a}
\langle J,W^\varepsilon_{++}[\psi]\rangle=
\int_{\mathbb{R}^3}d\eta\int_{\mathbb{T}^3}dk\widehat{\psi}^+
(k-\varepsilon\eta/2)^\ast\widehat{J}(\eta,k) \widehat{\psi}^+
(k+\varepsilon\eta/2)\,.
\end{equation}
We now choose a sequence of initial conditions $\psi^\varepsilon$
such that $\|\psi^\varepsilon\|\leq\textit{const}$ and such that
the initial Wigner function has a limit,
\begin{equation}\label{13.13a}
\lim_{\varepsilon\to 0}\langle
J,W^\varepsilon_{++}[\psi^\varepsilon]\rangle
=\int_{\mathbb{R}^3\times\mathbb{T}^3}J(r,k)\mu_0(drdk)\,.
\end{equation}
In addition one has to impose tightness in the sense that
\begin{equation}\label{13.13b}
\lim_{\mathbb{R}\to\infty}\lim_{\varepsilon\to
0}\sup\sum_{|x|>\mathbb{R}/\varepsilon}|\psi^\varepsilon_x|^2=0\,.
\end{equation}
At this level of generality $\mu_0(drdk)$ is a positive, bounded
measure. With $\psi^\varepsilon$ as initial datum the time-evolved
field is given by
\begin{equation}\label{13.14a}
\psi(t)=e^{-iH_\varepsilon t}\psi^\varepsilon\,.
\end{equation}
Clearly, the issue is to determine $\langle
J,W^\varepsilon[\psi(t/\varepsilon)]\rangle$ in the limit
$\varepsilon\to 0$.

To achieve the existence of the limit one needs two conditions on
the dispersion relation $\omega$.\smallskip\\
(i) The first condition we have met already in Section 5 and
requires $\omega$ to be a Morse function, meaning that all
critical points of $\omega$ are isolated and nondegenerate (no
zero eigenvalue in the quadratic approximation).\smallskip\\
(ii) The second condition is the crossing estimate, which refers
to the decay estimate of a particular oscillatory integral over
$\mathbb{T}^3\times\mathbb{T}^3$. It is too technical to be stated
explicitly here. The crossing estimate is verified for a few
particular dispersion relations \cite{Ch,ErSaY1,LuSp}. It is not
excluded that with improved technology the crossing estimate can
be reduced to the Morse property.

Before stating our result we have to explain what we mean by
solution of the Boltzmann equation (\ref{12.8}) with a measure as
initial condition. The standard method is to switch to the dual
equation and to prove that it is a contraction semigroup on
$C(\mathbb{R}^3\times\mathbb{T}^3,\mathbb{R})$, the space of
bounded and continuous functions, which follows from the key
observation that, since $\omega$ is Morse, the operator
\begin{equation}\label{13.15a}
B f(k)= 2\pi \mathbb{E}(\xi^2_0)\omega(k)^2\int_{\mathbb{T}^3}
\delta(\omega(k)-\omega(k')) f(k')dk'
\end{equation}
satisfies $\|Bf\|\leq c\|f\|$ in $C(\mathbb{T}^3)$ for some $c>0$.
In particular the total collision rate
\begin{equation}\label{13.16a}
\nu(k)=
2\pi\omega(k)^2\int_{\mathbb{T}^3}\delta(\omega(k)-\omega(k'))dk'
\end{equation}
is continuous, thus bounded. To the Boltzmann equation there is
associated the stochastic process $(r(t),k(t))$ with state space
$\mathbb{R}^3\times\mathbb{T}^3$. It is governed by
\begin{equation}\label{13.17a}
\frac{d}{dt}r(t)=\frac{1}{2\pi}\nabla\omega(k(t))\,,
\end{equation}
where $k(t)$ is a Markov jump process on $\mathbb{T}^3$ with jump
rate
$2\pi\mathbb{E}(\xi^2_0)\omega(k)^2\delta(\omega(k)-\omega(k'))dk'$.
We define the measure $\mu_t(drdk)$ as the joint distribution of
$(r(t),k(t))$ when started with $\mu_0$ as initial measure.

The following theorem is the main result of a joint paper with J.
Lukkarinen \cite{LuSp}.
\begin{theorem} Let $\omega$ be a Morse function and satisfy the
crossing estimate and let $\psi^\varepsilon\in \ell_2\oplus\ell_2$
be uniformly bounded and such that (\ref{13.13a}), (\ref{13.13b})
hold. Then
\begin{equation}\label{13.18a}
\lim_{\varepsilon\to 0} \mathbb{E}\big(\langle
J,W^\varepsilon(\psi(t/\varepsilon))\rangle\big)=
\int_{\mathbb{R}^3\times\mathbb{T}^3}\mu_t(drdk)J(r,k)\,,
\end{equation}
where $\mu_t$ is the solution of the Boltzmann equation
(\ref{12.8}) with initial datum $\mu_0$.
\end{theorem}
{\bf Remark:} The dispersion relations (\ref{4a.3}) and
(\ref{4a.4}) with $\omega_0>0$ are Morse and satisfy the crossing
estimate.\smallskip

Any description of the methods used in the proof would lead us too
far astray. Let me only emphasize that they are based on
techniques developed by Erd\"{o}s and Yau \cite{ErY}, see also
\cite{Er}, for estimating Feynman diagrams and for cutting
delicately the perturbation series in an $\varepsilon$-dependent
way. In a recent paper Chen \cite{Ch} considers the
Schr\"{o}dinger equation on a lattice with nearest neighbor
hopping and a random potential $V$ with $V(x)$, $x \in
\mathbb{Z}^3$, a collection of independent random variables. His
estimates greatly helped in our proof. Chen \cite{Ch1} also shows
that for his model the convergence of the Wigner function holds in
probability, which is a strong indication that the same property should hold 
for isotope disrder.

%%%%%%%%%%%%%%%%%%%%%%%%%%%%%%%%%%%%%

\section{Guide to the literature}\label{sec.14}
\setcounter{equation}{0}
\subsection{Phonon Boltzmann equation}\label{sec.14a}

I am not an expert in phonon physics and the guide reflects merely
my own reading. The focus is deliberately somewhat narrow and I
deal only with the rigorous derivation and a few basic properties
of the phonon Boltzmann equation.

The seminal paper on the subject is R. Peierls \cite{Pe} from
1929. He is the first one to write down the phonon Boltzmann
equation (\ref{8.16}). Nordheim \cite{No} follows a similar path
for weakly interacting quantum gases. Peierls' derivation consists
in a careful application of Fermi's golden rule. His argument,
with variations and modernized notation, has been repeated many
times. A standard reference is the Handbuch article by Leibfried
\cite{Le}. An excellent textbook discussion is Callaway \cite{Ca}.
I very much enjoyed the monograph by V.L.~Gurevich \cite{Gu}. He
also applies the Fermi golden rule but in addition discusses
extensively the physical conditions required for its applicability
in the derivation of the Boltzmann equation. As a standard, the
Fermi golden rule is introduced in the context of the spatially
homogeneous system. Spatial variation is simply added in the most
obvious way. A great advantage of the Wigner function formulation
is to incorporate spatial variation from the outset.

Since the most interesting aspects of phonon physics are related
to quantization, the classical anharmonic crystal tends to be
ignored. But in his basic article Peierls also treats the
classical system. Brout and Prigogine \cite{BrPr} provide a more
detailed account, which is summarized  in the book by I.~Prigogine
on nonequilibrium statistical mechanics \cite{Pr}. He and Peierls
argue that, through a random phase approximation, the joint
distribution of the $a(k)^\ast a(k)$ satisfies a diffusion
equation in the high-dimensional phase space. Reducing to the
one-particle distribution yields a nonlinear evolution equation
for $W(k,t)$, in spirit similar to the structure one has in the
Kac model of kinetic theory \cite{Ka,CaLo}. At the time such
reasoning was very fashionable. But its underlying assumptions are
rather dubious, see Appendix \ref{sec.16d}.

As regards to derivation from the microscopic Hamiltonian model
the next level is to improve on the Fermi golden rule, which
started with the work of van Hove \cite{vH} and lead into the
development of diagrammatic expansions in parallel with similar
techniques in quantum field theory. This is a vast area, still
active today. A very readable account with focus on weak coupling
and Boltzmann type transport equations is the slim monograph by
S.~Fujita \cite{Fu}. He discusses the impurity problem and
electron-electron collisions. But he could have treated phonons,
as well. Fujita immediatly employs Wigner functions as a matter of
fact, which makes one wonder who originally pushed this concept as
a tool for transport equations. In his famous paper Wigner
\cite{Wi} introduces the notion but then applies it to the
semiclassical limit of the quantum statistical partition function.

The importance of local stationarity has been stressed mostly in
the quarters of mathematical physics, since it is one central
property which needs to be established when proving the validity
of a macroscopic equation. Erd\"{o}s, Salmhofer, and Yau
\cite{ErSaY} discuss the strongly related problem of electron
collisions in the same spirit as done here. Benedetto {\it et al.}
\cite{BeCa} are more ambitious and, ignoring the issue of absolute
convergence, study the dominant terms of the time-dependent
perturbation series in the kinetic limit. Most likely, their
techniques extend to the present case.

The harmonic lattice with isotope disorder is in its structure rather
similar to a one-particle Schr\"{o}dinger equation with a weak
random potential. We refer to \cite{ErY,Er,Ch,Ch1,ErSaY1} for recent advances
and the derivation of the corresponding kinetic equation. Our
Theorem 14.1 relies on their work.

Bal,  Komorowski, and Ryzhik \cite{BKR} study the continuum wave equation with a weakly
disordered index of refraction. They consider a high frequency
approximation and prove that in this limit the Wigner function is
governed by (\ref{12.8}) with the jump collision operator
replaced by a spherical Laplacian, which turns out to by the small
angle approximation to the collision operator in (\ref{12.8}).

Compared to its famous sister the phonon Boltzmann equation has
received little mathematical attention, for no good reason. While
we expect that much of the technology developed in the context of
the Boltzmann equation carries over, we point out that the phonon case has two
simplifying features: The wave vector space is compact and more
importantly the velocity, $\nabla\omega(k)$, is uniformly bounded.
How far this will carry, only a detailed study can show. The
derivation of hydrodynamics should be more accessible, since the
Boltzmann equation has only a single conservation law and its corresponding nonlinear
diffusion equation (\ref{11.7}) has a global solution, say in a
finite macroscopic box with initial data bounded away from 0.

%%%%%%%%%%%%%%%%%%%%%%%%%%%%%%%%

\subsection{Energy transport in anharmonic
chains}\label{sec.14b} \setcounter{equation}{0}

Classical anharmonic chains are a challenging test ground for the
numerical integration of Newton's equation of motion ever since
the seminal work of Fermi, Pasta and Ulam \cite{FPU}. With
increasing computer power steady state current transport for chain
lengths up to $10^4$, in exceptional cases even $10^5$, are
reported. These studies mostly investigate strong anharmonicities
and are thus only loosely related to the kinetic theory discussed
here. For this reason we merely refer to a few review articles.
Jackson \cite{Ja} summarizes to work up to 1978, an authorative
2003 update being Lepri, Livi, and Politi \cite{LLP}. Bonetto,
Lebowitz and Rey-Bellet \cite{BLR} emphasize more theoretical
aspects, in particular large deviations and the fluctuation
theorem.

The numerical simulations available provide no clear evidence,
whether kinetic theory is applicable in dimension one (and two).
In the kinetic theory of gases collisions in one dimension are
degenerate, since particles just pass through each other. On the
other hand for lattice dynamics three phonon processes are
non-degenerate, as can be checked explicitly for the dispersion
relation (\ref{4a.4}). Therefore, in general, the collision
operator does not vanish. Ergodicity is more questionable. For our
standard example (\ref{4a.4}) at $\omega_0=0$ the components
$[-\frac{1}{2},0]$ and $[0,\frac{1}{2}]$ are not linked through
collisions. As $\omega_0$ increases these components shrink. In
their steady state the phonon current would not vanish. Other
couplings, four phonon processes, or thermal boundary drive could
restore ergodicity. Whether the microscopic model for small but
fixed anharmonicity has regular energy transport remains to be
studied. Only some loosely related results are available. Aoki and
Kusnezov \cite{AoK} numerically simulate the case $\omega_0=0$,
$\lambda=1$ and report good evidence for normal heat conduction,
i.e.~a steady state energy current proportional to $1/N$ with $N$
the chain length. For the same model the current-current momentum
and energy correlation functions are studied in \cite{L-DNG}. A
variety of other harmonic nearest neighbor chains with anharmonic
on-site potential is investigated in \cite{L}. Lefevre and
Schenkel \cite{LS} attempt to expand directly the steady state
probability distribution under thermal boundary conditions. They
report the term of order $\lambda$. From the perspective of
kinetic theory the term of order $\lambda^2$ would be related to
the chain length of order $\lambda^{-2}$.

%%%%%%%%%%%%%%%%%%%%%%%%%%%%%%%%%%%%%%%%%

\section{Appendix}\label{sec.16}
The proofs given below are due to J. Lukkarinen.
\subsection{Three phonon processes in case of nearest neighbor coupling
only}\label{sec.16a} \setcounter{equation}{0}

For nearest neighbor coupling the dispersion relation reads
\begin{equation}\label{16.1}
\omega(k)=\big(\omega^2_0+2\sum^3_{j=1}(1-\cos(2\pi
k^j))\big)^{1/2}\,.
\end{equation}
We prove that
\begin{equation}\label{16.2}
\omega(k)+\omega(q)-\omega(k+q)\geq \omega_0/2
\end{equation}
for all $q,k\in \mathbb{T}^3$. Therefore in this case three phonon
collisions are prohibited by energy conservation.\smallskip\\

We set $z=(z^1,z^2,z^3)$ and
\begin{equation}\label{16.3}
z^j(k)= i(\sqrt{a}-\frac{1}{\sqrt{a}}e^{-i2\pi k^j})\,,\;a>1\,.
\end{equation}
Then by direct computation $|z(k)|=\omega(k)$ with $\omega_0$
determined uniquely by $a$. We find
\begin{equation}\label{16.4}
|\omega(k+q)-\omega(k)|\leq \|z(k+q)|-|z(k)\|\leq|z(k+q)-z(k)|
\end{equation}
and
\begin{equation}\label{16.5}
|z(k+q)-z(k)|^2=\frac{1}{a}|z_0(q)|^2\leq |z_0(q)|^2
\end{equation}
with $z_0=z$ at $a=1$. Therefore
\begin{eqnarray}\label{16.6}
&&\omega(k)+\omega(q)-\omega(k+q)\geq
\omega(q)-|\omega(k+q)-\omega(q)|\nonumber\\
&&\hspace{40pt}\geq\omega(q)-|z_0(q)|=(\omega^2_0+|z_0(q)|^2)^{1/2}
-|z_0(q)|\geq\omega_0/2
\end{eqnarray}
for all $|z_0(q)|^2\geq 0$.

\subsection{Entropy as the logarithm of phase space volume}\label{sec.16c}

In \cite{GGL} Garrido, Goldstein, and Lebowitz argue that whenever
a suitable set of ``macrovariables" evolves in time according to
an autonomous deterministic law, then the entropy functional,
defined as the logarithm of the phase volume associated to
specified values of the macrovariables, is increasing in time. A
system of weakly interacting phonons should be no exception and we
will explain why.

Notationally it is convenient to choose the wave number torus as
$\mathbb{T}^3=[0,1]^3$. If the lattice volume is
$[1,\ldots,\ell]^3$, then the wave numbers are discretized as
$k\in(\mathbb{T}_\ell)^3=(\ell^{-1}[1,\ldots,\ell])^3$. We
partition the unit torus into cubes $\Delta_j$ of side length
$\delta$, $\delta M=1$, $j=1,\ldots,M^3$. Accordingly we set
\begin{equation}\label{16c.1}
H_j=\sum_{k\in\Delta_j\cap(\mathbb{T}_\ell)^3} a(k)^\ast a(k)
\end{equation}
as a function on phase space $(\mathbb{R}^6)^{\ell^3}$. The
$H_j$'s are the macrovariables. They are assumed to take a value
close to $\ell^3 e_j$ with
\begin{equation}\label{16c.2}
e_j=\delta^3\int_{\Delta_j}d^3k W(k)\,.
\end{equation}
Let ${\bf{e}}=(e_1,\ldots,e_{M^3})$. The corresponding region in
phase space is
\begin{equation}\label{16c.3}
\Gamma({\bf{e}},\delta,\nu)=\{(q,p)\in(\mathbb{R}^6)^{\ell^3}|\ell^3(e_j-\nu)\leq
H_j\leq\ell^3(e_j+\nu)\,,\;j=1,\ldots,M^3\}\,.
\end{equation}
Then, using the equivalence between mirocanonical and canonical
ensemble,
\begin{equation}\label{16c.4}
\lim_{\nu\to 0}\lim_{\ell\to\infty}\ell^{-3}\log
|\Gamma({\bf{e}},\delta,\nu)|=(\delta^3\sum^{M^3}_{j=1}\log e_j)
+\log\pi+1\,.
\end{equation}
If one now refines the partitioning into cubes by letting
$\delta\to 0$, one arrives at the entropy functional
\begin{equation}\label{16c.5}
\int_{\mathbb{T}^3} d^3k\big(\log W(k)+\log\pi+1\big)
\end{equation}
in accordance with (\ref{3.11}).

The quantum case is rather similar, once it is realized that the
operators from (\ref{16c.1}) are a family of commuting operators.
The conditions in (\ref{16c.3}) define a projection operator
$P({\bf{e}},\delta,\nu)$ on bosonic Fock space and
\begin{equation}\label{16c.6}
\lim_{\nu\to 0}\lim_{\ell\to\infty}\ell^{-3}\log \mathrm{tr}
P({\bf{e}},\delta,\nu)=(\delta^3\sum^{M^3}_{j=1}\big((1+e_j)\log
(1+e_j)-e_j\log e_j\big)\,.
\end{equation}
As before, upon refining the partition by letting $\delta\to 0$
one arrives at the entropy functional
\begin{equation}\label{16c.7}
\int_{\mathbb{T}^3} d^3k\big((1+W(k))\log (1+W(k))-W(k)\log
W(k)\big)
\end{equation}
in accordance with (\ref{7.8b}).

As to be expected on general grounds \cite{GGL}and as confirmed by
Propositon \ref{9.prop1}, the thus defined entropy is increasing
in time when $W(k,t)$ evolves according to the phonon Boltzmann
equation.

\subsection{The Brout-Prigogine equation}\label{sec.16d}

In the context of wave turbulence, over recent years the validity
of the Boltzmann transport equation has been scrutinized with the
aim to understand the necessity for corrections \cite{ZS,NNB,CLN}.
One part of the enterprise are numerical simulations testing the
validity of Gaussian local statistics \cite{CLNP,CLN-2}. In these
works the authors follow the Brout-Prigogine scheme
\cite{BrPr,Pr}. Since it differs from our approach, to comment on
their method might be instructive.

We consider the finite volume
$\Lambda=[1,\ldots,\ell]^3\subset\mathbb{Z}^3$. With periodic
boundary conditions our Hamiltonian reads
\begin{equation}\label{16d.1}
H= \frac{1}{2}\sum_{x\in\Lambda}p^2_x +
\frac{1}{2}\sum_{x,y\in\Lambda}\alpha^\mathrm{p}(x,y)q_x q_y +
\frac{1}{3}\sqrt{\varepsilon}\sum_{x\in\Lambda}q^3_x\,,
\end{equation}
where $\alpha^\mathrm{p}$ are the periodized elastic constants and
includes $\omega^2_0$. We now rotate $q$, and $p$, such that
$\alpha^\mathrm{p}(x,y)$ becomes diagonal. It has the eigenvalues
$\omega^2_k$ with $k\in\Lambda^\ast=(\ell^{-1}[1,\ldots,\ell])^3$,
the dual lattice. If $\widetilde{q}_k$, $\widetilde{p}_k$ denotes
the new coordinates and momenta, we further switch canonically to
action-angle variables through
\begin{equation}\label{16d.2}
q_k=(J_k/\omega_k)^{1/2}\cos\alpha_k\,,\;
\widetilde{p}_k=(\omega_k J_k)^{1/2}\sin\alpha_k\,,
\end{equation}
$0<J_k$, $\alpha_k\in 2\pi\mathbb{T}$. In action-angle variables
the Hamiltonian becomes
\begin{equation}\label{16d.3}
H=\sum_{k\in\Lambda^\ast}\omega_k J_k+\sqrt{\varepsilon}
H_1(J,\alpha)\,.
\end{equation}
The precise form of $H_1$ is easily worked out, but not needed for
our summary. The equations of motion are then
\begin{eqnarray}\label{16d.4}
\dot{\alpha}_k= \omega_k +\sqrt{\varepsilon}
\frac{\partial}{\partial J_k}H_1(J,\alpha)\,,\nonumber\\
\dot{J}_k=-\sqrt{\varepsilon}
\frac{\partial}{\partial\alpha_k}H_1(J,\alpha)\,.
\end{eqnarray}
Clearly, the $\omega$'s are the fast variables while the actions
change slowly.

We impose the initial distribution, $\rho(0)$, on phase space
which evolves under the flow (\ref{16d.4}) to $\rho(t)$. $\rho(0)$
is taken to depend only on $J$, the \textit{random phase
approximation}, and one is interested in the distribution of slow
variables at the kinetic time $\varepsilon^{-1}t$,
\begin{equation}\label{16d.5}
\rho^\varepsilon_t(J)=
\prod_{k\in\Lambda^\ast}\{(2\pi)^{-1}\int^{2\pi}_0
dx_k\}\rho(J,\alpha,\varepsilon^{-1}t)\,.
\end{equation}
Brout and Prigogine use second order perturbation theory for the
Liouvillean, which suggests that $\rho^\varepsilon_t(J)$ evolves
approximately by a diffusion process. The computation is very
readably explained in [28, pp.~36-60] and there is no need to
repeat. As net result they obtain a diffusion process on
$(\mathbb{R}_+)^{\ell^3}$ with generator
\begin{eqnarray}\label{16d.6}
Lf(J)=\gamma\sum_{k,k',k''\in\Lambda^\ast}(\omega_k\omega_{k'}\omega_{k''})^{-1}
\delta(\omega_k+\omega_{k'}-\omega_{k''})\delta(k+k'-k'')\nonumber\\
\big(\frac{\partial}{\partial J_k}+\frac{\partial}{\partial
J_{k'}}-\frac{\partial}{\partial J_{k''}}\big)J_k J_{k'}J_{k''}
\big(\frac{\partial}{\partial J_k}+\frac{\partial}{\partial
J_{k'}}-\frac{\partial}{\partial J_{k''}}\big)f(J)\,.
\end{eqnarray}

(\ref{16d.6}) is a system of coupled diffusion processes. If we
consider one representative triple,
$(J_1,J_2,J_3)\in(\mathbb{R}_+)^3$, then the diffusion process
$(J_1(t),J_2(t),J_3(t))$ moves along the line
$\{\big(J_1(0),J_2(0),J_3(0)\big)+\lambda(1,1,-1)$,
$\lambda\in\mathbb{R}\}$. The diffusion process never exits the
domain $(\mathbb{R}_+)^3$, since the diffusion coefficient,
$J_1J_2J_3$ vanishes sufficiently fast towards the boundary.

According to (\ref{16d.6}) the first moment evolves as
\begin{equation}\label{16d.7}
\frac{d}{dt}\langle J_k\rangle_t= \langle LJ_k\rangle_t\,.
\end{equation}
Taking $\ell\to\infty$ and assuming the factorization $\langle
J_{k'}J_{k''}\rangle_t=\langle J_{k'}\rangle_t \langle
J_{k''}\rangle_t$ one arrives at a closed equation for $\langle
J_k \rangle_t$. As a check on consistency, it indeed agrees with
the Boltzmann transport equation (\ref{4.7}).

The tricky part of the argument is the diffusion approximation
(\ref{16d.6}). For fixed $\ell$, the limit $\varepsilon\to 0$ in
(\ref{16d.4}) is covered by the perturbation theory for integrable
systems, see e.g.~\cite{AKN} Chapter 5. Even if the initial phases
are assumed to be random, there is simply no diffusion
approximation in sight. The motion of the angles is
quasi-periodic, thus much too regular for the purpose of
diffusion. One is forced to take with $\varepsilon\to 0$
simultaneously $\ell\to\infty$. Kinetic scaling requires
$\ell=\mathcal{O}(\varepsilon^{-1})$, which means to enter murky
waters. It remains to be seen whether there is some intermediate
scale on which (\ref{16d.6}) is a valid approximation.

To my understanding, the transformation to action-angle variables
easily misses the central physical mechanism for the validity of
the kinetic description. It is the wave propagation in physical
space, and its good spatial mixing properties, which ensures that
even in presence of a small nonlinearity the wave field retains
approximately the Gaussian statistics.

\subsection{Solutions to (\ref{9.21})}\label{sec.16b}

We set
\begin{equation}\label{16b.1}
\partial_\alpha\partial_\beta\psi(k)=A_{\alpha\beta}\,,\;
\partial_\alpha\partial_\beta\psi(k')=\widetilde{A}_{\alpha\beta}\,,\;
\partial_\alpha\partial_\beta\omega(k)=B_{\alpha\beta}\,,\;
\partial_\alpha\partial_\beta\omega(k')=\widetilde{B}_{\alpha\beta}\,.
\end{equation}
Then (\ref{9.21}) transcribes to
\begin{equation}\label{16b.2}
A_{\alpha\gamma}\widetilde{B}_{\beta\delta}+\widetilde{A}_{\alpha\delta}B_{\beta\gamma}=
A_{\beta\gamma}\widetilde{B}_{\alpha\delta}+\widetilde{A}_{\beta\delta}B_{\alpha\gamma}
\end{equation}
and we have to find out all possible solutions under the condition
that $B$ and $\widetilde{B}$ are invertible. We multiply in
(\ref{16b.2}) with $(B^{-1})_{\gamma\gamma'}$ and
$(\widetilde{B}^{-1})_{\delta\delta'}$ and sum over $\gamma$,
$\delta$. Let us define
\begin{equation}\label{16b.3}
C=AB^{-1}\,,\quad \widetilde{C}=\widetilde{A}\widetilde{B}^{-1}\,.
\end{equation}
Changing $\gamma'$, $\delta'$ back to $\gamma$, $\delta$ yields
\begin{equation}\label{16b.4}
C_{\alpha\gamma}\delta_{\beta\delta}+
\widetilde{C}_{\alpha\delta}\delta_{\beta\gamma}
=C_{\beta\gamma}\delta_{\alpha\delta}+
\widetilde{C}_{\beta\delta}\delta_{\alpha\gamma}\,.
\end{equation}

In (\ref{16b.4}) we choose indices $\alpha\neq\beta\neq\gamma$,
where it is used that $d\geq 3$, and we set $\delta=\alpha$, resp.
$\delta=\beta$. Then
$C_{\alpha\beta}=c_\alpha\delta_{\alpha\beta}$. Correspondingly
from $\alpha\neq\beta\neq\delta$ and $\gamma=\alpha$, resp.
$\gamma=\beta$, it follows that
$\widetilde{C}_{\alpha\beta}=\widetilde{c}_\alpha\delta_{\alpha\beta}$.
Thus
\begin{equation}\label{16b.5}
c_\alpha\delta_{\alpha\gamma}\delta_{\beta\delta}+
\widetilde{c}_\alpha\delta_{\alpha\delta}\delta_{\beta\gamma}=
c_\beta\delta_{\beta\gamma}\delta_{\alpha\delta}+
\widetilde{c}_\beta\delta_{\beta\delta}\delta_{\alpha\gamma}\,.
\end{equation}
Setting $\alpha=\beta=\gamma=\delta$ one concludes
$c_\alpha=\widetilde{c}_\alpha$ and setting $\alpha=\beta$,
$\alpha=\gamma$, $\beta=\delta$ one concludes $c_\alpha=c_\beta$,
$\alpha\neq\beta$. Combining both identities, there exist some
constant $a$ such that
\begin{equation}\label{16b.6}
C_{\alpha\beta}=a \delta_{\alpha\beta}\,,\quad
\widetilde{C}_{\alpha\beta}=a \delta_{\alpha\beta}
\end{equation}
and consequently, using (\ref{16b.3}),
\begin{equation}\label{16b.7}
A=aB\,,\quad \widetilde{A}=a \widetilde{B}\,.
\end{equation}

%%%%%%%%%%%%%%%%%%%%%%%%%%%%%%%%%%%%%%

\end{document}